\documentclass[preprint]{aastex}
\slugcomment{}

\shorttitle{The WHIM towards PG\,1259+593} \shortauthors{Richter et al.}

\begin{document}

\title{{\it FUSE} and {\it STIS} Observations of the Warm-Hot Intergalactic
Medium towards PG\,1259+593$^1$}

\author{Philipp Richter\altaffilmark{2},
Blair D. Savage\altaffilmark{3},
Todd M. Tripp\altaffilmark{4,5}
\and
Kenneth R. Sembach\altaffilmark{6}
}

\altaffiltext{1}{Based partly on observations with the NASA/ESA {\it Hubble Space
Telescope} obtained at the Space Telescope Science Institute, which is
operated by the Association of Universities for Research in Astronomy, Inc., under
NASA contract NAS\,5-26555.}
\altaffiltext{2}{Institut f\"ur Astrophysik und Extraterrestrische Forschung,
Universit\"at Bonn, Auf dem H\"ugel 71, 53121 Bonn, Germany} 
\altaffiltext{3}{Department of Astronomy, University of Wisconsin-Madison,
475 N. Charter St., Madison, WI 53706, USA} 
\altaffiltext{4}{Princeton University Observatory, Peyton Hall, Ivy Lane, Princeton, NJ\,08544, USA}
\altaffiltext{5}{Department of Astronomy, University of Massachusetts, Amherst, MA 01003, USA} 
\altaffiltext{6}{Space Telescope Science Institute, 3700 San Martin Drive, Baltimore, MD 21218, USA}

\begin{abstract}

We use {\it Far Ultraviolet Spectroscopic Explorer} (FUSE) and 
{\it Space Telescope Imaging Spectrograph} (STIS) spectra to study intergalactic 
absorption towards the quasar PG\,1259+593 ($z_{\rm em}=0.478$),
with a particular emphasis on the warm-hot intergalactic medium (WHIM).
The combined FUSE and STIS spectrum of PG\,1259+593 
covers the full wavelength range between $905$ and $1730$ \AA\, at a spectral 
resolution of $\sim 25$ km\,s$^{-1}$ for the FUSE bandpass ($\lambda 
\leq 1180$ \AA) and $\sim 7$ km\,s$^{-1}$ for the STIS range ($\lambda > 1150$ \AA). 
The signal-to-noise ratios (S/N) per resolution element are $\sim 10-30$
(FUSE) and $\sim 7-17$ (STIS).
We identify 135 intergalactic absorption lines with equivalent widths
$\geq 10$ m\AA, tracing 78 absorption components in 72 Ly\,$\alpha$/Ly\,$\beta$
absorption-line systems.
Metal-line absorption by species such as C\,{\sc iii},
C\,{\sc iv}, O\,{\sc iii}, O\,{\sc iv}, O\,{\sc vi}, and Si\,{\sc iii} is clearly
detected in four systems, and is possibly seen in four additional cases.
We study the distribution and physical properties of the WHIM as sampled by
O\,{\sc vi} and intrinsically broad Ly\,$\alpha$ lines.
The number of intervening O\,{\sc vi} absorbers for 
equivalent widths $W_{\lambda} \geq 24$ m\AA\, is
$3-6$ over an unobscured redshift path of $\Delta z\approx 0.368$.
This implies a number density of O\,{\sc vi} systems, $dN_{\rm OVI}/dz$,
of $\sim 8-16$ above this equivalent width limit along this sight line. A seventh
intervening O\,{\sc vi} absorber is possibly detected 
with $W_{\rm r}($O\,{\sc vi})$\approx15$ m\AA.
The range of $dN_{\rm OVI}/dz=8-16$ for $W_{\lambda} \geq 24$ m\AA\, 
is consistent with estimates from other sight lines, 
supporting the idea that intervening intergalactic O\,{\sc vi} absorbers contain an 
substantial fraction of the baryonic mass in the low-redshift Universe. We identify 
a number of broad Ly\,$\alpha$ absorbers with large Doppler parameters 
($b\sim40-200$ km\,s$^{-1}$) and low column densities ($N$(H\,{\sc i})$< 10^{14}$ cm$^{-2}$). 
For pure thermal broadening, these widths correspond to temperatures of
$\sim 1\times10^5-3\times10^6$ K.
While these broad absorbers could be caused by blends of multiple, unresolved lines, 
continuum undulations, or by kinematic flows and Hubble broadening, 
we consider the possibility that some of these
 features are single-component, thermally broadened Ly\,$\alpha$ lines. 
These systems could represent WHIM absorbers that are too weak, too metal-poor, 
and/or too hot to be detected in O\,{\sc vi}. If so, their widths and their frequency 
in the PG\,1259+593 spectrum imply that these absorbers trace an even larger fraction 
of the baryons in the low-redshift Universe than the O\,{\sc vi} absorbing systems.
A thermal Doppler broadening explanation for one of these broad features is supported
by the probable detection of O\,{\sc vi} near the velocity
of a broad Ly\,$\alpha$  and Ly\,$\beta$ absorber with an O\,{\sc vi} 
line width $\sim 4$ times smaller
than for H\,{\sc i}.

\end{abstract}

\section{Introduction}

The analysis of absorption lines in the local and distant intergalactic 
medium (IGM) towards extragalactic background sources such as
quasars (QSOs) and active galactic nuclei (AGNs)
has become an important research area to investigate the gaseous environment of 
galaxies and galaxy-clusters, and to study the large scale structure, the chemical 
evolution and the baryonic content of the IGM.
The most sensitive tracer to follow the distribution and evolution
of the IGM from high to low redshifts is the Ly\,$\alpha$ line of neutral hydrogen. 
A large number of Ly\,$\alpha$ lines occur along a typical QSO
sight line, resulting in the ``Ly\,$\alpha$ forest'' (e.g., Rauch 1998; Lu et al.\,1996). 
For the low-redshift Universe, the observation of intergalactic Ly\,$\alpha$ absorption 
requires space-based instrumentation, such as
the former {\it Goddard High Resolution Spectrograph} (GHRS)
and the more recently deployed {\it Space Telescope Imaging Spectrograph} (STIS), both
installed on the {\it Hubble Space Telescope} (HST).
For higher redshifts ($z>1.6$), ground based 8\,m-class telescopes such as Keck
and the {\it Very Large Telescope} (VLT)
provide a wealth of information about the Ly\,$\alpha$ forest
in the early Universe (e.g., Kim et al.\,2002). It is generally accepted 
that at $z=3$ almost all of the baryons in the Universe are located in the highly-ionized 
Ly\,$\alpha$ forest (Rauch et al.\,1997; 
Weinberg et al.\,1997), whereas at lower redshifts an increasing
fraction of the baryons may reside in a low-density, shock-heated component at 
temperatures between $10^5$ to $10^7$ K, the so-called warm-hot intergalactic medium (WHIM).
In the local Universe (i.e., near $z=0$), the WHIM may contain
as much as 30 to 40 percent of the baryons (e.g., Cen \& Ostriker 1999).
In comparison, the Ly\,$\alpha$ forest at $z=0$ probably contains about 
20 percent of the baryons (Penton, Shull \& Stocke 2000). Detecting the WHIM 
phase at low redshifts is a challenging task.
Absorption spectroscopy of the O\,{\sc vi} doublet 
($\lambda\lambda 1031.9, 1037.6$) currently 
provides one of the best diagnostics to study the 
warm phase ($10^5-10^6$ K) of the WHIM at low redshift. 
However, observing O\,{\sc vi} in the low-redshift IGM 
requires space-based high-resolution spectrographs
such as STIS and the {\it Far Ultraviolet Spectroscopic Explorer} (FUSE)
together with a substantial amount of observing time
(e.g., Tripp, Savage \& Jenkins 2000; Savage et al.\,2002).
Due to these instrumental restrictions, the number
of sight-lines that have been analyzed for O\,{\sc vi} absorption at $z<1$
unfortunately is still very limited. So far, intervening WHIM O\,{\sc vi} 
absorption for $z<1$ in FUSE and STIS data has been
found towards H\,1821+643 (Tripp, Savage \& Jenkins 2000; 
Oegerle et al.\,2000) , PG\,0953+415 
(Tripp \& Savage 2000; Savage et al.\,2002), and
PKS\,0405$-$123 (Chen \& Prochaska 2000). 
Together with the non-detections or 
marginal and weak detections reported for the
sight lines towards PG\,0804+761 (Richter et al.\,2001a)
and 3C\,273 (Sembach et al.\,2001), the current
data imply a number of intervening O\,{\sc vi} absorbers
per unit redshift of $dN_{\rm OVI}/dz=14^{+9}_{-6}$ for restframe equivalent 
widths $W_r \geq 50$ m\AA\, at 
$\langle z \rangle =0.09$ (Savage et al.\,2002). Assuming that the average 
metallicity of these absorbers is 0.1 solar, the cosmological mass density in the
units of the current critical density is $\Omega_b($O\,{\sc vi}$)\geq 0.002 h_{75}^{-1}$. 
Clearly, additional observations with FUSE and STIS are desired to
improve the statistics on the intervening O\,{\sc vi} absorbers, and to gain new 
insights about their physical properties
(e.g., their ionization properties) and their metal abundances.
For those absorbers that cannot be detected in
O\,{\sc vi} absorption (e.g., if the overall column density is too low, 
the metallicity is too low, or the temperature is too high), other tracers have 
to be used to study the warm-hot intergalactic gas phase. X-ray observations with
{\it Chandra} and {\it XMM-Newton} (e.g., Rasmussen, Kahn, \& Paerels 2003; Fang et al.\,2002)
hold the prospect of improving
our understanding of the distribution and physical 
properties for the million-degree phase of the WHIM, which 
may contain the dominating fraction of the baryons at low redshifts. 
However, detecting
X-ray signatures such as O\,{\sc vii} and
O\,{\sc viii} absorption from the low-redshift IGM
is difficult with current instrumentation, and
so far the information about the intergalactic X-ray forest
is relatively sparse.
For high-metallicity, high-column density absorbers,
one may search for
FUV absorption by Ne\,{\sc viii} $\lambda\lambda 
770.4, 780.3$. 
But due to the relatively low cosmic abundance of Ne,
Ne\,{\sc viii} is not expected to be a sensitive tracer
for the WHIM in the low-redshift Universe.
A promising approach
to study the WHIM at low redshifts is to search for broad Ly\,$\alpha$ lines. 
For pure thermal broadening, Ly\,$\alpha$ absorbers with Doppler parameters 
exceeding $b\approx40$ km\,s$^{-1}$ sample WHIM gas with
temperatures $T>10^5$ K. There are some compelling detections of 
broad Ly\,$\alpha$ lines at low redshifts (e.g., Tripp et al.\,2001), but
generally broad Ly\,$\alpha$ lines are difficult to identify and analyze due
to continuum placement uncertainties, line-blending problems, and relatively
low signal-to-noise ratios in existing data.

In this paper we investigate intervening intergalactic absorption by H\,{\sc i}, 
O\,{\sc vi} and other species with FUSE and STIS in direction of the
bright QSO PG\,1259+593 ($V=15.84$; $z_{\rm em}=0.478$; $l=120\fdg6$, 
$b=+58\fdg1$). FUSE and STIS data of PG\,1259+593 have been extensively used to 
study abundances and physical conditions in Galactic halo gas 
in this direction (Richter et al.\,2001b; Collins, Shull \& Giroux 2003;
Sembach et al.\,2004; Fox et al.\,2004). With more than
$600$ ks of integration time it represents one of the prime objects of the FUSE mission.
The line of sight towards PG\,1259+593 samples several strong Ly\,$\alpha$ systems, 
as shown by low-resolution {\it Faint Object Spectrograph} (FOS) data
(Bahcall et al.\,1993). The goal of this paper
is to present an overview of the intergalactic absorption towards PG\,1259+593, 
to identify the main absorption systems, to perform a first, basic analysis of these 
absorbers, and to search for the WHIM absorbers that may contain an important 
fraction of the baryonic mass in the low-redshift Universe.
The outline of this paper is the following: in
\S2 we review the observations and the data reduction
process. The analysis of intergalactic absorption
is presented in \S3. In \S3 we also describe the properties of the 
individual absorption systems and derive
statistical information about the IGM towards PG\,1259+593. 
We discuss our results in \S4 and summarize our study in \S5.

\section{Observations and Data Reduction}

\subsection{FUSE Observations}

FUSE observations of PG\,1259+593 were carried out 
between February 2000 and March 2001 in various individual observation-runs. 
For a detailed description
about the FUSE instrument and its performance 
in space see Moos et al.\,(2000) and Sahnow et al.\,(2000).
Table 1 provides basic information
about the individual FUSE data sets. PG\,1259+593
was observed through the large (LWRS, $30\farcs0 \times 30\farcs0$)
aperture in the photon-address mode, which stores
X/Y location, arrival time, and pulse-height of each detection as a photon list. 
The 236 exposures of PG\,1259+593 provide a total of $\sim 630$ ks of integration time. 
The raw data were processed with the standard FUSE reduction pipeline ({\tt CALFUSE} v2.2.2). 
Extra care was taken to screen the raw data for possible particle event bursts, Earth limb
avoidance, South Atlantic Anomaly passage, and pulse height distribution constraints. 
Summed exposure lists for
each of the four FUSE channels (LiF\,1, LiF\,2, SiC\,1, SiC\,2) were created and 
corrected for geometric distortions, Doppler shifts, grating motions, astigmatism, 
detector backgrounds, and scattered light effects. The data were further flux calibrated 
and wavelength calibrated to a heliocentric velocity scale. The zero point 
of the FUSE wavelength scale was adjusted to that of the (more accurate)
STIS wavelength scale by comparing the velocity scale of narrow interstellar
absorption lines from atomic
species such as Fe\,{\sc ii} and Si\,{\sc ii} in both the FUSE and the
STIS wavelength range (see Sembach et al.\,2004 for details). 
The average flux in the FUSE spectrum, 
which covers the wavelength range between 905 and 1187 \AA,
is $\sim 2\times10^{-14}$ erg\,cm$^{-2}$\,s$^{-1}$\, \AA$^{-1}$. 
Note that the current FUSE flux calibration for PG\,1259+593 does not account for 
the occurrence of dark horizontal stripes in the 
detector images that reduce the total flux in specific
wavelength regions (see Fig.\,1). The spectral resolution in the FUSE data
is $\sim22-25$ km\,s$^{-1}$ at a rebinned pixel size of $\sim 7.5$ km\,s$^{-1}$. The signal-to-noise 
ratio (S/N) in the night+day FUSE data of PG\,1259+593 data varies between 
$10$ and $30$ per spectral resolution element. 
A more detailed description of the FUSE data reduction of 
PG\,1259+593 is provided in Sembach et al.\,(2004).

\subsection{STIS Observations}

STIS observations of PG\,1259+593 were obtained
in January and December 2001 (see Table 1) using
the intermediate-resolution (FWHM$\sim 7$ km\,s$^{-1}$) echelle 
far-UV grating (E140M) and the $0\farcs2 \times 0\farcs06$ slit. The STIS 
data cover the wavelength range between $1150$ and $1729$ \AA\, 
with small gaps between the echelle orders at wavelengths $>1634$ \AA. 
The pixel size in the data is $\sim 3.5$ km\,s$^{-1}$. Detailed
information about the STIS instrument is provided
by Woodgate et al.\,(1998), Kimble et al.\,(1998) and Leitherer et al.\,(2002).
The PG\,1259+593 STIS data were reduced following the
individual steps described by Tripp et al.\,(2001). The data reduction 
includes a two-dimensional scattered light correction 
(Landsman \& Bowers 1997; Bowers et al.\,1998). Individual 
extracted flux-calibrated spectra were co-added together 
(weighted by their inverse variances averaged over a large region
of the spectrum)
to produce the final composite spectrum in
a heliocentric velocity scale. Hot pixels that occasionally
occur in the calibrated STIS data have been corrected for by 
interpolating between adjacent pixels.
The resulting S/N per resolution element in the STIS data 
varies between $\sim 7$ and $17$. The highest S/N region 
in the STIS data occurs near $1400$ \AA.

\subsection{Data Analysis}

The combined FUSE and STIS spectrum of PG\,1259+593 is shown in Fig.\,1.
More than 200 absorption features are seen
in the FUSE and STIS data of PG\,1259+593; these features are identified in 
Figs.\,2$-$12. Wavelengths and oscillator strengths ($f$ values) have been 
adopted from the compilations of Morton (2003, 1991) and Verner et al.\,(1994);
they are listed in Table 2.
Since both resolution and S/N is varying along the spectrum, typical $3\sigma$ detection limits
range from $\sim 15$ to $\sim 50$ m\AA. In a first step we have separated intergalactic
absorption lines from local foreground absorption by the interstellar medium in
the disk and halo of the Milky Way. The interstellar spectrum (FUSE and STIS) in
the direction of PG\,1259+593 is discussed in detail by Richter et al.\,(2001b)
and Sembach et al.\,(2004). Interstellar absorption consists of three main velocity 
components located near $v_{\rm LSR}=-2$, $-55$ and $-129$ km\,s$^{-1}$ 
(Sembach et al.\,2004), representing local Milky Way disk gas, gas from the 
Intermediate Velocity Arch (IV Arch), and gas from high-velocity cloud Complex C, 
respectively (note that $v_{\rm LSR}=v_{\rm helio}+10.5$ km\,s$^{-1}$ in this direction). 
Each of these velocity components have to be considered carefully for possible
blending problems with intergalactic absorption.
Stronger interstellar absorption features (with equivalent widths $\geq 30$ m\AA, typically)
have been marked with star symbols in the spectrum plots of PG\,1259+593 in Figs.\,2$-$12.
After dealing with the interstellar absorption features, we have identified 135 intergalactic
absorption lines from 78 individual absorption components in 72 different absorption systems (see next
section for details). We have measured
wavelengths and redshifts by fitting Voigt profiles (convolved with Gaussian instrumental
profiles corresponding to the spectral resolution of the two instruments; see \S2.1 and \S2.2) to
the absorption lines. Normalized velocity profiles have been created by fitting low-order
polynomials to the (generally slightly varying) continuum flux of PG\,1259+593 and transforming
each absorption profile into a restframe velocity scale. Restframe equivalent widths, $W_{\rm r}$,
have been measured for each line by a pixel integration over the normalized velocity profile.
Column densities have been derived by three different methods: if multiple lines from a
species are available in a system (e.g., multiple H\,{\sc i} Lyman series lines), we have
constructed curves of growth (COG) from the measured equivalent widths and derived
logarithmic column densities, log $N$, and Doppler-parameters ($b$ values). 
For these multi-line systems, the various
Lyman series lines are located in both
the higher-resolution STIS wavelength band and the
lower-resolution FUSE regime, thus in regions where
different instrumental line spread
functions apply.
Here, the COG analysis (which requires only the
equivalent width for each line as input parameter)
represents a very robust method
to determine column densities and $b$ values for
this inhomogeneous set of absorption lines.
Moreover, the COG method allows to directly visualize
the influence of outliers (e.g., lines that are partly
blended) for deriving log $N$ and $b$ for a multi-line
system, and thus provides an important tool to estimate
realistic uncertainties for these parameters.
In some systems,
we have fitted single lines from selected ions to curves of growth with $b$ values previously
derived for other species.
In addition, we have used the apparent optical depth (AOD) method (see Savage \& Sembach 1991
for a detailed description) in a number of cases to derive logarithmic column densities by
integrating over the normalized velocity profiles of
mostly unsaturated lines. Finally, we have determined log $N$ and $b$ by 
Voigt-profile fitting of high S/N IGM lines located in the
higher resolution STIS wavelength range. This method was preferred to
derive column densities and $b$ values for H\,{\sc i} in the
metal-line systems and the Ly\,$\alpha$ forest, except for those absorbers
that consist of a large number of Lyman series lines
and for which the COG method was used.

Our measurements are compiled in Tables $3-6$. 
Formal $1\sigma$ errors for $W_{\rm r}$, log $N$,
and $b$ are
based on photon-statistics and continuum placement constraints.
The combined use of the different methods for deriving $N$ allows to check upon
the possible existence of systematic errors that cannot be accounted for in the formal
error estimate for each method. 
Such possible systematic errors could be unexpected continuum undulations,
unresolved and unidentified line blending, unknown instrumental features, and
fixed-pattern noise. Values for log $N$ derived by the different methods
usually agree well within their formal $2\sigma$ error range (see Table 3).
However, for the Ly\,$\alpha$ and Ly\,$\beta$ forest lines a reliable identification
and analysis of weak and broad absorption features is particularly difficult if such
effects are present. In addition to the formal errors provided in Tables 3, 5 and 6
we therefore give an additional ``reliability status'' 
for those lines for which we see evidence that the
estimates for log $N$ and $b$ (and their formal errors) could be incorrect due to one
of the effects listed above. Typical problematic cases are: stronger lines that show
asymmetries or weak wings, weak lines that have unidentified features and undulations
in the adjacent continuum, and apparently broad lines that show evidence that they are
composed of several, unresolved sub-components. In Table 3 we flag those lines with
the note ``uncertain'' in the last column. Similarly, we list in 
the seventh column of Table 5 the status of a
Ly\,$\alpha$ forest line as either OK or UC (uncertain).

\section{Intergalactic Absorption}

\subsection{Overview}

For the 135 intergalactic absorption lines detected towards PG\,1259+593  we
have identified 72 corresponding absorption systems at redshifts between
$0.002$ and $0.436$. Note that some of the identifications have to be
considered as tentative due to the fact that some features have low S/N and/or they
occur in regions where line blending problems do not allow unambiguous
identifications. These lines have been marked with
questions marks in Figs.\,2$-$12. Six of the 72 systems consist of
two sub-components each, so that the spectrum consists of 78 IGM absorption components in total.
For these sub-components (in Tables 3, 4 and 5 given as ``Component A'' and 
``Component B'') we have measured column densites and $b$ values separately.
Of the identified 72 systems, four to eight 
are metal line systems detected in one or more lines of
C\,{\sc iii}, C\,{\sc iv}, O\,{\sc iii}, O\,{\sc iv}, O\,{\sc vi}, Si\,{\sc iii}, 
and possibly Ne\,{\sc viii}. These systems will be discussed in detail in \S3.2. 
Rest-frame equivalent widths, column densities,
$b$ values, and other information for each system are listed in
Table 3. Table 4 summarizes the column densities of the metal-line absorbers
(note that all metal line systems detected
towards PG\,1259+593 have
H\,{\sc i} column densities log $N$(H\,{\sc i})$\geq13.6$).
Fourteen additional absorption systems are detetected
solely in H\,{\sc i} in two or more lines from the Lyman series 
(Ly\,$\alpha -$ Ly\,$\epsilon$), but without corresponding metal lines. Also for
these systems we list restframe equivalent widths, column densities,
etc. in Tables 3 and 4, while their properties are discussed
in \S3.4. Finally, we find 50 absorbers detected only in one single line, which
we generally assume to be Ly\,$\alpha$. Since
the given wavelength limit ($\lambda \leq 1729$ \AA) does not allow
us to detect Ly\,$\alpha$ absorption for redshifts $0.422\leq z \leq0.478$, some
of the lines between 1458\AA\, and 1516\AA\, identified as Ly\,$\alpha$ 
may be actually Ly\,$\beta$ belonging to Ly\,$\alpha$ absorption outside the 
wavelength range of the STIS data. Without having additional high-resolution spectral 
data for $\lambda>1729$ \AA, however, we are not able to clarify
the line identification for this redshift range at this point. The Ly\,$\alpha$/Ly\,$\beta$ 
forest is discussed in detail in \S3.4 and \S3.5.
Redshifts, logarithmic H\,{\sc i} column densities and $b$ values of the 50 systems 
seen solely in Ly\,$\alpha$ absorption are given in Table 5.
There is a good qualitative match
between the high-resolution STIS data presented
here and the low resolution (FWHM$\sim 1-2$ \AA) FOS data of PG\,1259+593 
(Bahcall et al.\,1993). Bahcall et al.\,identified eight strong Ly\,$\alpha$ absorbers 
(their Table 6) along the total redshift path towards PG\,1259+593. 
These identifications coincide with the absorption systems at
$z=0.04606$, $0.08933$, $0.21949$, $0.22313$, $0.28335$, $0.29236$,
$0.43148$ and $0.43569$
detected in the STIS spectrum (see Table 4).

\subsection{Metal-Line Systems}

\subsubsection{The System at $z=0.00229$}

This system (see Fig.\,13) has the lowest redshift of all absorbers and is most likely
related to the galaxy UGC\,08146 ($z=0.00224$; see also \S3.6.2). The absorber is 
detected in Ly\,$\alpha$ and Ly\,$\beta$, and
possibly in O\,{\sc vi} $\lambda 1031.9$. The Ly\,$\alpha$ absorption of this systems 
lies in the wing of the strong interstellar Ly\,$\alpha$ absorption and therefore has a 
relatively low S/N. A profile fit yields log $N$(H\,{\sc i})$=13.57\pm0.10$ and 
$b=42.1\pm4.4$ km\,s$^{-1}$. 
Using the relation between the Doppler parameter, $b$, and the temperature, $T$,

\begin{equation}
b=0.129\sqrt{\frac{T}{A}}\,{\rm km\,s}^{-1}
\end{equation}

\noindent
(with $A$ being the atomic weight for the element),
this $b$ value corresponds to $T\leq1.1\times10^5$ K.
The temperature could be significantly lower than $1.1\times10^5$ K
if effects like turbulent gas motions, gas flows and unresolved sub-component
structure contribute to the total breadth of the line
\footnote{We will use equation (1) throughout this paper to
estimate upper limits for $T$ from measurements of
Doppler parameters.}.
O\,{\sc vi} $\lambda 1031.9$ absorption is possibly
detected with log $N$(O\,{\sc vi})$=13.73\pm0.14$. 
Using the formal error calculation, the significance
of the line is $2.9\sigma$ (see \S2.3 and Table 3); however,
the absorption appears to be exceptionally broad
(ranging from approximately $-110$ to $+120$ km\,s$^{-1}$)
and we thus cannot exclude that this feature is an unexpected
continuum undulation rather than O\,{\sc vi} at $z=0.00229$.
We regard this case as a tentative intervening O\,{\sc vi} detection. 
If the feature is produced by O\,{\sc vi}, the absorption may have a 
multicomponent structure with the part of the absorption between $-100$ and $+50$
km\,s$^{-1}$ associated with the H\,{\sc i} Ly\,$\alpha$
absorption, and the more extended
absorption from $+50$ to $+120$ km\,s$^{-1}$ occurring where H\,{\sc i} absorption
is not detected. Such multicomponent O\,{\sc vi} and H\,{\sc i} absorbers have
been seen in STIS spectra of the QSO H\,1821+643 ($z_{\rm em} = 0.297$) by Tripp
et al.\,(2000).
There is a possible H\,{\sc i} absorption component near $-240$ km\,s$^{-1}$ 
seen in Ly\,$\alpha$, but its significance is low ($\sim 2.7\sigma$) due to the 
low S/N in the vicinity of the ISM
Ly\,$\alpha$ absorption. At this radial velocity, the feature would have 
$z\approx0.0015$ or $v_{\rm rad}=+450$ km\,s$^{-1}$ (in the heliocentric 
velocity frame), and thus might correspond to gas located in the Local Group. 
Unfortunately, possible O\,{\sc vi} $\lambda1031.9$ absorption at this velocity
is totally blended by Ly\,$\beta$
at $z=0.00760$ (see next subsection).

\subsubsection{The System at $z=0.00760$}

The absorption system at $z=0.00760$ (see Fig.\,13) is detected in 
H\,{\sc i} Ly\,$\alpha$, Ly\,$\beta$, and possibly O\,{\sc vi} $\lambda 1031.9$.
From a profile fit of the Ly\,$\alpha$ absorption we obtain 
log $N$(H\,{\sc i})$=14.05\pm0.05$ and $b=34.6\pm2.0$ km\,s$^{-1}$. The
$b$ value implies  $T\leq7.2\times10^4$ K (see previous subsection).
The Ly\,$\alpha$ absorption shows a wing at positive velocities,
suggesting the presence of a weak second component
near $+60$ km\,s$^{-1}$. Weak O\,{\sc vi} $\lambda 1031.9$ 
absorption is possibly
detected in this system near $1039.8$ \AA\, at a column
density of log $N($O\,{\sc vi}$)=13.06\pm0.04$ (AOD method). 
The formal significance
of this detection is $3.8\sigma$ for the
LiF\,1A channel and $1.8\sigma$ for the
LiF\,2B channel
(see Table 3). However, the absorption unfortunately 
occurs in a wavelength region of the FUSE instrument 
where detector features are commonly observed
(in particular in the LiF\,1A channel).
With an equivalent
width of $W_{\rm r}\approx 15$ m\AA\,for 
the O\,{\sc vi} $\lambda 1031.9$ line, 
no detectable absorption is expected (and observed) to occur
in the weaker O\,{\sc vi} $\lambda 1037.6$ line which
could confirm the presence of O\,{\sc vi} 
at $z=0.00760$. Due to the very small equivalent width
of the possible O\,{\sc vi} absorption 
and the possible presence of detector features in the
region where O\,{\sc vi} at $z=0.00760$ is expected we 
regard this system as a tentative intervening 
O\,{\sc vi} absorber.

\subsubsection{The System at $z=0.04606$}

This is one of the the two more complex metal-line systems 
towards PG\,1259+593 (together with the system at $z=0.21949$). 
Absorption is seen in the first eight
lines of the H\,{\sc i} Lyman series (Ly\,$\alpha$ to Ly\,$\theta$),
and in C\,{\sc iii}, Si\,{\sc iii}, C\,{\sc iv}, and O\,{\sc vi} (Fig.\,14 and Table 3). 
A two-component structure is present 
(see, for instance, the absorption profiles of H\,{\sc i}
Ly\,$\beta$ and C\,{\sc iii}), with
the stronger component at $0$ km\,s$^{-1}$ in the
$z=0.04606$ restframe (Component A)
and the weaker one near $+95$ km\,s$^{-1}$ (Component B).
Both components are present in H\,{\sc i} Ly\,$\alpha -$Ly\,$\delta$,
C\,{\sc iii}, and O\,{\sc vi} $\lambda 1031.9$, but
for Si\,{\sc iii} and  C\,{\sc iv} $\lambda 1548.2$
line blends hamper the analysis of the
positive-velocity component.
The profiles of the H\,{\sc i} lines reveal internal substructure within
Component A.  Component B appears to be a combination of a narrow and
broad absorption components.
With an H\,{\sc i} column density of log $N$(H\,{\sc i})$=15.81\pm0.08$
and a $b$ value of $33.2^{+6.5}_{-5.1}$ km\,s$^{-1}$ (from a curve-of-growth
analysis; see Table 3)
Component A represents the strongest IGM Ly\,$\alpha$ absorber 
along this sight line. Having eight H\,{\sc i} lines available in this system and
a two-component structure that introduces significant
uncertainties for the Voigt-profile fitting, 
we prefer the curve-of-growth technique to obtain values for log $N$ and $b$
for the H\,{\sc i} in this
absorption system. The logarithmic column densities in Component A for the two
intermediate ions C\,{\sc iii} and Si\,{\sc iii} are $13.80\pm0.04$ and $12.74\pm0.07$,
respectively (AOD method).
The C\,{\sc iv} absorption 
(log $N$(C\,{\sc iv})$=13.61\pm0.08$; AOD) appears to be
quite narrow, and is offset towards positive velocities
(near $+25$ km\,s$^{-1}$) compared to the low ions and the H\,{\sc i}
($0$ km\,s$^{-1}$). Note that
unresolved saturation may possibly be affecting these C\,{\sc iii}, Si\,{\sc iii},
and C\,{\sc iv} column density estimates.
The O\,{\sc vi} $\lambda 1031.9$ absorption in
Component A (log $N$(O\,{\sc vi})$=13.68\pm0.05$; AOD) is 
broader than the C\,{\sc iv} absorption
and merges at positive velocities into an almost equally strong O\,{\sc vi} absorption
in Component B (see below). The O\,{\sc vi} $\lambda 1037.6$ 
line unfortunately has very low S/N (only the less sensitive SiC\,1A channel
is available for this line; see Fig.\,4) and thus does not provide 
any further information on the O\,{\sc vi} absorption in either Component A or Component B.
The $b$ value for the H\,{\sc i} Lyman series in Component A ($33.2$ km\,s$^{-1}$)
suggests $T\leq6.5\times10^4$ K.
The presence of C\,{\sc iii}, Si\,{\sc iii}, C\,{\sc iv}, and O\,{\sc vi} absorption
implies that Component A is a multi-phase absorber, possibly having a 
cooler interior surrounded by a high-temperature envelope. 
Due to the complexity of this system and the relatively 
low S/N in the data we refrain from performing 
a more detailed modelling of Component A.

The H\,{\sc i} column density for Component B is
log $N$(H\,{\sc i})$=14.56\pm0.15$ (COG), thus significantly lower than for Component A.
While the lines of Si\,{\sc iii} $\lambda 1206.5$ and  C\,{\sc iv} 
$\lambda 1548.2$ are blended by other IGM lines, C\,{\sc iii} absorption
is clearly present (log $N$(C\,{\sc iii})$=13.42\pm0.04$; AOD method). Also
the somewhat noisy C\,{\sc iv} $\lambda 1550.8$ line shows a weak absorption component near
$+95$ km\,s$^{-1}$. This feature, however, corresponds only to a $1.7\sigma$ significance,
so that we only can place an upper limit of log $N$(C\,{\sc iv})$\leq 13.64$.
O\,{\sc vi} $\lambda 1031.9$ absorption in Component B
(log $N$(O\,{\sc vi})$=13.63\pm0.06$; AOD) is nearly as strong
as in Component A (again, no significant absorption is seen
in the much noisier O\,{\sc vi} $\lambda 1037.6$ line; see above). 
The values of $N($H\,{\sc i}$)/N($O\,{\sc vi}) in Components A and B are 135 and 8.5,
respectively. This ratio has been observed to range from $\sim 0.1$
to $100$ which suggests that although H\,{\sc i} and O\,{\sc vi} are observed to be
coupled kinematically, the two species often arise in quite different
gaseous phases (Shull 2003).

\subsubsection{The System at $z=0.21949$}

This is the second of the two more complex 
metal-line absorption systems (the $z=0.04606$ system
being the other). Absorption is seen in eight lines of the H\,{\sc i} Lyman
series (Ly\,$\alpha-$Ly\,$\epsilon$ and Ly\,$\eta-$Ly\,$\iota$; Ly\,$\zeta$ is blended
by Galactic N\,{\sc i} $\lambda 1134.3$), as well as in
C\,{\sc iii}, O\,{\sc iii}, Si\,{\sc iii}, O\,{\sc iv}, and
O\,{\sc vi} (Fig.\,16; note that H\,{\sc i} Ly\,$\iota$ has not been included in the plot).
Absorption by N\,{\sc iv} $\lambda 765.1$ is possibly also detected.
However,  N\,{\sc v} $\lambda\lambda 1238.8,1242.8$ and 
Si\,{\sc iv}  $\lambda\lambda 1393.8, 1402.8$
are not detected with relatively large upper limits to
their rest-frame equivalent widths that do not provide
any useful information.
The strong Ly\,$\alpha$ line has a $-50$ to $-100$ km\,s$^{-1}$ wing,
suggesting another absorption component at negative velocities.
This feature is too weak to be present in any other of the
lines from the H\,{\sc i} Lyman series. The left wing
of H\,{\sc i} Ly\,$\beta$ is blended by Galactic S\,{\sc ii} absorption.
The H\,{\sc i} column density of the main component (at $0$ km\,s$^{-1}$)
is log $N$(H\,{\sc i})$=15.25\pm0.06$ (COG method; see Table 3 for details on
the individual methods), thus being the second strongest IGM absorber
towards PG\,1259+593. From a Voigt profile fit of the Ly\,$\beta$ line we
obtain $b=32.3\pm1.4$ km\,s$^{-1}$, implying $T\leq6.3\times10^4$ K.
The logarithmic column densities of C\,{\sc iii} (ionization potential is $\sim 48$ eV) and
O\,{\sc iii} (ionization potential is $\sim 55$ eV) are $13.62\pm0.13$ (Voigt profile
fit) and $13.82\pm0.06$ (AOD method), respectively.
For O\,{\sc iv} we derive a column 
density of log $N$(O\,{\sc iv})$=14.27\pm0.05$ (AOD method).
Unresolved saturations may affect the C\,{\sc iii} and 
O\,{\sc iv} absorption,  so that the true values of $N$(C\,{\sc iii}) 
and $N$(O\,{\sc iv}) could be somewhat larger than those listed.
The Si\,{\sc iii} absorption at $z=0.21949$ is weak and relatively narrow. 
From a profile fit we obtain log $N$(Si\,{\sc iii})$=12.04\pm0.07$ and
$b=7.9\pm2.0$ km\,s$^{-1}$ (see Table 3).
N\,{\sc iv} absorption is marginally detected at a $1.9 \sigma$ level near $933$ \AA, thus
in a region where the FUSE spectrum is relatively noisy. We derive a $3\sigma$ upper limit
for the N\,{\sc iv} column density of log $N$(N\,{\sc iv})$\leq 13.35$.
While the absorption of the intermediate ions
(C\,{\sc iii}, O\,{\sc iii}, Si\,{\sc iii}, O\,{\sc iv}, and N\,{\sc iv}) is like H\,{\sc i}
centered at zero velocities, the O\,{\sc vi} shows a very different
absorption pattern. Two O\,{\sc vi} absorption components are visible in the 
stronger O\,{\sc vi} $\lambda 1031.9$ line
near $0$ and $-50$ km\,s$^{-1}$. The negative velocity component possibly 
relates to the negative velocity wing seen in the Ly\,$\alpha$ absorption.
The weaker O\,{\sc vi} $\lambda 1037.6$ line 
is consistent with this two-component structure. The double-component O\,{\sc vi}
absorption is very symmetric, the entire structure
having a center velocity near
$-25$ km\,s$^{-1}$. It could possibly be related to
an expanding shell or a symmetric outflow. A two-component
Voigt profile fit yields identical column densities for both components 
(log $N$(O\,{\sc vi})$=13.68\pm0.06$), and very similar $b$ values
($16.2\pm2.4$ and $15.5\pm1.5$ km\,s$^{-1}$; see also Table 3). The total
O\,{\sc vi} column density is log $N$(O\,{\sc vi})$\approx 14.0$. For pure thermal
broadening, the individual $b$ values correspond to temperatures of $T\approx2.5\times10^5$
and $2.3\times10^5$ K, respectively, thus very close to the peak temperature for O\,{\sc vi}
in collisional ionization equilibrium 
($2.8\times10^5$ K; Sutherland \& Dopita 1993).

With the simultaneous occurrence of O\,{\sc iii}, O\,{\sc iv}, and O\,{\sc vi} it is of
interest to see if the main absorption component in this system can be
modeled by assuming the gas is in photoionization equilibrium with the
background UV ionizing radiation.  For the modeling we use the ionization 
equilibrium code CLOUDY (v94.00; Ferland 1996) to compute
column densities of various ions through a slab illuminated
with the Haardt \& Madau (1996) UV radiation background from QSOs
and AGNs appropriate for a redshift of $\sim0.2$.  The calculation assumes
solar relative abundances from Anders \& Grevessse (1989) for all elements 
except oxygen and carbon for which we adopt the new results from
Allende Prieto, Lambert \& Asplund (2001, 2002). Grain heating or cooling are
ignored and
the H\,{\sc i} column density in 
the slab is taken to be log $N$(H\,{\sc i})$ = 15.25$. 
Since the column density of O\,{\sc iv} is rather uncertain, we concentrate on
the implications of log $N$(O\,{\sc iii})$ = 13.82\pm0.06$ and
log $N$(O\,{\sc vi})$ =13.68\pm0.06$
in the component near $0$ km\,s$^{-1}$. These O\,{\sc iii} and O\,{\sc vi}
column densities suggest an ionization parameter of log $U=-1.5$ where
$U=n_{\gamma}/n_{\rm H}$ is the ratio of 
ionizing photon density to gas density. The resulting
oxygen abundance is log (O/H) $-$ log (O/H)$_{\sun}=[$O/H$]=-1.25\pm0.10$.
The model predicts $\sim0.2$ dex more O\,{\sc iv} than is estimated from
the probably saturated O\,{\sc iv} line.
For log $U=-1.5$, the expected amount of Si\,{\sc iii} assuming a solar O 
to Si abundance  ratio is 1.0 dex less than the observed value
of log $N$(Si\,{\sc iii})$=12.04\pm0.07$. This unrealistic abundance result
suggests the absorption component near $0$ km\,s$^{-1}$ either is not
Si\,{\sc iii} absorption at $z=0.21949$ (but another unidentified IGM line), or,
more likely, it is Si\,{\sc iii} at $z=0.21949$ but
tracing a multiphase mixture of absorbing structures. Disentangling
the different contributions to each ion from each phase can not be 
reliably done with these relatively low S/N spectra which contain both
weak absorption lines (O\,{\sc iii}, O\,{\sc vi}, and Si\,{\sc iii}) and very strong lines
(C\,{\sc iii}, O\,{\sc iv}), which are probably saturated.
Because of the difficulties introduced by the probable multiphase
nature of the main absorption component, the abundance for oxygen and
the value of log $U$ estimated above are not likely to be correct.

More can be learned about the O\,{\sc vi} component near $-50$ km\,s$^{-1}$ with
log $N$(O\,{\sc vi})$=13.68\pm0.06$ since O\,{\sc iv} is only marginally detected at that
velocity and O\,{\sc iii} and C\,{\sc iii} are not detected. Integrating 
between $-90$ and $-35$ km\,s$^{-1}$ we find that the 
weak O\,{\sc iv} feature near $-50$ km\,s$^{-1}$
has $W_{\rm r}\approx13$ m\AA\,($1\sigma$ evidence) which 
corresponds to log $N$(O\,{\sc iv})$\approx 13.43$. If the
O\,{\sc vi} and the marginally detected O\,{\sc iv} are
produced by photoionization, the required value of the ionization parameter is 
log U$\approx-0.9$. The hydrogen
absorption possibly associated with this highly ionized absorber near $-50$ km\,s$^{-1}$ is
visible as the broad negative velocity wing in the Ly\,$\alpha$ profile. If
we add an H\,{\sc i} component with $-50$ km\,s$^{-1}$ to the principal H\,{\sc i}
absorption near zero velocities and fit the wing of the Ly\,$\alpha$ profile, we
obtain log $N$(H\,{\sc i})$\approx 13.5$ and $b\approx40$ km\,s$^{-1}$ for the
$-50$ km\,s$^{-1}$ component. This very uncertain value of $b$ implies
$T\sim10^5$ K if the broad feature is thermally broadened. Given the
uncertainties, the line breadth is large enough to provide support to the 
collisional ionization explanation for the origin of O\,{\sc iv} and O\,{\sc vi}.

Observations of
C\,{\sc iv} (unfortunately located redwards
of the wavelength range of the currently available STIS data) would be very useful 
to disentangle the ionization conditions in this interesting
absorber in more detail.
No Ne\,{\sc viii} $\lambda 770.4$ is detected in the $z=0.21949$ absorber at a
limiting logarithmic column density level of
log $N$(Ne\,{\sc viii})$\leq 13.94$ ($3\sigma$).
In collisional ionization equilibrium O\,{\sc vi} and Ne\,{\sc viii} peak in
abundance at $2.8\times10^5$ K and $7.0\times10^5$ K, respectively (Sutherland \& Dopita
1993). Comparing the amount of O\,{\sc vi} and Ne\,{\sc viii} therefore can provide
temperature constraints. In our ionization discussions we will assume
a solar abundance ratio for O/Ne and have taken the solar O and
Ne abundances from Allende Prieto, Lambert \& Asplund (2001) and Anders \& Grevesse
(1989), respectively.
Relating the limit for Ne\,{\sc viii} to the total O\,{\sc vi} column density
in both components implies $N$(O\,{\sc vi})$/N$(Ne\,{\sc viii})$\geq0.1$ and
$T\leq5.6\times10^5$ K, if the gas is in collisional ionization equilibrium.

\subsubsection{The System at $z=0.22313$}

This system is detected in H\,{\sc i} Ly\,$\alpha$ and Ly\,$\beta$,
and possibly in O\,{\sc vi} $\lambda 1037.6$ 
(Fig.\,17). O\,{\sc vi} $\lambda 1031.9$ may also be present,
but it lies in the vicinity of other strong IGM
lines (e.g., Si\,{\sc iii} $z=0.04606$ and C\,{\sc iii} $z=0.29236$, see
also Fig.\,7), which partly overlap the O\,{\sc vi} absorption (Fig.\,17).
The H\,{\sc i} Ly\,$\alpha$ absorption is well fitted by a single-component
absorber with log $N$(H\,{\sc i})$=13.92\pm 0.03$ and $b=34.8\pm1.1$ km\,s$^{-1}$
(Voigt-profile fitting).
Within the error range, the AOD method and COG fitting of the two Lyman
lines result in very similar
column densities (see Table 3). The O\,{\sc vi} column density is log $N=13.99\pm0.11$,
derived from O\,{\sc vi} $\lambda 1037.6$ via the AOD method. For pure
thermal broadening, the $b$ value from the H\,{\sc i} profile fit yields 
$T\leq7.3\times10^4$ K. The O\,{\sc vi} $\lambda 1037.6$ line
appears to
be quite broad, ranging from approximately $-90$ to $+60$ km\,s$^{1}$
(Ly\,$\alpha$: approximately $-60$ to $+60$ km\,s$^{-1}$).
However, the O\,{\sc vi} absorption is weak, so that the determined
O\,{\sc vi} velocity range is very sensitive to the
continuum placement. Thus, the O\,{\sc vi} line width is too uncertain to
draw meaningful conclusions
about the physical properties of the absorber. In view of the relatively
low $b$ value of the H\,{\sc i} absorption it is plausible that 
H\,{\sc i} absorption and O\,{\sc vi} absorption occur in physically
distinct regions. Although the O\,{\sc vi} $\lambda 1037.6$ absorption
has a formal significance of $\sim 4\sigma$, we cannot completely 
exclude that this broad absorption feature is caused by a 
small-scale undulation in the continuum flux. Further it is possible 
that this feature is caused by a broad Ly\,$\alpha$ absorption at 
$z=0.04417$ rather than being O\,{\sc vi} at $z=0.22313$. For these 
reasons, we can claim only a tentative detection of 
O\,{\sc vi} absorption in this system.

There is a $1.4\sigma$ absorption feature present precisely at the wavelength where
Ne\,{\sc viii} $\lambda 770.4$ at $z=0.22313$ 
would be expected (see lower-most right panel in
Fig.\,17). Due to the low significance, however, 
we cannot judge whether this
feature is real or not.
We derive a $3\sigma$ upper limit of log $N$(Ne\,{\sc viii})$\leq13.75$.
The observed limit for the 
absorber at $z=0.22313$ of
$N$(O\,{\sc vi})$/N$(Ne\,{\sc viii})$>1.7$ implies $T<5\times10^5$ K
for the O\,{\sc vi} absorber if the ionization is caused by electron
collisions in gas close to collisional ionization equilibrium
with a solar ratio of oxygen to neon.

\subsubsection{The System at $z=0.25971$}

This system is seen in H\,{\sc i} Ly\,$\alpha$ and 
Ly\,$\beta$, O\,{\sc vi} $\lambda\lambda 1031.9, 1037.6$, and possibly
in Ne\,{\sc viii} $\lambda 770.4$ (Fig.\,18). Although the H\,{\sc i} 
column density is relatively small (log $N$(H\,{\sc i})$=13.84\pm0.12$
from profile fitting; see Table 3 for details),
the Ly\,$\alpha$ absorption is quite broad,
ranging approximately from $-60$ to $+60$ km\,s$^{-1}$ (see Fig.\,18)
and looks flat bottomed suggesting two blended
H\,{\sc i} absorption components of nearly
equal strength. Also the O\,{\sc vi} $\lambda 1031.9$ absorption 
shows weak evidence for such a double-component structure.
However, the S/N is formally not high enough to convincingly claim
the existence of two (ore more) individual absorption components
and we thus will adopt 
a single absorption component in the following.
The one-component profile fit of the Ly\,$\alpha$ line yields 
$b=40.5\pm4.9$ km\,s$^{-1}$, equivalent
to $T\leq10^5$ K
\footnote{Note that the temperature is much smaller
if the absorber has two or more sub-components.}.
No intermediate ions such as C\,{\sc iii} and O\,{\sc iii}
are detected. The O\,{\sc vi}
absorption (log $N$(O\,{\sc vi})$=13.84\pm0.07$; AOD; see Table 2) 
seems to span a velocity range roughly similar
to that of the H\,{\sc i}. A reliable estimate of
the O\,{\sc vi} line width by way of profile fitting, however, is not 
possible because the O\,{\sc vi} absorption is relatively weak
and the continuum in the vicinity of the O\,{\sc vi}
absorption appears to have some small-scale undulations
(see Fig.\,18)

There is a possible detection of Ne\,{\sc viii} $\lambda 770.4$
in this system (see lower-most right panel in Fig.\,18) at
a $1.9\sigma$ level. More data would be required to investigate 
whether this absorption feature is real.
At this point, we only can place an upper
limit of log $N$(Ne\,{\sc viii})$\leq13.96$.
The $3\sigma$ limit on
$N$(O\,{\sc vi})$/N$(Ne\,{\sc viii})$>0.76$ implies $T<6\times10^5$ K for
gas in collisional ionization equilibrium with solar O and Ne abundance
ratios. If instead we assume the possible Ne\,{\sc viii} feature observed
with a $1.9\sigma$ significance is real, the derived value of log $N$(Ne\,{\sc viii})
is $\sim13.8$ which is similar to the value for O\,{\sc vi}. These two species will
have the same abundance in a hot collisionally ionized plasma 
with solar abundance ratios if $T$ is near $5.6\times10^5$ K.
However, this is
inconsistent with the value of $T\leq10^5$ K estimated from the H\,{\sc i} line
assuming the H\,{\sc i} line breadth is from pure thermal Doppler broadening.
If Ne\,{\sc viii} actually is detected, the value of
$N$(O\,{\sc vi})$/N($Ne({\sc viii}) would suggest that
the gas is not in collisional ionization equilibrium or
the bulk of the H\,{\sc i} absorption is formed in a cooler gas phase.

\subsubsection{The System at $z=0.29236$}

Three H\,{\sc i} Lyman lines (Ly\,$\alpha$, Ly\,$\beta$, and Ly\,$\gamma$) 
are detected in this relatively strong system
together with metal absorption from
C\,{\sc iii}, O\,{\sc iii}, and O\,{\sc iv} (Fig.\,19).
O\,{\sc vi} and Ne\,{\sc viii} are not detected.
The Ly\,$\alpha$ line shows a very weak second absorption component 
near $-80$ km\,s$^{-1}$. The main component has an H\,{\sc i} column density of
log $N$(H\,{\sc i})$=14.47\pm0.05$, and the H\,{\sc i} $b$ value 
is $26.2\pm1.6$ km\,s$^{-1}$, as derived from 
a profile fit of the Ly\,$\beta$ line (see Table 3). This $b$ value corresponds
to a temperature of $\leq4.1\times10^4$ K. Unfortunately,
the C\,{\sc iii} line in the STIS part of the spectrum
is partly blended by other IGM lines (see Figs.\,7 \& 19), so that
it is difficult to reliably determine its behaviour
at negative velocities.
For the intermediate ions C\,{\sc iii}, O\,{\sc iii}, and O\,{\sc iv}
we derive logarithmic column densities of
$13.18\pm0.07$, $13.80\pm0.06$ and $14.16\pm0.04$ (AOD method)
for the main component, respectively.
The value for C\,{\sc iii} is very uncertain because of
the blending problems. In contrast, the measurements for H\,{\sc i} , 
O\,{\sc iii}, and O\,{\sc iv} appear quite reliable.
The relatively
low $b$ value, the presence of various intermediate ion lines, 
as well as the lack of any high ions indicate 
that this system consists mainly of cooler
gas with temperatures $T\leq4.1\times10^4$ K.
If the gas in this system is in photoionization equilibrium, we can
use the well determined values of 
$N$(O\,{\sc iii}) and $N$(O\,{\sc iv}) to reliably
estimate the ionization parameter and thereby obtain an estimate of
the oxygen abundance in this low redshift system. For the system at
$z=0.29236$ we use the Haardt \& Madau (1996) UV radiation field for
$z=0.3$ and take the H\,{\sc i} column density in the 
slab to be log $N$(H\,{\sc i})$=14.5$. The column densities
log $N$(O\,{\sc iii})$=13.80\pm0.06$ and log $N$(O\,{\sc iv})$=14.16\pm0.04$
imply log $[N$(O\,{\sc iii})$/N$(O\,{\sc iv})$]=-0.36\pm0.07$
which constrains
the value of the ionization parameter log $U$ to be $-1.68\pm10$. With this
value of log $U$, the model yields the observed O\,{\sc iii}
and O\,{\sc iv} column
densities for [O/H]$=-0.5\pm0.1$. These numbers decrease to [O/H]$=
-0.74\pm0.1$ if we instead use the older solar abundance from Anders \&
Grevesse (1989). Using the uncertain C\,{\sc iii} column density of
log $N$(C\,{\sc iii})$=13.18\pm0.07$, where 
the error does not adequately account for the
uncertainty in the line blending correction, we obtain [C/H]$=-1.0\pm0.1$, 
implying [C/O]$=-0.5\pm0.1$ in this absorption system. The
listed errors do not allow for possible systematic errors associated
with the photoionization modeling assumptions such as the geometry of
the absorber and the shape of the ionizing background. Subsolar 
values of C/O are commonly found in low metallicity dwarf galaxies 
(Garnett et al.\,1995) and the data point for this QSO absorption system
fits on the observed trend of log (C/O) versus log (O/H) shown in Figure
3 of Garnett et al.\,(1995) after adjustment to the same reference 
abundances. In view of the substantial uncertainty in the C\,{\sc iii}
measurement we will not discuss the various nucleosynthetic 
explanations for such a behavior (see Garnett et al.\,1995). However,
our measurements for this system illustrate the potential of UV and
far-UV observations for providing interesting information on the 
nucleosynthetic  history of the gas in QSO systems at low redshift.
Although blending has created problems in this one case for C\,{\sc iii}, the
blending problem is much less severe at low redshift than in the Ly\,$\alpha$
forest at high redshift.
The large ionization parameter log $U=-1.68$ required to fit the 
observations implies the gas density in the system is low with a total
(H$^++$H$^0$) hydrogen density $n_{\rm H}=1.8\times 10^{-5}$ cm$^{-3}$.
At this density, the
absorber must have a thickness of 34 kpc in order to achieve the total
O\,{\sc iii} and O\,{\sc iv} column densities observed.

\subsubsection{The System at $z=0.31978$}

This is system is detected solely in H\,{\sc i}
(Ly\,$\alpha$ and Ly\,$\beta$) and possibly 
O\,{\sc vi} ($\lambda 1031.9$; see Fig.\,19).
The Ly\,$\alpha$ absorption is very broad,
spanning from approximately $-150$ to $+100$ km\,s$^{-1}$, 
while the candidate O\,{\sc vi} $\lambda 1031.9$ line is quite
narrow and well defined. 
The weaker O\,{\sc vi}
$\lambda 1037.6$ line is heavily blended by
interstellar Ni\,{\sc ii} $\lambda 1370.1$ absorption in the Complex C
component near $+20$ km\,s$^{-1}$ (in the $z=0.31978$ restframe), 
and possibly also by a broad, weak Ly$\alpha$ absorbers 
near $-100$ km\,s$^{-1}$. However, a multi-component fit of
the entire area including Ni\,{\sc ii} and a broad
Ly$\alpha$ component indicates that there
is enough absorption to fit O\,{\sc vi} $\lambda 1037.6$
into this pattern at the strength that would be
expected from the stronger O\,{\sc vi} $\lambda 1031.9$ line.

Single-component Voigt profile fits to the Ly\,$\alpha$ and
O\,{\sc vi} $\lambda 1031.9$ lines yield redshifts of
$0.319775\pm000029$ and $0.319731\pm000009$, respectively.
The redshift difference of $0.000044\pm0.000030$
corresponds to a velocity
difference of $\Delta v=10.0\pm 7.0$ km\,s$^{-1}$
in the restframe of the system. Although the data suggest a small
velocity shift for the O\,{\sc vi} absorption compared to H\,{\sc i},
the size of the shift is only $1.4\sigma$. 
In addition, evidence of relative wavelength calibration
errors as large as $\sim$1 pixel have been found in some STIS E140M
spectra (Tripp et al.\,2004, in preparation). Although wavelength calibration errors
this large are quite rare in E140M spectra (Leitherer et al.\,2002), it
is possible that the O\,{\sc vi}$-$H\,{\sc i} offset could be a
calibration artifact.
Because of
substantial continuum placement uncertainty, the Ly\,$\beta$
measurement has a redshift uncertainty $\sim3$ times larger
than for Ly\,$\alpha$. Therefore, Ly\,$\beta$ is not helpful
in establishing if the apparent small velocity shift between
H\,{\sc i} and O\,{\sc vi} is real.
There is a suggestion of a two component structure to
the H\,{\sc i} Ly\,$\alpha$ absorption with a weak second component
appearing near $-120$ km\,s$^{-1}$ in the $z=0.31978$ rest frame.
However, the appearance of the feature is very strongly
influenced by the adopted position of the uncertain continuum
on the blue wing of the Ly\,$\alpha$ line. Because of the
uncertain continuum, we cannot claim this possible
second component is real.
The one component Voigt profile fits to the Ly\,$\alpha$ and
O\,{\sc vi} $\lambda 1031.9$ lines yield
$b($H\,{\sc i}$)=74.4\pm8.7$ km\,s$^{-1}$,
$b($O\,{\sc vi}$)=19.3\pm4.2$ km\,s$^{-1}$, and log 
$N($H\,{\sc i}$)=13.98\pm0.06$. The ratio of
these $b$ values is $3.85\pm0.97$ which is consistent
with the expected ratio of 4 if these two species
existed in the same hot gas with the line broadening
dominated by thermal Doppler broadening.
Assuming pure thermal Doppler broadening $b($H\,{\sc i}$)$
implies $T=(3.3^{+0.8}_{-0.7})\times10^5$ K while
$b($O\,{\sc vi}$)$ implies $T=(3.6^{+1.7}_{-1.4})\times10^5$ K.
Both values of $T$ are close to the
temperature of $2.8\times10^5$ K where
O\,{\sc vi} peaks in abundance if collisional
ionization equilibrium prevails.
An upper limit for the temperature in the absorber
can be determined from the absence of Ne\,{\sc viii}
in the system. A quite stringent column density limit of
log $N($Ne\,{\sc viii}$)\leq13.57$ can be placed due to
the good S/N in the FUSE range where Ne\,{\sc viii} absorption would
be expected to occur (near 1017 \AA, see Table 3 and Figure 19).
With log $N($O\,{\sc vi}$)=13.44\pm0.06$
from the AOD method we obtain
$N($O\,{\sc vi}$)/N($Ne\,{\sc viii}$)\geq0.74$ for the
absorber. If the gas in the system
has a solar O to Ne abundance ratio and has a physical
state close to collisional ionization equilibrium, this
ionic ratio implies $T\leq6\times10^5$ K based on the
collisional ionization equilibrium calculations of Sutherland \&
Dopita (1993). A similar approach can be used to determine
a lower limit to $T$ from the absence of O\,{\sc iv}
$\lambda 787.7$ absorption with a
a $3\sigma$ limit for log $N($O\,{\sc iv}$)\leq13.30$
(see Table 3). With
$N($O\,{\sc iv}$)/N($O\,{\sc vi}$)\leq 0.73$ the lower
limit on $T$ is $2\times10^5$ K.

Due to its simplicity and its suggested high-temperature characteristics, 
this system represents an interesting case for a more detailed
modelling of the physical conditions.
The large H\,{\sc i} $b$ value
and the presence of O\,{\sc vi} already suggest
that this absorption system consists mainly
of hot gas at temperatures around $3\times10^5$ K. 
However, due to the fact that the O\,{\sc vi} $\lambda 1037.6$ 
absorption is completely blended, we cannot exclude
that the line that we had identified as
O\,{\sc vi} $\lambda 1031.9$ at $z=0.31978$ is not O\,{\sc vi} but 
just another weak (and narrow) Ly\,$\alpha$ absorber or another unknown IGM line.
From a conservative point of view, we therefore have to regard this system
as a tentative intervening O\,{\sc vi} absorber.
If the relatively narrow line we identify as O\,{\sc vi}
really is O\,{\sc vi}, the
measurements imply that the broad profiles 
of Ly\,$\alpha$ and Ly\,$\beta$ probably are
dominated by thermal Doppler broadening at $T\sim3\times10^5$ K since the
narrow O\,{\sc vi} line, the broad Lyman lines, as well 
as the limits for Ne\,{\sc viii} and O\,{\sc iv} yield nearly the same gas temperature.
With the assumption that this absorbing gas is in collisional ionization 
equilibrium we can draw several additional interesting conclusions. 
The ionization fraction $N$(H\,{\sc i})$/N$(H$_{\rm total}$) at
$T=3\times10^5$ K
is $1.5\times10^{-6}$ (Sutherland \& Dopita 1993). 
With log $N($H\,{\sc i}$)=13.98\pm0.06$ this implies 
$N$(H$_{\rm total})=6.4\times10^{19}$ cm$^{-2}$.
Similiarly, $N$(O\,{\sc vi})$/N$(O$_{\rm total})=0.20$
at $T=3\times10^5$ K and therefore
$N$(O$_{\rm total})=1.4\times10^{14}$ cm$^{-2}$ and
$N$(O$_{\rm total})/N$(H$_{\rm total})=2.1\times 10^{-6}$. 
With a solar value of (O/H)$=4.9\times 10^{-4}$
(Allende Prieto, Lambert \& Asplund 2001), these total column
density estimates  imply that the O abundance in this O\,{\sc vi} system is
$\sim 4.3\times10^{-3}$ solar.
The result is very sensitive to the adopted value of $T$. For example,
for $T=2\times10^5$ or $T=4\times10^5$ K, the inferred oxygen abundance
increases to $8.9\times10^{-3}$ and $7.7\times10^{-3}$ times solar, respectively.
The derived metallicity for this absorber could change substantially
if the H\,{\sc i} and O\,{\sc vi} component structure is more complex than we have
assumed in our analysis.

While the estimates depend on the validity of the
assumption of collisional ionization equilibrium in the absorbing gas,
it is noteworthy that low density hot plasmas with such low abundances
actually cool rather slowly since the principal heavy element coolants
(C, N, and O)  have low abundance.
An abundance as low as $\sim 4 \times10^{-3}$ solar, if correct, 
has important implications since the estimated baryonic content of the O\,{\sc vi}
systems scales inversely with the assumed oxygen abundance.
Savage et al.\,(2002) estimated that the gas in low redshift O\,{\sc vi} systems
contribute  $\sim0.002$ to the cosmological closure density assuming the
average oxygen abundance in O\,{\sc vi} systems is $0.1$ solar. This
contribution is comparable to that found in galaxies but $\sim20$ times
smaller than the total contribution for baryons estimated from the 
Cosmic Background radiation or from big bang nucleosynthesis. However, 
if the typical oxygen abundances in the O\,{\sc vi} systems are 10
to 25 times smaller, the estimate for the baryonic content of these
systems would increase by factors of 10 to 25.

\subsection{O\,{\sc vi} Absorbers}

The analysis of the metal-line systems towards
PG\,1259+593 reveals the presence of at least three 
intervening O\,{\sc vi} systems that are 
clearly detected in either one ($z=0.04606$) or 
both O\,{\sc vi} lines ($z=0.21949$ and $z=0.25971$). We tentatively detect
O\,{\sc vi} in four additional systems ($z=0.00229, 0.00760, 0.22313, 0.31978$), 
but for these systems the data do not allow us to claim a firm detection.
Therefore, we assume the number of intervening O\,{\sc vi} absorbers 
towards PG\,1259+593 to be $3-7$ in total. Six of these lines
have $W_{\rm r}\geq 25$ m\AA, the seventh has $W_{\rm r}=15$ m\AA.

The available wavelength range for intervening O\,{\sc vi} absorption 
towards PG\,1259+593 is from $\sim 1032$ to $\sim 1495$ \AA. The latter wavelength 
separates intervening O\,{\sc vi} absorption with radial
velocites $>6000$ km\,s$^{-1}$ away from the quasar from systems possibly 
associated with the QSO (``associated systems''). The wavelength
range corresponds to a O\,{\sc vi}
redshift path of $\Delta z\approx 0.449$.
In this regime,
the S/N is roughly constant, and we estimate
the typical lower equivalent width limit 
to detect intervening O\,{\sc vi} absorption 
to be $W_{\rm r}=24$ m\AA\, ($3\sigma$ restframe equivalent width limit).
However, the many intergalactic and interstellar absorption
lines in this wavelength range obscure a significant 
portion of this redshift path, so we have to correct $\Delta z$ 
for this blocking effect. Integrating the blocking for
each individual line between $1032$ and $1495$ \AA, 
we find a total blocking of $\Delta z_{\rm B}=0.085$, so 
that the total unobscured redshift path
for O\,{\sc vi} absorption is $\Delta z=0.364$.
In this study we assume the
presence of the stronger of the two O\,{\sc vi} absorption 
lines to be sufficient to claim at least a tentative
detection of an intervening O\,{\sc vi} system,
so that the blocking correction does not have to be applied twice. 
Thus, the number of intervening O\,{\sc vi} systems 
with $W_{\rm r}\geq 24$ m\AA\, ($3\sigma$)
per unit redshift towards PG\,1259+593 is
$dN_{\rm OVI}/dz=8-16$ 
\footnote{Note that a this 
equivalent width limit the weak tentative detection 
of O\,{\sc vi} at $z=0.00760$ ($W_{\rm r}\approx15$m\AA)
is not included in our estimate for $dN_{\rm OVI}/dz$.}.
At this sensitivity limit,
this value is consistent with previous studies
(e.g., Savage et al.\,2002).

It often difficult to determine if O\,{\sc vi} systems are collisionally
ionized or photoionized (e.g., Tripp et al. 2001, 2002). Consequently,
it is valuable to compare the observed statistics of these O\,{\sc vi} lines
to the predicted statistics from theoretical work in order to test
whether the O\,{\sc vi} absorbers are generally consistent with the expected
properties of the low-redshift WHIM.  The number of intervening O\,{\sc vi}
absorbers per unit redshift is one of the most readily measured
statistics, both in the theoretical models and the real observations.
We find that the O\,{\sc vi} $dN/dz$ values that we observe toward PG1259+593
and other sight lines (Savage et al.\,2002) are in agreement with the
predictions of hydrodynamic simulations of cosmological structure
growth (Cen et al.\,2001; Fang \& Bryan 2001; Chen et al.\,2003), but
only if the mean metallicity of the IGM is not too high.  Chen et
al. (2003) show that if the metallicity of the IGM were elevated to
solar metallicity, for example, then their predicted $dN/dz$ for O\,{\sc vi}
absorbers would be significantly higher than our observed values.
However, with $Z\sim 0.1\,Z_{\odot}$, the observed $dN/dz$ is close to
the predicted value from Chen et al. model. Since the baryonic content
of the O\,{\sc vi} systems is inversely proportional to the metallicity of the
gas, the conclusion that $Z<Z_{\odot}$ in O\,{\sc vi} absorbers bolsters
the finding that these systems contain an substantial fraction of the
baryons at the present epoch. The observations also indicate that at
least some O\,{\sc vi} absorbers arise in photoionized gas, and this is
predicted by the hydrodynamic models as well.  At this juncture, the
simulations and observations appear to be in good agreement with
regard to the O\,{\sc vi} absorber statistics.

\subsection{The Ly\,$\alpha$/Ly\,$\beta$ Forest}

Of the systems that are detected solely in H\,{\sc i}, 
14 show absorption by at least two lines of 
the H\,{\sc i} Lyman series, including one
system ($z=0.43569$) for which the Ly\,$\alpha$ 
absorption has been identified in the FOS spectrum
of PG\,1259+593 (Bahcall et al.\,1993). Redshifts, H\,{\sc i} column densities, 
and $b$ values are listed in Table 3. Velocity plots
are presented in Figs.\,13, 15, 17, 
18, 20 and 21. 50 systems are seen only in one single 
line, which we assume to be Ly\,$\alpha$ (see discussion
in \S 3.1). Redshifts, column densities and $b$ values for these systems 
(as derived by using Voigt-profile fitting)
are given in Table 5.
We have given every Ly\,$\alpha$ line 
listed in Table 5 a ``reliability flag'' (Table 5, seventh column) 
to indicate the reliability of
the listed line parameters (equivalent width, 
column density, Doppler parameter) for each line
(see \S2.3 for details).
Including the eight metal-line
systems discussed in the previous section, the line
of sight towards PG\,1259+593 consists of 72 intervening intergalactic 
absorption line systems (78 absorption components), for which a
short statistical analysis is presented in the following sections.

\subsubsection{$dN/dz$ for Ly\,$\alpha$ Absorbers}

The wavelength range for possible intergalactic 
Ly\,$\alpha$ absorption is $\sim 511$ \AA\,
($\sim 1218-1729$ \AA, corresponding to $\Delta z\approx0.420$), and 
there are 77 Ly\,$\alpha$ lines detected in this range. 
It is misleading, however, to derive
the number of Ly\,$\alpha$ absorbers per unit
redshift, $dN_{\rm Ly\,\alpha}/dz$, for this redshift range given 
the fact that the S/N in the STIS data
decreases significantly toward higher wavelengths, and some of the 
Ly\,$\alpha$ identifications thus are doubtful.
A roughly constant S/N of $\sim 12$ per resolution 
element is provided for the range between 1218 and 1525\AA\,
($\Delta z \approx 0.253$). In this range, 47 Ly\,$\alpha$ lines
are seen above an limiting rest-frame equivalent width of $W_{\lambda}\geq30$ m\AA.
For this sensitivity limit,
we find $dN_{\rm Ly\,\alpha}/dz=190\pm28$ for the 
sight line towards PG\,1259+593, after
applying a blocking correction
of $\Delta z_{\rm B}\approx0.006$ for 
strong ISM lines to the above redshift path.
This value is consistent with the low-redshift 
$dN_{\rm Ly\,\alpha}/dz$ distribution presented
by Penton, Shull \& Stocke (2000; their Fig.\,8)
for their GHRS sample of 15 extragalactic targets.
Recent studies (e.g., Janknecht, Baade \& Reimers 2002; Kim et 
al.\,2002) suggest that a break in redshift evolution of the
Ly\,$\alpha$ forest (characterized by the power law
$dN/dz=(1+z)^{\gamma}$) appears at $z\approx 1$. For high-column
density systems with $13.65\leq$\,log $N\leq 16.00$ there
is an abrupt change in $\gamma$ from $\sim 2.4$ for $z>1$ to
$\sim 0$ for $z\leq1$. This break is less evident for weaker
systems with $13.10\leq$\,log $N\leq 14.00$, where the
evolution at higher redshift is slower
($\gamma \sim 1.4$ for $z>1$; Janknecht, Baade \& Reimers 2002).
The redshift evolution of the Ly\,$\alpha$ forest
density usually is attributed to the combined
(counteracting) effects of a decreasing
ionizing background flux together with a decreasing
gas density due to expansion of the
Universe.
Our data yields log $dN/dz \sim 2.0$ for
Ly\,$\alpha$ absorbers with $13.6\leq$\,log $N\leq16.0$ and
log $dN/dz \sim 1.7$ for systems with $13.1\leq$\,log $N\leq14.0$
in our selected redshift range $z\leq 0.25$, thus
consistent with the previous low redshift data 
(Weymann et al.\,1998; Penton, Shull \& Stocke 2000).

Interestingly, there is no absorber directly associated 
with the quasar PG\,1259+593 itself. While the Ly\,$\alpha$
line at or near $z=0.478$
lies outside the STIS range studied here, intrinsic Ly\,$\beta$ 
absorption would be expected near $1516$ \AA, but no
absorption is seen around this wavelength (see Fig.\,10).
The closest IGM line bluewards is at $1508.963$, which we 
identify as Ly\,$\alpha$ $z=0.24126$ (see Table 5). If this 
line would be intrinsic Ly\,$\beta$ rather 
than intervening Ly\,$\alpha$, it would already be
more than $1300$ km\,s$^{-1}$ away from the quasar.
Also the low-resolution FOS data of PG\,1259+593 (Bahcall et al.\,1993)
show no evidence for a signficant absorption line near $1797$ \AA\, where
intrinsic Ly\,$\alpha$ absorption would be expected. A broad Ly\,$\alpha$
emission from the QSO itself, however, is clearly seen at this 
wavelength (Bahcall et al.\,1993).

\subsubsection{Distribution of $b$ Values and Column Densities}

Fig.\,22, upper left panel, shows the number 
distribution of $b$ values for the 
absorbers (and their sub-components), given in $10$ km\,s$^{-1}$ wide bins.
The distribution plotted with the thin line represents
the number distribution of all absorption components for which
$b$ values have been measured; the distribution plotted with
the thick line shows the number distribution of $b$ values
excluding the cases that are flagged as ``uncertain'' (UC; see
Tables 3, 5 and \S2.3).
The distribution of $b$ values peaks around $25$ km\,s$^{-1}$, very similar
to what is found in the larger HST sample presented by
Penton, Shull \& Stocke (2000).
The distribution shows a wing
towards higher $b$ values. This wing extends to $100$ km\,s$^{-1}$ in
the distribution of the more reliable $b$ values (thick line), 
and to $210$ km\,$^{-1}$ for the total sample (thin line).
Most of the high $b$ values with $100\leq b\leq210$ km\,s$^{-1}$
most likely are caused by unresolved sub-component structure in lines, continuum
undulations, and other effects that may produce broad spectral features.
Thus, part of the wing feature in the $b$
value distribution certainly is artificial in nature
(see also discussion in Penton, Shull \& Stocke, 2000).
However, there are seven lines with $b\geq50$ km\,s$^{-1}$ that
do not show evidence that such effects are responsible
for the large breadth of the line. These lines thus may
represent intrinsically broad Ly\,$\alpha$ absorbers;
they will be discussed below.
Examples for Voigt profiles fitted to broad Ly\,$\alpha$ absorbers
are presented in Figs.\,23 and 24.

In the upper right panel of Fig.\,22 we have plotted the distribution of
column densities for the 78 absorption components, divided into bins of 
$0.4$ for log $N$(H\,{\sc i}), starting
at log $N$(H\,{\sc i}$)=12.6$. Similar to the $b$ value distribution, the thin line
shows the distribution for log $N$ for the total sample, whereas the thick line
shows the log $N$ distribution excluding the uncertain cases.
This distribution has its maxiumum
in the range between log $N$(H\,{\sc i})$=13.0-13.8$.
Note that the 
determination of log $N$(H\,{\sc i}) depends on the
adopted $b$ value for each individual system. At the same
equivalent width, a lower $b$ value would 
result in a higher column density. Therefore, the 
distribution of column densities and the distribution of
$b$ values are coupled by the curve-of-growth analysis (thus by the
Voigt function). The observed distribution indicates a
peak at column densities slightly higher than found
in the sample by Penton, Shull \& Stocke (2000), although 
the $b$ value distribution is very similar.

In the lower panel of Fig.\,22 we have plotted Doppler 
parameters versus logarithmic column densities. Systems for which
$b$ values and column densities can be reliably determined 
(status is 'OK')
are shown as filled circles; systems for which
the values of log $N$ and $b$ are uncertain (status 'UC') are
are shown as
open circles. Most of the data points with $b<60$ km\,s$^{-1}$ follow 
the trend for a small increase of the $b$ values with significantly increasing
column densities. Such a trend is expected assuming that
the higher column-density absorbers arise from clouds
that have higher volume densities and higher temperatures 
(e.g., Dav\'e \& Tripp 2001). There appears to be, however, a second branch 
in this plot for $b\geq60$ km\,s$^{-1}$, indicating absorption systems
that exhibit a strong increase in $b$ to very high Doppler parameters
whereas the increase in $N$ is only modest. Most of the uncertain, 
broad features lie on this branch (open circles), although there are 
a few systems with reliable
values for log $N$ and $b$ (filled circles). If this branch is not 
an artifact caused by line blending, continuum undulations and/or other effects, 
we may have found a new, independent class of low-column
density H\,{\sc i} absorbers with large to very large Doppler parameters. 
If thermal line broadening would be the dominating broadening 
mechanism for these systems, they may trace WHIM absorbers in the $10^5$ to $10^6$ K 
temperature range. Such WHIM absorbers naturally would have very 
low neutral gas fractions (Sutherland \& Dopita 1993)
and therefore low H\,{\sc i}
column densities. We consider the reliability and 
nature of these
broad absorbers in the following section.

\subsection{Broad Ly\,$\alpha$ Absorbers}

From all Ly\,$\alpha$ lines listed in Tables 3 and 5 we are particularly 
interested in those with $b$ values
exceeding $40$ km\,s$^{-1}$, as they may sample intervening 
WHIM absorbers with temperatures above $T\sim10^5$ K. However, it is 
likely that next to thermal broadening other effects like line blends, 
unresolved sub-components, low S/N, kinematic flows, Hubble
broadening and continuum undulations contribute to the population
of broad absorbers along the line of sight towards PG\,1259+593. For the 
single-line Ly\,$\alpha$ absorbers listed in Table 5 we have analyzed in detail
the shape and other properties of the absorbers in order to sort 
out those lines for which there is evidence 
that their width is caused not by thermal broadening but
by something else.
Some of the absorbers show typical signatures for unresolved 
sub-components, as indicated by asymmetric line shapes
or positive/negative velocity wings. These cases have been marked accordingly in
the last row of Table 5. Moreover, for many of the broader lines listed
in Table 5 the estimate for $b$ is very sensitive
to the choice of the continuum level. This, together with
the relatively low S/N for some of these lines,
limits the credibility of the identification of
broad lines with $b$ values above $40$ km\,s$^{-1}$, although the formal 
$1\sigma$ errors derived from Voigt profile fitting 
(see Figs.\,23, 24, and Table 5) may be only a few percent.
From Tables 3 and 5 we find nine broad absorption
systems detected solely in H\,{\sc i} Ly\,$\alpha$ (Table 5) and/or
Ly\,$\beta$ (Table 3) with $b$ values larger than
$40$ km\,s$^{-1}$ for which a) we can determine $b$ with a 
reasonable accuracy (i.e., their status given in Table 5 is 'OK'),
and b) which do not show any evidence
for unresolved sub-structure or blending. 
These absorbers are listed in Table 6. Table 6 also
lists the two O\,{\sc vi} systems that have
H\,{\sc i} $b$ values larger than $40$ km\,s$^{-1}$ and
that do not show evidence for further sub-component structure
(see \S3.2). It is 
important to note at this point that the current 
STIS data are not accurate enough to {\it exclude} that line blending, 
sub-component structure, or continuum undulations contribute
to the large widths of the absorbers given in Table 6. We can only 
claim that the current data {\it show no evidence}
that these lines are affected by such processes.
We thus regard these
systems as possible candidates for thermally broadened WHIM aborbers
with temperatures $T\geq10^5$ K. Despite the
remaining uncertainty about their nature, we want to discuss their physical
and statistical properties in the following to
explore, what importance the possible existence of such systems 
has for our knowledge about the WHIM and its 
baryonic content. All of the candidates listed
in Table 6 have $40 \leq b \leq 100$ km\,s$^{-1}$, sampling
a temperature regime (for pure thermal broadening) of approximately 
$1\times10^5$ to $6\times10^5$ K. Thus, none of the really
broad lines with $b\geq 100$ km\,s$^{-1}$
from Table 5 are considered to be sufficiently securely
detected (see also Fig.\,24, lower-most panel).
This has to do mostly with the difficult
continuum placement for the particularly broad and shallow features.
As these broad absorbers (if real) may contain
a large amount of the baryonic mass in the IGM, we will 
- nevertheless - discuss the importance of these lines
for $\Omega_b$ in the following.
In principle, the broad Ly\,$\alpha$ lines may sample WHIM 
absorbers a) that have total column densities too 
low for O\,{\sc vi} being detectable, b) that have total 
column densities comparable to existing O\,{\sc vi} systems but 
have very low oxygen abundances, or c) that have 
large total column densities but temperatures
too high ($T>5\times10^5$ K) for O\,{\sc vi} to be 
present unless the total hydrogen column density
would be very large.
The latter type of absorbers is particularly interesting, 
as such WHIM clouds would contain large amounts of
baryonic matter and could be
observed in X-ray  O\,{\sc vii}/O\,{\sc viii}
absorption (``X-ray forest''; e.g., Rasmussen, Kahn \& Paerels 2003;
Fang et al.\,2002).
We expect the broad Ly\,$\alpha$
absorbers to arise in shock-heated, highly-ionized gas clouds 
(e.g., Cen \& Ostriker 1999), so the first step 
towards estimating their baryonic content is to determine
their ionization fraction. In the following, we 
assume collisional ionization equilibrium (CIE) and use
the models presented by Sutherland \& Dopita (1993). 
For the cooler and denser clouds with $T<5\times10^5$ K
the assumption of CIE may be somewhat problematic, as such gas 
may cool rapidly. However, the cooler absorbers 
also have the lowest baryon content, so their uncertainty 
contributes only little to the total uncertainty for deriving 
the overall baryonic content of the broad Ly\,$\alpha$ lines. 
For the higher temperature Ly\,$\alpha$ absorbers the assumption 
of CIE should be reasonably accurate.
For the temperature regime between $10^5$ and
$10^7$ K the ionization correction parameter
calculated by Sutherland \& Dopita (1993), 
log $f_{\rm H}=-$log (H$^0/($H$^0+$H$^+))\approx$
log (H$^+/$H$^0$), can be approximated by a simple polynomial:

\begin{equation}
{\rm log}\,f_{\rm H}={\rm log}\,f_{\rm H}(T)\approx -13.9 + 5.4\,{\rm log}\,T - 0.33\,
({\rm log}\,T)^2.
\end{equation}

\noindent
From equation (1), it follows that log $T\approx$\,{\rm log}\,(60\,$b^2$) for hydrogen, which
can be set into equation (2) to derive log $f_{\rm H}$ for an individual absorber 
directly as a function of its $b$ value (under the assumption that thermal broadening is
the dominating broadening process). 
The cosmological mass density of the broad Ly\,$\alpha$ absorbers, 
$\Omega_b$(BL) (in units of the todays
critical density, $\rho_{\rm c}$), can be given as 

\begin{equation}
\Omega_b{\rm (BL)}=\frac{\mu\,m_{\rm H}\,H_0} 
{\rho_{\rm c}\,c\,\Delta X}\,\sum_{i}\,f_{\rm H,i}\,N({\rm H\,I})_i \\
\approx 1.667\times10^{-23}\,\Delta X^{-1}\,\sum_{i}\,f_{\rm H,i}\,N({\rm H\,I})_i\,,
\end{equation}

\noindent
where $\mu=1.3$ corrects for the presence of helium, $m_{\rm H}=1.673 \times
10^{-27}$ kg is the proton mass, $H_0=75$ km\,s$^{-1}$\,Mpc$^{-1}$ 
and $\rho_{\rm c}=3H_0\,^2/8\pi G$. 
$\Delta X$ is the 
distance interval in which the broad Ly\,$\alpha$ absorbers are located.
The index $i$ is running from 1 to $n$, where $n$ is the number
of broad Ly\,$\alpha$ aborbers in the distance interval $\Delta X$.
For a redshift range from $z_{\rm min}$ to $z_{\rm max}$
and with $q_0=0$ we can write

\begin{equation}
\Delta X=0.5\,\{[(1+z_{\rm max})^2]-[(1+z_{\rm min})^2-1]\}.
\end{equation}

\noindent
However, a blocking correction has to be included 
for $\Delta X$ to account for the regions in which
possible broad Ly\,$\alpha$ absorbers may remain unnoticed
due to blending with other ISM or IGM lines.
For PG\,1259+593, the available redshift path to detect broad Ly\,$\alpha$
absorbers is $\Delta z = z_{\rm max} \approx 0.420$. 
The total blocking from
other intergalactic and interstellar lines for this redshift range sums up
to $\Delta z_{\rm B}=0.065$, and we estimate 
$\Delta X=0.430$ from equation (4).
If we take the eight candidates detected solely 
in H\,{\sc i} Ly\,$\alpha$ listed in Table 6
(ignoring the one broad Ly\,$\beta$ system at 
$z=0.43569$ for which Ly\,$\alpha$
lies outside the STIS wavelength range),
we find a number of broad Ly\,$\alpha$ aborbers
per unit redshift of $dN_{\rm BL}/dz \approx 23$ toward PG\,1259+593. 
Note that we do
not consider the two additional O\,{\sc vi} systems listed in Table 6 at
this point, as they are treated separately to constrain
$dN_{\rm O\,VI}/dz$ and the resulting $\Omega_b$(O\,{\sc vi})
(see also \S3.3).
For each of the eight Ly\,$\alpha$ absorbers we can derive $f_{\rm H,i}$ from
equation (2), as listed in Table 6 (seventh column). As we formally cannot
exclude that processes other than thermal broadening contribute to
the width of the lines given in Table 6, we conservatively list 
the values for $f_{\rm H,i}$ and $T$ as upper limits.
Together with the measured column densities,
$N$(H\,{\sc i}$)_i$, we thus obtain $\Omega_{b}$(BL)$\leq0.0031\,h_{75}^{-1}$. 
This limit is higher than what
is currently estimated for $\Omega_{b}$ from gas residing in intervening
O\,{\sc vi} absorbers ($\Omega_{b}$(O\,{\sc vi})$\geq 0.002\,h_{75}^{-1}$
at $\langle z \rangle =0.09$
for O\,{\sc vi} equivalent widths $\geq 50$ m\AA\, and an average 
metallicity of 0.1 solar; see Savage et al.\,2002).
It is important to note that
$\Omega_{b}$(O\,{\sc vi})
depends on the assumed metallicity, while $\Omega_{b}$(BL) 
is determined independently of the metallicity. However, although
we have very carefully selected our sub-sample of broad Ly\,$\alpha$
lines, we are heavily biased against absorbers with large $b$ values, 
which have the highest ionization correction and 
which thus contain most of the baryonic mass. All the absorbers listed
in Table 6 have $b$ values less than $100$ km\,s$^{-1}$, because 
the shallow, broad absorbers with higher $b$ values listed in Table 5 
are most uncertain. However, if only one of the very broad features 
listed in Table 5 is a real thermally broadened H\,{\sc i}
absorber, the baryonic mass density derived
from the presence of this
{\it single} absorption system (if representative for the 
given redshift path) would be immense  - it would
have a larger baryonic content
than the combined contributions of all nine systems
not containing O\,{\sc vi} listed in Table 6.
For future observations it is therefore of crucial interest to 
investigate whether similar broad absorbers are seen also toward
other low-redshift quasars, and to evaluate their frequency 
and physical properties.

\subsection{Association with Known Galaxies/Galaxy-Clusters} 

\subsubsection{Overview}

To further investigate the properties and nature of
the detected metal line systems, it is of great interest to 
find possible connections between the absorption line systems 
and galaxies/galaxy-groups that are located in the general field 
of the background source PG\,1259+593. 
We have searched
the NED data archive ({\tt http://nedwww.ipac.caltech.edu})
for a first inspection to find nearby galaxies and galaxy-groups 
for which redshift data have been published. 
Moreover, Chen et al.\,(2001) provide information
for two galaxies at $z=0.19620$ and $z=0.24120$ in the immediate 
vicinity of the PG\,1259+593 sight line (angular separation $<2$ arcmin)
based on HST WFPC2 data. 
Within a search radius 
of $60\farcm0$ around the center position of PG\,1259+593, 
NED finds 16 objects (14 galaxies and two galaxy groups)
that have redshifts
$z\leq 0.478$. We have listed these objects together with their 
coordinates and redshifts in Table 7. 
The most nearby absorbers found towards PG\,1259+593
and their impact parameters, $\rho_{75}$ (assuming 
$H_0=75$ km\,s$^{-1}$\,Mpc$^{-1}$ and $q_0=0$)
are also listed (Table 7, columns seven and eight).
Eight absorption systems have nearby galaxies with impact parameters
$\rho_{75}\leq 600$ kpc, three have $\rho_{75}\leq 300$ kpc.
For the future we are planning additional observations using the WIYN telescope
to investigate in detail the relation between individual absorption-line
systems and galaxies/galaxy-groups in the
field of PG\,1259+593.

\subsubsection{UGC\,08146}

A very close match in redshift is found for
the galaxy UGC\,08146 ($z=0.00224$) and the
absorption system at $z=0.00229$ (see \S3.2.1), which 
shows H\,{\sc i} absorption and possibly absorption 
by O\,{\sc vi}. It thus appears plausible that the 
$z=0.00229$ absorption occurs in the halo or nearby intergalactic 
environment of UGC\,08146. In Fig.\,25 we show
the sky positions of PG\,1259+593 and UGC\,08146 on an 
image taken from the {\it Digitzed Sky Survey} (DSS).
For UGC\,08146 we have overlayed H\,{\sc i} 21cm contours
from data of the {\it Westerbork Synthesis Radio Telescope} 
(WSRT; from Rhee \& von Albada 1996). Assuming
$H_0=75$ km\,s$^{-1}$\,Mpc$^{-1}$, 
the redshift of $z=0.00224$
($v_{\rm r}=672$ km\,s$^{-1}$) corresponds to a
distance of $\sim 9.0\,h^{-1}_{75}$ Mpc. Five arcmin in
Fig.\,25 therefore are equivalent to $\sim 13$ kpc
at the distance of UGC\,08146. The angular separation
between UGC\,08146 and the line of sight to
PG\,1259+593 is $\sim 21$ arcmin,
so that the projected distance is only $\sim 55\,h^{-1}_{75}$ kpc. If we
take the small redshift difference of $\Delta z=5\times10^{-5}$ 
($\Delta v=15$ km\,s$^{-1}$) into account and assume a
"pure" Hubble flow (see, e.g., Stocke et al.\,1995), we derive
a cloud-galaxy distance of $\sim 207\,h^{-1}_{75}$ kpc. The sight line
towards PG\,1259+593 thus appears to pass through the
immediate intergalactic
environment of UGC\,08146.

UGC\,08146 is a dwarf spiral, classified as Sc(d) (e.g., Stil \&
Israel 2002). The optical diameter of UGC\,08146 in the blue 
is $\sim 9$ kpc (defined as $25^{\rm th}$ magnitude isophote, including
an extinction correction), while the H\,{\sc i} effective 
diameter (enclosing 50 percent of the H\,{\sc i} mass) is
$21$ kpc or more than twice as large (Rhee \& von Albada 1996). 
The undisturbed nature
of the H\,{\sc i} profile and the lack of any other nearby galaxies
suggests that UGC\,08146 is relatively isolated. Stil \& Israel (2002)
propose that UGC\,08146 may be loosely associated with two
other dwarf galaxies, DDO\,123 and UGC\,7544.
If these galaxies are part of a loose galaxy group, the detected 
H\,{\sc i} and the possible O\,{\sc vi} absorption at $z=0.00229$ may sample
one or more filaments in the hot intragroup gas. The feature that 
we have (tentatively) identified as O\,{\sc vi} $\lambda 1031.9$ absorption
in the $z=0.00229$ absorber (see \S3.2.1 and Fig.\,13) is very broad, 
ranging from $-100$ to $+120$ km\,s$^{-1}$. This is much broader 
than what would be expected for pure thermal broadening for O\,{\sc vi} 
based on the H\,{\sc i} $b$ value of $\sim 42$ km\,s$^{-1}$. For pure 
thermal broadening one expects $b$(O\,{\sc vi})=$b$(H\,{\sc i})$/4\approx
10.5$ km\,s$^{-1}$ because of the 16 times higher atomic weight of oxygen.
However, the (very uncertain) width for O\,{\sc vi} rather corresponds
to $b>150$ km\,s$^{-1}$.
Thus, if real, the
O\,{\sc vi} might trace very hot intragroup gas with a large 
velocity dispersion, while the H\,{\sc i} samples somewhat 
cooler gas that is more confined. A value of
$b>150$ km\,s$^{-1}$ for hot intragroup gas would be in good agreement 
with typical line-of-sight
velocity dispersions observed in loose groups of
galaxies (e.g., Tucker et al.\,2000).
Additional UV and optical data would be very helpful to further 
investigate the possible connection between 
the $z=0.00229$ absorber and UGC\,08146.

\subsubsection{Galaxies and Galaxy Groups near $z=0.009$}

Six objects listed by NED (four galaxies and two
galaxy groups) have redshifts between $0.00858$ and
$0.00953$.
They all could be related to the absorption system at
$z=0.00760$, which is detected in H\,{\sc i} and
possibly O\,{\sc vi} (see \S3.2.2). Impact parameters
for the $z=0.00760$ system and these objects vary between
491 kpc (the galaxy UGC 08046) and 622 kpc (the
galaxy group Mahtessian 185). The absorption at
$z=0.00760$ thus may arise in the intragroup medium
of WBL 425 or Mahtessian 185 (see Table 7), and/or is related to
the intergalactic environment of one of the
individual galaxies listed in Table 7. 

\subsubsection{1259+5920 (+0270$-$0313)}

The elliptical or S0 galaxy 1259+5920 (+0270$-$0313)
at $z=0.19670$ lies only $\sim0.7$ arcmin away
from the line of sight towards PG\,1259+593 (Chen et al.\,2001),
and most likely is associated with the
two-component Ly\,$\alpha$/Ly\,$\beta$ absorber at $z=0.19620$
(see Table 3). The impact parameter is 
$\rho_{75}=120$ kpc, suggesting that the $z=0.19620$
absorption occurs in the halo of 1259+5920 (+0270$-$0313)
or in its nearby intergalactic
environment.  

\subsubsection{1259+5920 ($-$0234+0685)}

The late-type spiral galaxy 1259+5920 ($-$0234+0685) lies close to to the
direction of PG\,1259+593 (angular separation $\sim 1.2$ arcmin)
and has a redshift ($z=0.24120$) very similar to that
of the broad Ly\,$\alpha$ absorber at $z=0.24126$
(see also Chen et al.\,2001). We
derive an impact parameter of $\rho_{75}=248$ kpc.
The weak but very broad Ly\,$\alpha$ absorption 
($b($H\,{\sc i}$)=89.1\pm6.9$ km\,s$^{-1}$) thus may
arise in an extended halo of 1259+5920 ($-$0234+0685), or
in an WHIM filament in the vicinity of this galaxy.

\section{Concluding Remarks}

As the temperature of the intergalactic medium 
undergoes a significant change from
high to low redshifts, a large fraction of the baryonic matter 
in the local Universe is expected to reside in the WHIM phase - a gas 
phase that is particularly difficult to detect.
The combined FUSE and STIS spectrum of PG\,1259+593 
represents one of the few high-quality UV and
FUV data sets
currently available to study the distribution
and physical properties of the low-redshift WHIM and the 
local Ly\,$\alpha$ forest. However, the PG\,1259+593 data 
also show how difficult the analysis of the local IGM is 
due to the limitations in S/N and spectral resolution in these space-based data.
We have detected three intervening O\,{\sc vi} absorption line systems towards
PG\,1259+593 and have found four additional candidates for 
which we cannot claim a firm detection.
With a number of intervening O\,{\sc vi} systems per
unit redshift of $dN_{\rm OVI}/dz=8-16$ for equivalent widths
$W_{\rm r}\geq24$ m\AA, the sight line towards PG\,1259+593
adds important information about the distribution of
intervening O\,{\sc vi} absorbers
in the local Universe.
Savage et al.\,(2002) summarize previous FUSE and STIS absorption line
measurements of intervening O\,{\sc vi}
in the low-redshift IGM, including the lines of sight 
toward H\,1821+643, PG\,0953+415, PG\,0804+761, and 
3C\,273 (Tripp, Savage \& Jenkins 2000; Oegerle et al.\,2000;
Tripp \& Savage 2000; Richter et al.\,2001a; Sembach et al.\,2001).
Six O\,{\sc vi} systems are detected above
an equivalent width limit of $50$ m\AA\, for the O\,{\sc vi} $\lambda 1031.9$
line, suggesting that $dN_{\rm O\,VI}/dz=14^{+9}_{-6}$ for $\left<z\right>=0.09$, 
and therefore $\Omega_b($O\,{\sc vi}$)\geq0.002\,h_{75}^{-1}$
for an average metallicity of $0.1$ solar (for details
see Savage et al.\,2002). The line of sight towards PG\,1259+593 
consists of $3-5$ more intervening O\,{\sc vi} systems 
with $W_{\rm r}\geq50$ m\AA, in agreement with
the previous observations and $dN_{\rm O\,VI}/dz$ estimates. 
The problem of having to separate
clearly detected O\,{\sc vi} systems from the many tentative 
absorbers yet remains somewhat unsatisfying, and 
more accurate sight-line analyses appear
to be necessary to reduce the yet considerable 
uncertainty for $\Omega_b($O\,{\sc vi}$)$.

The FUSE and STIS absorption line
measurements suggest that intervening O\,{\sc vi}
absorption lines reliably trace the WHIM in the low-redshift 
Universe, although some O\,{\sc vi} may also
arise in low-density, photoionized gas (see Savage et al.\,2002). 
We note at this point that the collisional ionization
interpretation for most of the weak O\,{\sc vi} absorbers
holds only
for the low-redshift Universe, where a significant fraction
of the intergalactic gas is expected to reside in the WHIM phase
(Cen \& Ostriker 1999) at $T\geq10^5$ K.
At higher redshifts
($z\sim 2$), a larger fraction of the O\,{\sc vi} absorbing systems 
apparently traces intergalactic structures that are photoionized
by a hard UV background. This is
evident from the narrow line widths ($b_{\rm O\,VI}\leq
10$ km\,s$^{-1}$) of some
of the weaker O\,{\sc vi} absorbing structures
at high redshifts with log $N$(H\,{\sc i}$)\leq 15$
(Bergeron et al.\,2002; Carswell, Schaye \& Kim 2002).
Such small $b$ values are
inconsistent with temperatures $T\geq10^5$ K.
However, the results of Simcoe, Sargent, \& Rauch (2002)
imply that collisional ionization has to be taken into account
for the higher-column density O\,{\sc vi} systems at $z\sim2$
(log $N$(H\,{\sc i}$)\geq 15$),  
as this gas
possibly traces shock-heated material from supernova-driven
galactic winds.
An important but yet unanswered question is,
what baryonic mass fraction of the WHIM at low $z$ is not
detected in studies that concentrate
solely on O\,{\sc vi} absorption ? WHIM clouds
that have intrinsically low gas column densities, and/or low metallicities, 
and/or very high temperatures ($\geq 10^{5.5}$ K) 
are difficult to detect in O\,{\sc vi} absorption.
Other high ions such as Ne\,{\sc viii} may trace the WHIM in
a higher temperature regime, but
our study (and other measurements) provide no evidence that there exists
a significant number of detectable intervening 
Ne\,{\sc viii} absorbers at low redshifts.
Towards PG\,1259+593 we have not found significant 
($\geq 3\sigma$) Ne\,{\sc viii} absorption
in any of the absorption systems.
The possibility to detect
Ne\,{\sc viii} absorption in FUSE and STIS data, 
however, is probably low due to
the relatively low cosmic
abundance of neon compared to oxygen (log (Ne/O)$_{\sun}=-0.61$; 
Allende\,Prieto, Lambert, \& Asplund 2001, 2002; Anders \& Grevesse 1989)
and the lower S/N of the FUSE data in the spectral region where the Ne\,{\sc viii}
lines are accessible.
In two systems ($z=0.22313$ and
$z=0.25971$) Ne\,{\sc viii} absorption may be present
at low statistical confidence ($\leq 1.9\sigma$);
if more Ne\,{\sc viii} data from other
sight lines will become available, one may find evidence for 
Ne\,{\sc viii} absorption in WHIM absorbers by stacking together a 
larger number of Ne\,{\sc viii} velocity profiles. 
Observations of the X-ray absorption in the
lines of O\,{\sc vii}, O\,{\sc viii} and other 
species are becoming an important way to trace
hot gas in the local intergalactic medium 
(Rasmussen, Kahn \& Paerels 2003;
Fang et al.\,2002; Nicastro et al.\,2002).
But even with the most modern X-ray
observatories, such as {\it Chandra} and {\it XMM-Newton}, observations 
of intervening intergalactic X-ray absorbers represent a 
challenging task, and so far the number of detected X-ray absorption
features in the IGM is quite low. Thus, good statistics
about the low-redshift X-ray forest will not be 
readily available in the near future.

A promising approach to trace the WHIM at temperatures
up to $T\sim10^6$ K
is to search for broad, shallow Ly\,$\alpha$
absorbers. With the FUSE and
STIS data presented here we possibly
have found for the first time evidence that
at least part of the WHIM up to $T\sim10^6$ K may be
traced by Ly\,$\alpha$ absorption.
Support for the
detectability of gas this hot in Ly\,$\alpha$
comes from X-ray observations.
Studies of the immediate intergalactic
environment of the Milky Way suggest the possible presence of an WHIM
filament with an O\,{\sc vii} column
density of $\sim 10^{16}$ cm$^{-2}$ 
(see Rasmussen, Kahn, \& Paerels 2003). If this O\,{\sc vii} 
column density is representative for a WHIM absorber, 
and if we further assume a
temperature of $10^6$ K together with collisional ionization 
equilibrium (Sutherland \& Dopita 1993) and a 
metallicty of the gas of 0.1 solar, we
would expect to see Ly\,$\alpha$ absorption
at the level of log $N$(H\,{\sc i})$\approx13.5$ and 
$b\approx130$ km\,s$^{-1}$, in line with
some of the broader absorbers towards PG\,1259+593 listed
in Table 5. Depending on the column density and
temperature distribution of the WHIM absorbers, it may be that
only a small fraction of these systems can be detected
in H\,{\sc i} absorption. There could be, however, also 
a number of WHIM systems that are somewhat cooler than the
X-ray absorbing systems, and thus have larger
neutral gas column densities and possibly O\,{\sc vi} absorption. 
The relatively strong and broad $z=0.31978$ absorber towards PG\,1259+593 could be
an example for such a ``transition system''. The
H\,{\sc i} $b$ value ($\sim 74$ km\,s$^{-1}$), the probable 
detection of O\,{\sc vi} absorption and the resulting
O\,{\sc vi} $b$ value ($\sim 19$ km\,s$^{-1}$), as
well as the limits derived for O\,{\sc iv} and Ne\,{\sc viii} 
are all consistent with a WHIM absorber
at $T\sim3\times 10^5$ K. This
system thus may represent the best observational evidence that some broad
Ly\,$\alpha$ absorbers indeed trace the WHIM.

Dav\'e \& Tripp (2001) have performed hydrodynamic
simulations to produce artifical absorption line spectra to compare
the characteristics of the simulated Ly\,$\alpha$ absorption 
with STIS absorption line data for the low-redshift 
IGM towards PG\,0953+415 and H\,1821+643.
They use an automated Voigt profile fitting technique 
to determine column densities and $b$ values for both artifical 
and real spectra, finding excellent agreement between these two data 
sets. As shown in their Fig.\,4, they do not find a population of
very broad  Ly\,$\alpha$ absorbers that would trace the WHIM
at $T>3\times10^5$ K. However, it is not clear what sensitivity 
the automated fitting routine has for very broad, shallow absorption features.
Since artificial and real data are fitted in an identical manner, 
the routine possibly could miss weak absorbers with large
Doppler parameters, mainly because such features would be treated as
continuum undulations and removed in the normalization process.
Thus, a population of weak, broad Ly\,$\alpha$ lines
may remain unnoticed
in the simulated spectra and the STIS data of PG\,0953+415 
and H\,1821+643 using this technique.
A careful re-inspection of these data is necessary
to clarify this point. Additional high S/N observations with STIS
will be important to investigate 
whether the broad features we have detected
in the line of sight towards PG\,1259+593 really arise 
from high-temperature WHIM absorbers, or whether they only 
represent a mixture of instrumental effects and unresolved 
absorption components. Having available a larger number
of sight lines
with different background continua clearly would help to tackle 
some of the systematic effects that hamper the non-ambiguous 
identification of broad, shallow absorption features. Also, 
combined X-ray and FUV observations will help to answer the 
question of whether broad Ly\,$\alpha$ features can serve as a
reliable tracer for gas with temperatures up to $\sim10^6$ K.
If so, studying broad Ly\,$\alpha$ absorbers in high
S/N FUV spectra could be a big step forward
to understand the distribution and physical properties of 
the WHIM at low redshifts and its baryonic content.
 
\section{Summary}

We have analyzed intergalactic absorption in FUSE and 
STIS FUV spectra of the QSO PG\,1259+593 ($z_{\rm em}=0.478$).\\
\\
\indent
1. We identify 135 intergalactic absorption lines at 
$W_{\rm r} \geq 10$ m\AA, including $4-8$ metal 
line systems that are detected in one or more lines
of C\,{\sc iii}, C\,{\sc iv},
O\,{\sc iii}, O\,{\sc iv}, O\,{\sc vi}, and Si\,{\sc iii}. 
We determine equivalent widths and column densities
for 72 absorption line systems.
The overall distribution of Doppler parameters ($b$ values) and column
densities for the local Ly\,$\alpha$ forest towards PG\,1259+593
is consistent with results from previous studies
considering lines with $b\leq 40$ km\,s$^{-1}$.

2. We discuss physical conditions in the metal-line systems. 
The two strongest absorption systems at $z=0.04606$ and $0.21949$ 
consist of multiple absorption components with different
gas-phases. For the single-component system at $z=0.29236$ the
analysis of H\,{\sc i}, C\,{\sc iii}, O\,{\sc iii}, and O\,{\sc iv} 
absorption implies abundance ratios 
of log $[$O/H$]=-0.5$ and log $[$C/O$]=-0.5$. This is comparable 
with what is found in low-metallicity dwarf galaxies.

3. The number of intervening O\,{\sc vi} absorbers along this 
line of sight is $3-6$ for a typical equivalent width limit of 
$W_{\lambda}\geq24$ m\AA. Together with an unobscured
redshift path of $\Delta z \approx 0.368$ this number 
implies $dN_{\rm O\,VI}/dz \sim 8-16$ in this direction. 
At this equivalent width limit, the finding is consistent 
with $dN_{\rm O\,VI}/dz$ results from previous absorption 
line studies. The result supports the idea that intervening 
O\,{\sc vi} systems trace a significant fraction of
the baryons in the local Universe residing
in the warm phase of the warm-hot intergalactic medium (WHIM).

4. A broad Ly\,$\alpha$ absorption
system ($b\approx 74$ km\,s$^{-1}$)
is found at $z=0.31978$. The system is
possibly also detected in O\,{\sc vi} $\lambda 1031.9$ 
absorption. This absorber may represent a collisionally 
ionized intergalactic gas cloud at a temperature of $\sim 3\times10^5$ K 
and a very low but uncertain oxygen abundance, (O/H), of
$4.3\times10^{-3}$ solar.

5. We detect a large number of broad Ly\,$\alpha$ lines 
along the line of sight towards PG\,1259+593 with Doppler 
parameters $b\geq40$ km\,s$^{-1}$ and relatively low 
column densities (log $N$(H\,{\sc i})$< 14$). If thermal 
broadening determines the large 
widths of these lines, their $b$ values correspond to 
temperatures between $10^5$ and several $10^6$ K. 
Although some of these lines may be composites
of several blended, unresolved absorption features,
caused by continuum undulations, or are produced
by gas flows and Hubble broadening, we speculate that
at least some of these features represent WHIM absorbers
that are not detected in O\,{\sc vi} absorption.
From a sub-set of eight well-detected broad Ly\,$\alpha$ lines (BL)
with $40$ km\,s$^{-1}<b<100$ km\,s$^{-1}$ we 
estimate $dN_{\rm BL}/dz\approx 23$
and $\Omega_b=0.0031\,h_{75}^{-1}$. If the
line widths are dominated by thermal Doppler 
broadening, then these absorbers 
would contain a large fraction of the
baryons in the low-redshift IGM.

\acknowledgments

This work is based on data obtained for the
the Guaranteed Time Team by the NASA-CNES-CSA FUSE
mission operated by the Johns Hopkins University. 
Financial support has been provided by NASA contract 
NAS5-32985. 
The STIS observations of PG1259+593 were obtained for HST GO program
8695, with financial support from NASA grant GO-8695.01-A from the Space
Telescope Science Institute.
P.R. was supported by the research grant RI\,1124 from
the {\it Deutsche Forschungsgemeinschaft}.
T.M.T received support from NASA Long Term Space Astrophysics
grant NAG5-11136.
P.R. acknowledges the opportunity to visit 
the Osservatorio Astrofisico di Arcetri,
Florence, Italy between spring 2002 and fall 2003; a susbtantial part of
this paper was written at this institute.

\clearpage
\includegraphics{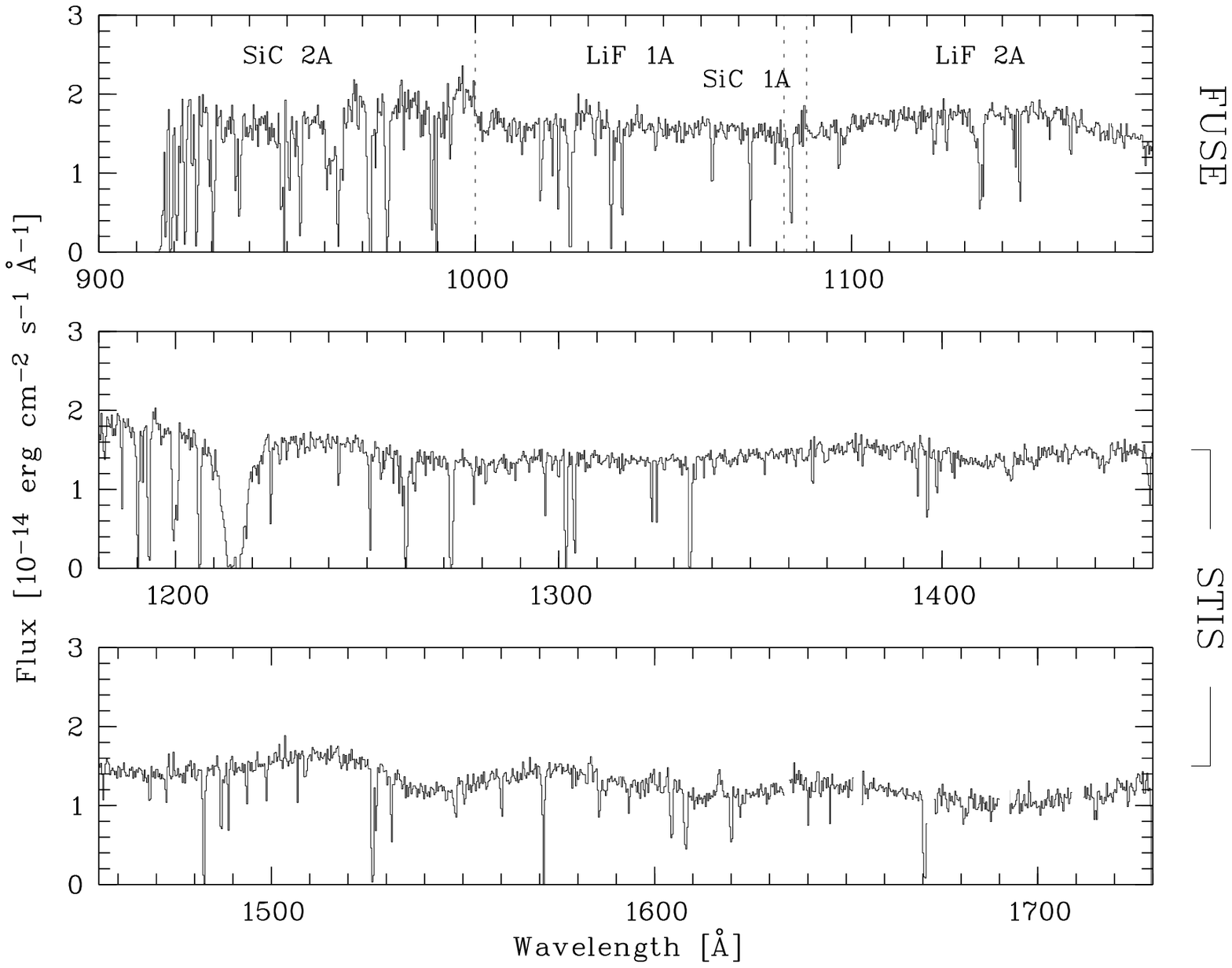}
\figcaption[f01.eps]{Combined FUSE and STIS spectrum of PG\,1259+593.
For this plot, the data have been binned to $0.3$ \AA\, wide
pixels. Note that most
of the narrow interstellar and intergalactic absorption features are
not resolved at this bin size. At $\lambda>1634$ \AA\,the STIS data have
several small gaps between the individual echelle orders. At the 
junction between the FUSE and the STIS data at 1180 \AA\, the absolute
flux in the FUSE LiF\,2A channel is inconsistent with the (correct) flux in the 
STIS data ($F_{\rm STIS}\approx 1.5\,F_{\rm FUSE}$) because of
dark horizontal stripes in the two-dimensional FUSE detector images
(``worms'') that reduce the total flux in the respective wavelength regions
(in this case for $\lambda > 1160$ \AA) and that are not accounted 
for in the FUSE flux calibration of
PG\,1259+593.}

\clearpage
\newpage
\includegraphics{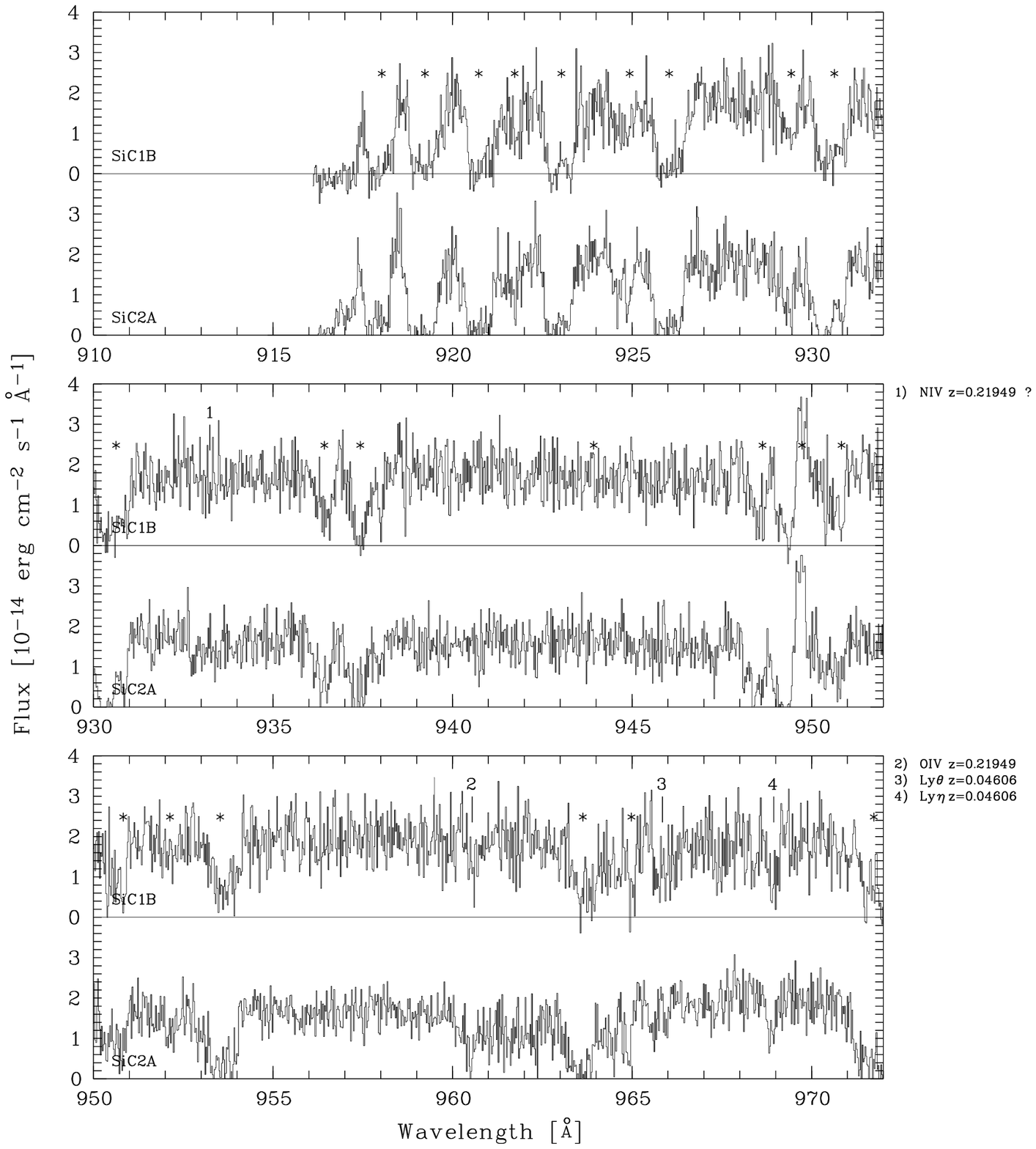}
\figcaption[f02.eps]{FUV spectrum of PG\,1259+593 (Figs.\,2-12), 
including FUSE data
($916-1188$ \AA) and STIS data ($1180-1729$ \AA). For these plots, the data
have been binned to $0.025$ \AA\, wide pixels, except for the SiC\,1A 
and SiC\,2B data, which have been binned to $0.1$ \AA\, wide pixels.
IGM lines are indicated with tic marks and numbers above the spectrum and
are identified on the right hand side;
absorption features from local interstellar lines are labeled with a star symbol.
Uncertain lines and apparently broad Ly\,$\alpha$ absorbers are marked
with `?' and 'B', respectively (see \S2.3 for more details).}

\clearpage
\newpage
\includegraphics{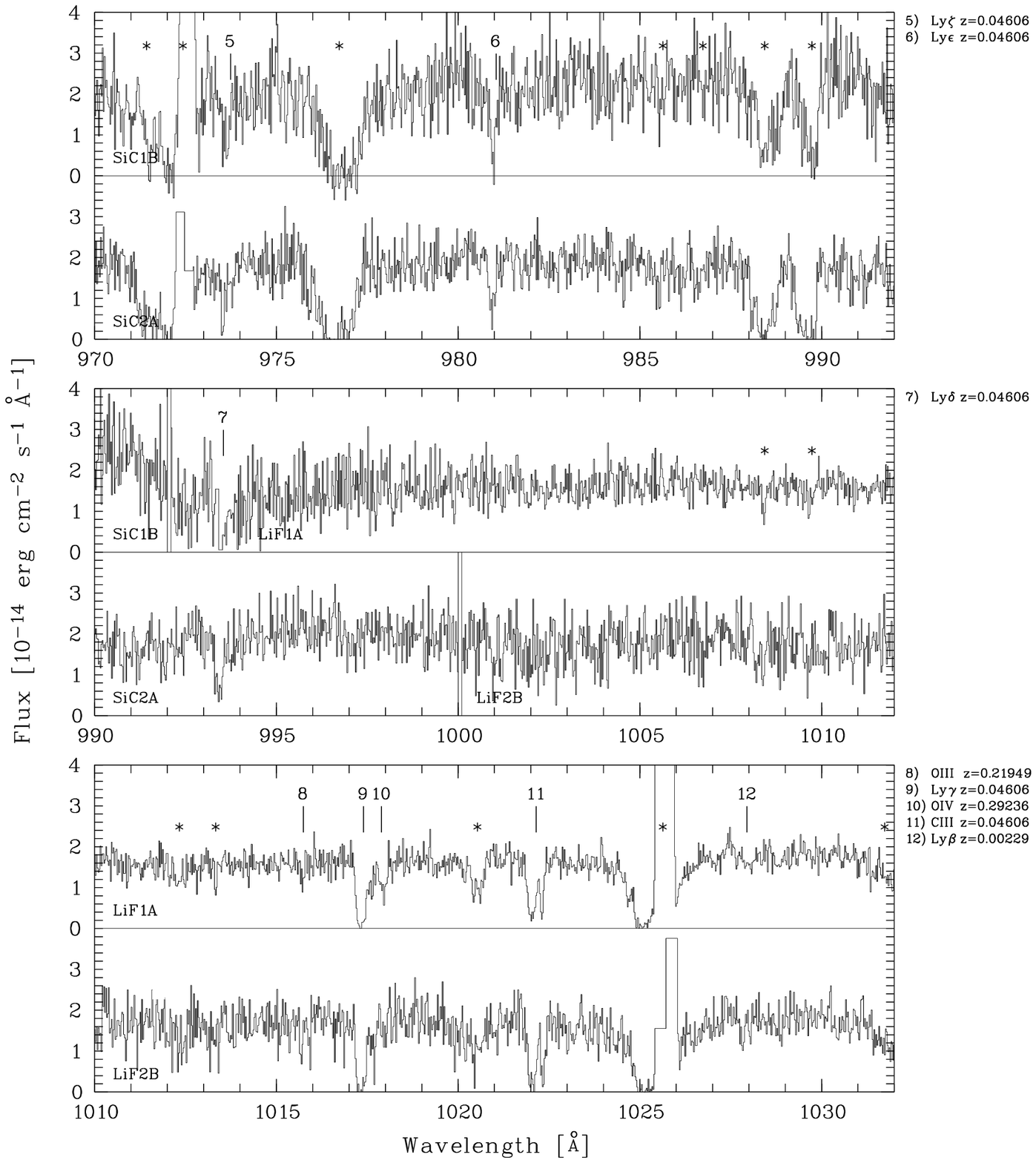}
\figcaption[f03.eps]{FUV spectrum of PG\,1259+593 (see caption to Fig.\,2).}

\clearpage
\newpage
\includegraphics{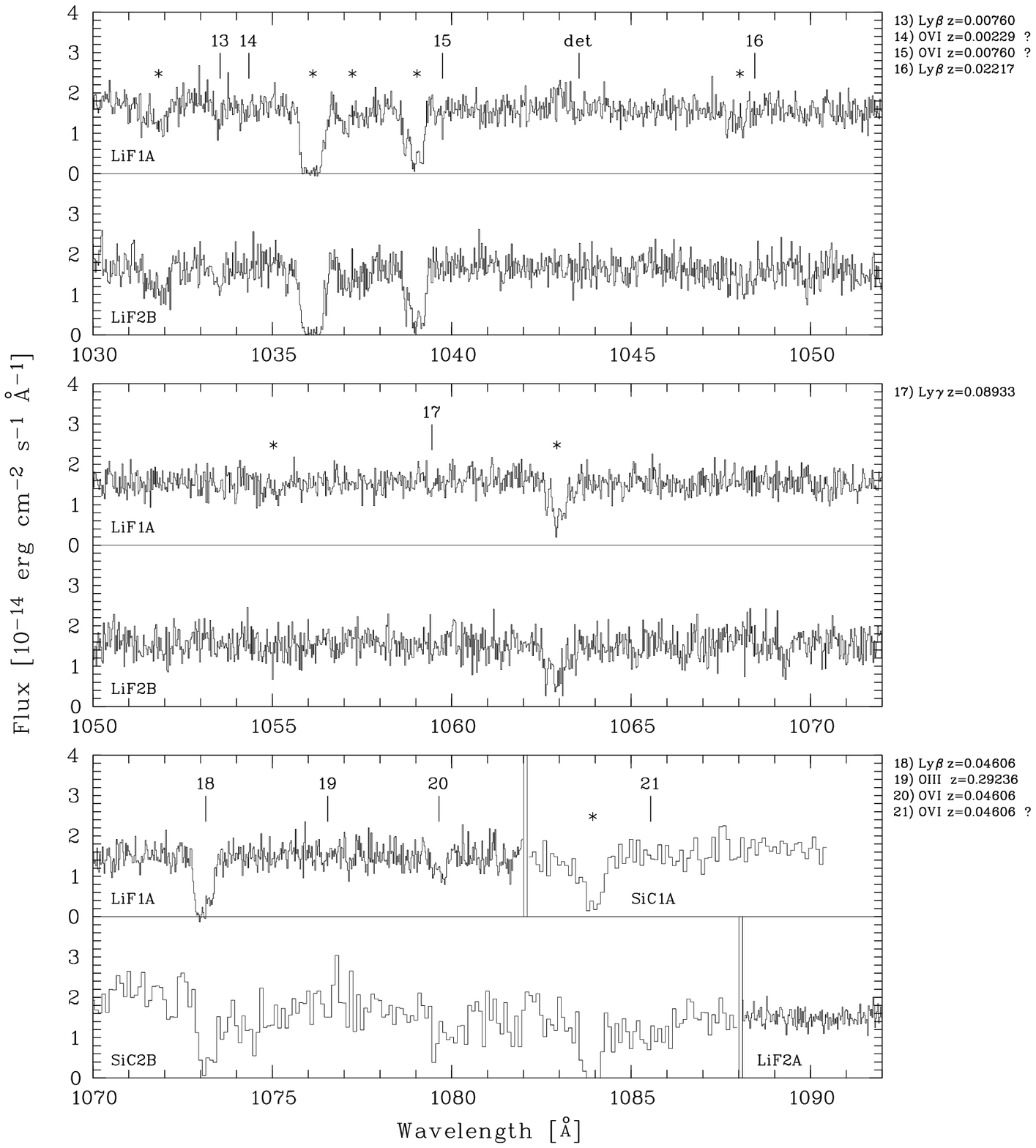}
\figcaption[f04.eps]{FUV spectrum of PG\,1259+593 (see caption to Fig.\,2).}

\clearpage
\newpage
\includegraphics{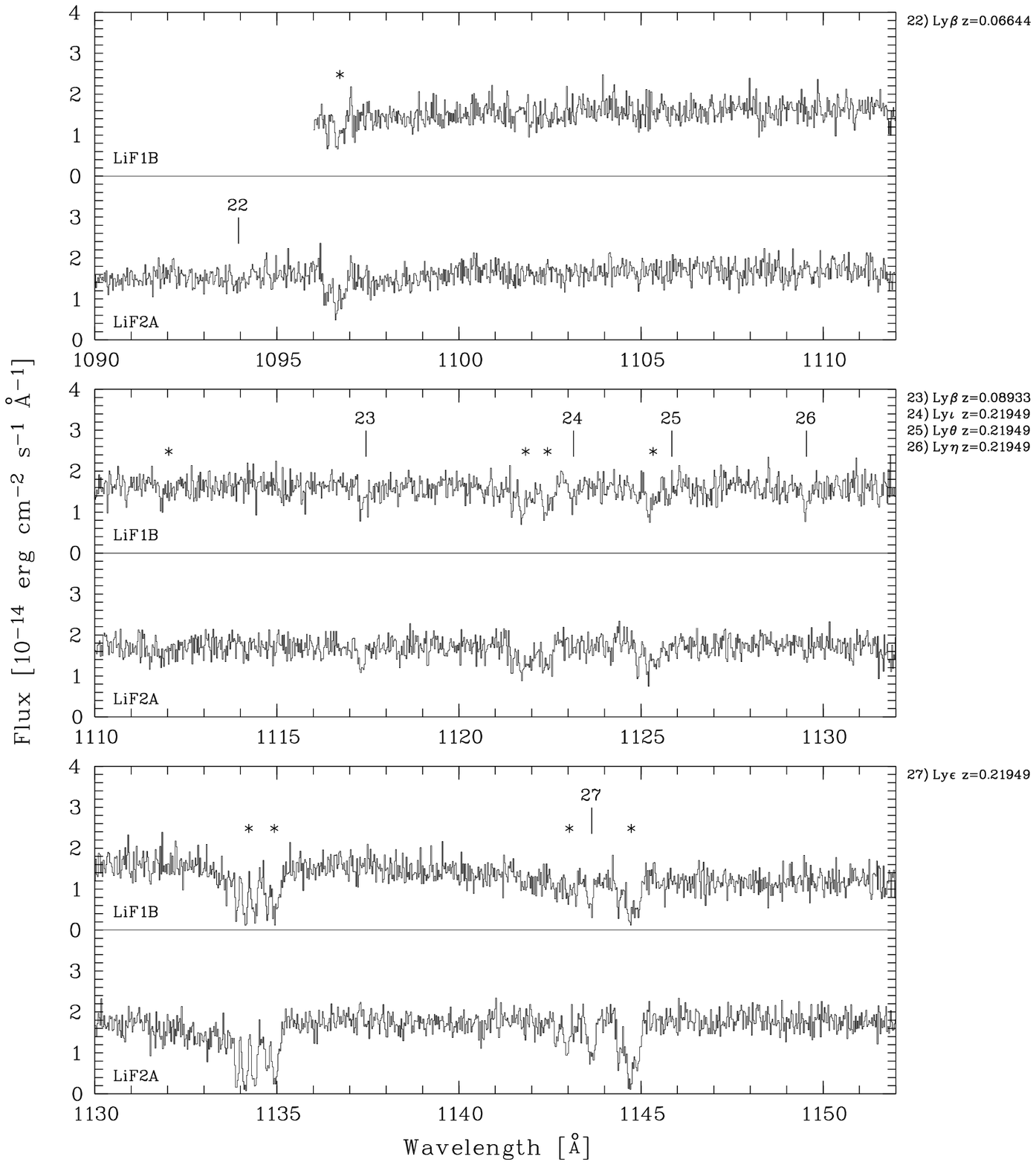}
\figcaption[f05.eps]{FUV spectrum of PG\,1259+593 (see caption to Fig.\,2).}

\clearpage
\newpage
\includegraphics{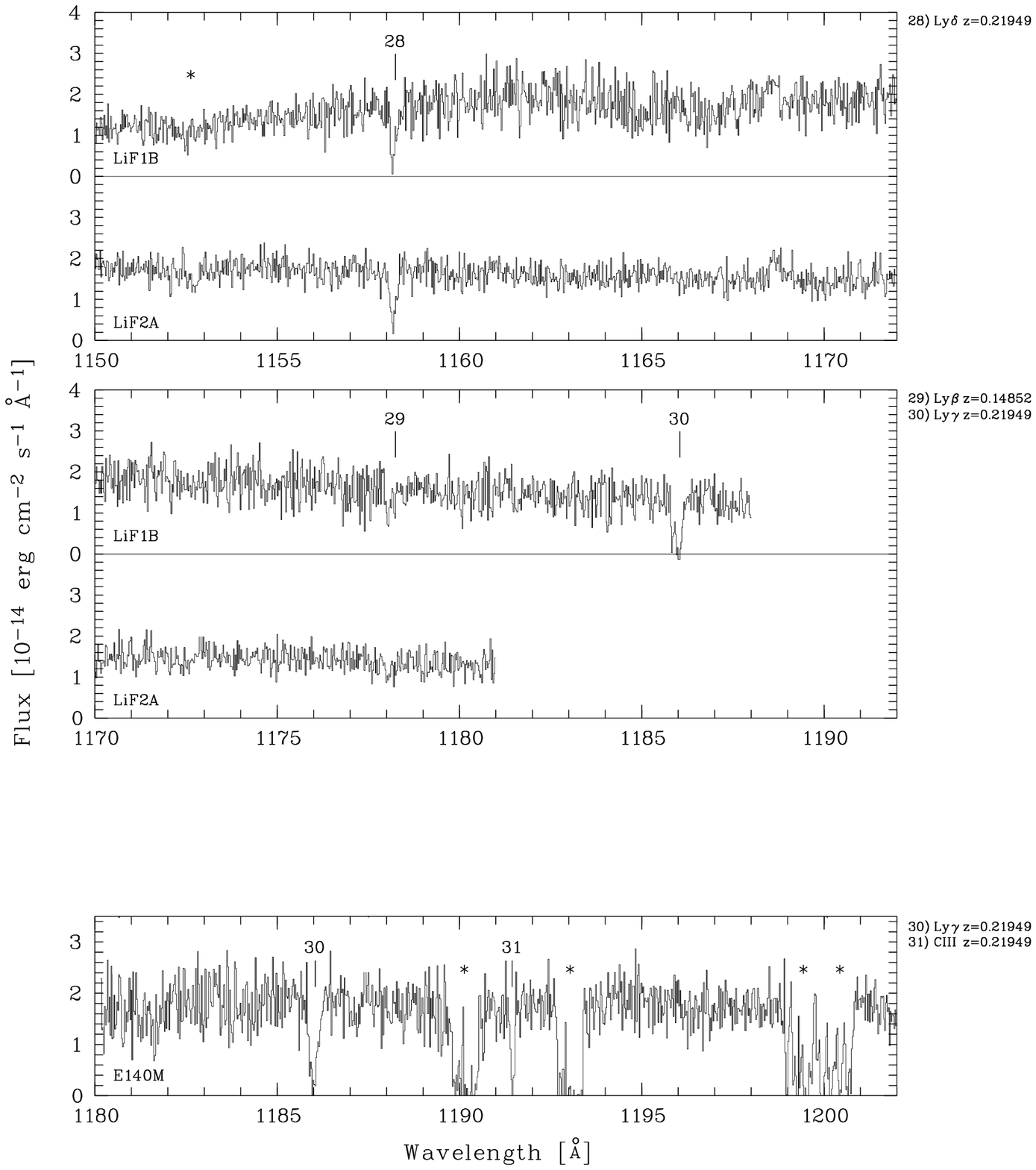}
\figcaption[f06.eps]{FUV spectrum of PG\,1259+593 (see caption to Fig.\,2).}

\clearpage
\newpage
\includegraphics{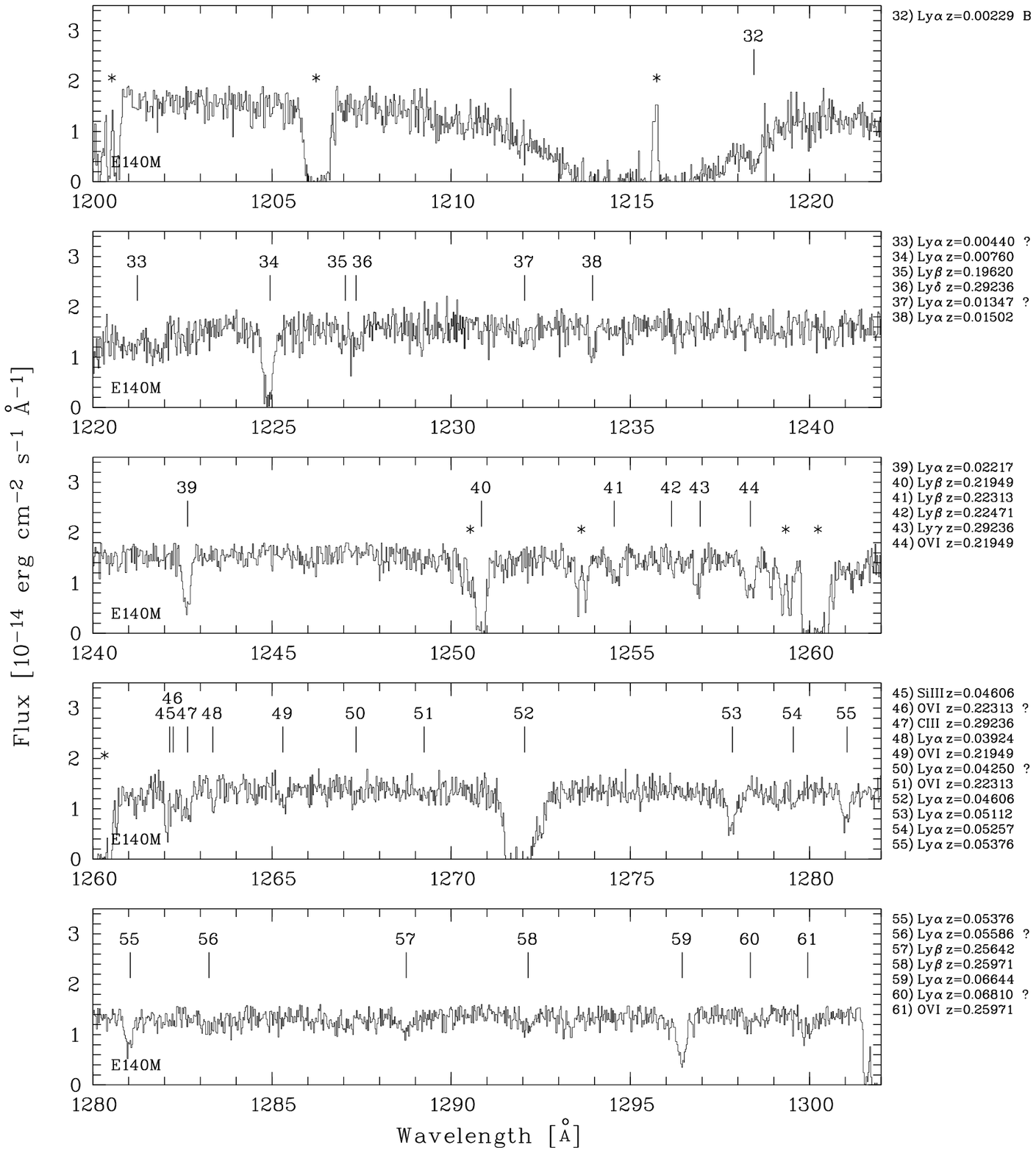}
\figcaption[f07.eps]{FUV spectrum of PG\,1259+593 (see caption to Fig.\,2).}

\clearpage
\newpage
\includegraphics{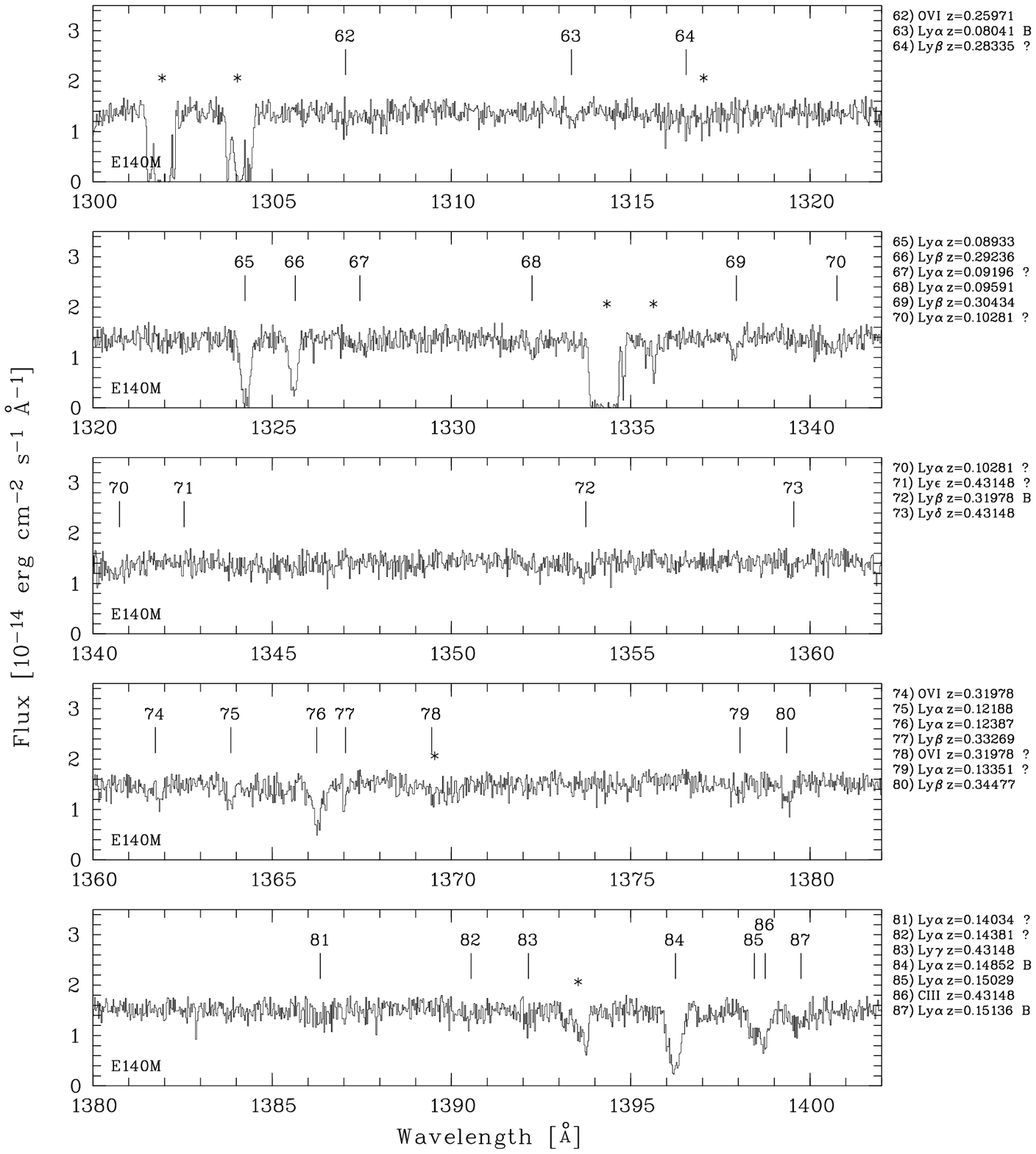}
\figcaption[f08.eps]{FUV spectrum of PG\,1259+593 (see caption to Fig.\,2).}

\clearpage
\newpage
\includegraphics{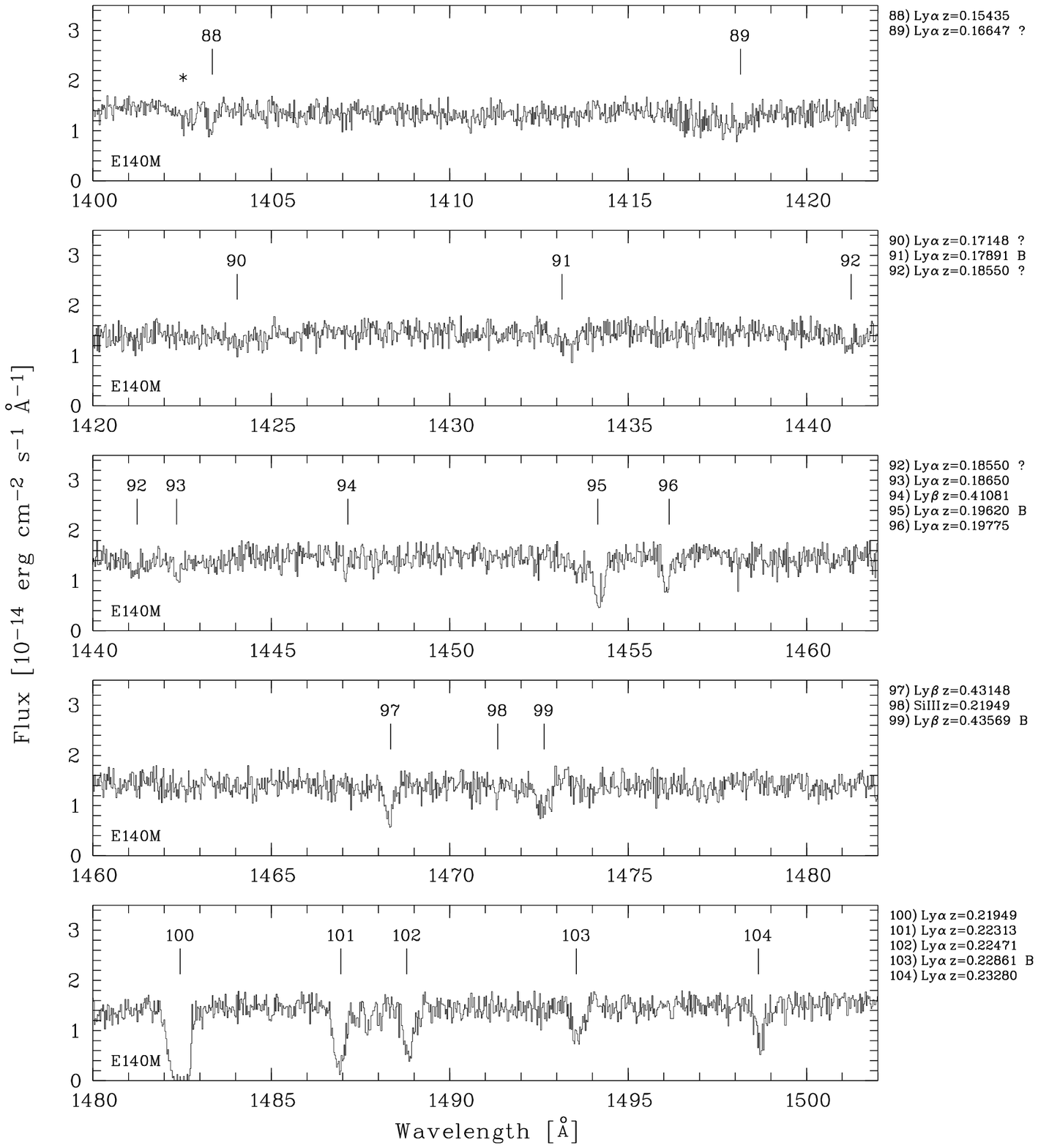}
\figcaption[f09.eps]{FUV spectrum of PG\,1259+593 (see caption to Fig.\,2).}

\clearpage
\newpage
\includegraphics{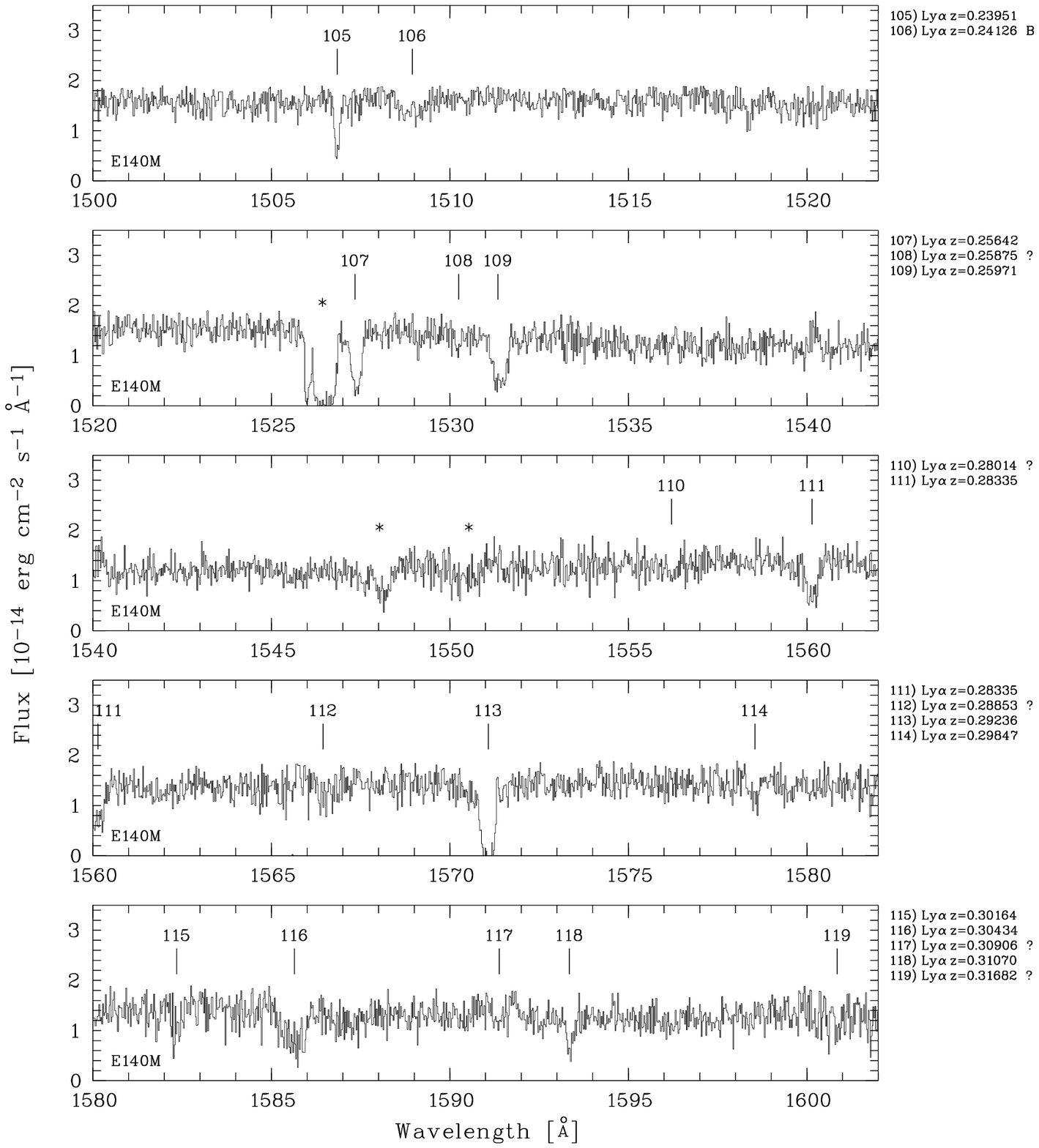}
\figcaption[f10.eps]{FUV spectrum of PG\,1259+593 (see caption to Fig.\,2).}

\clearpage
\newpage
\includegraphics{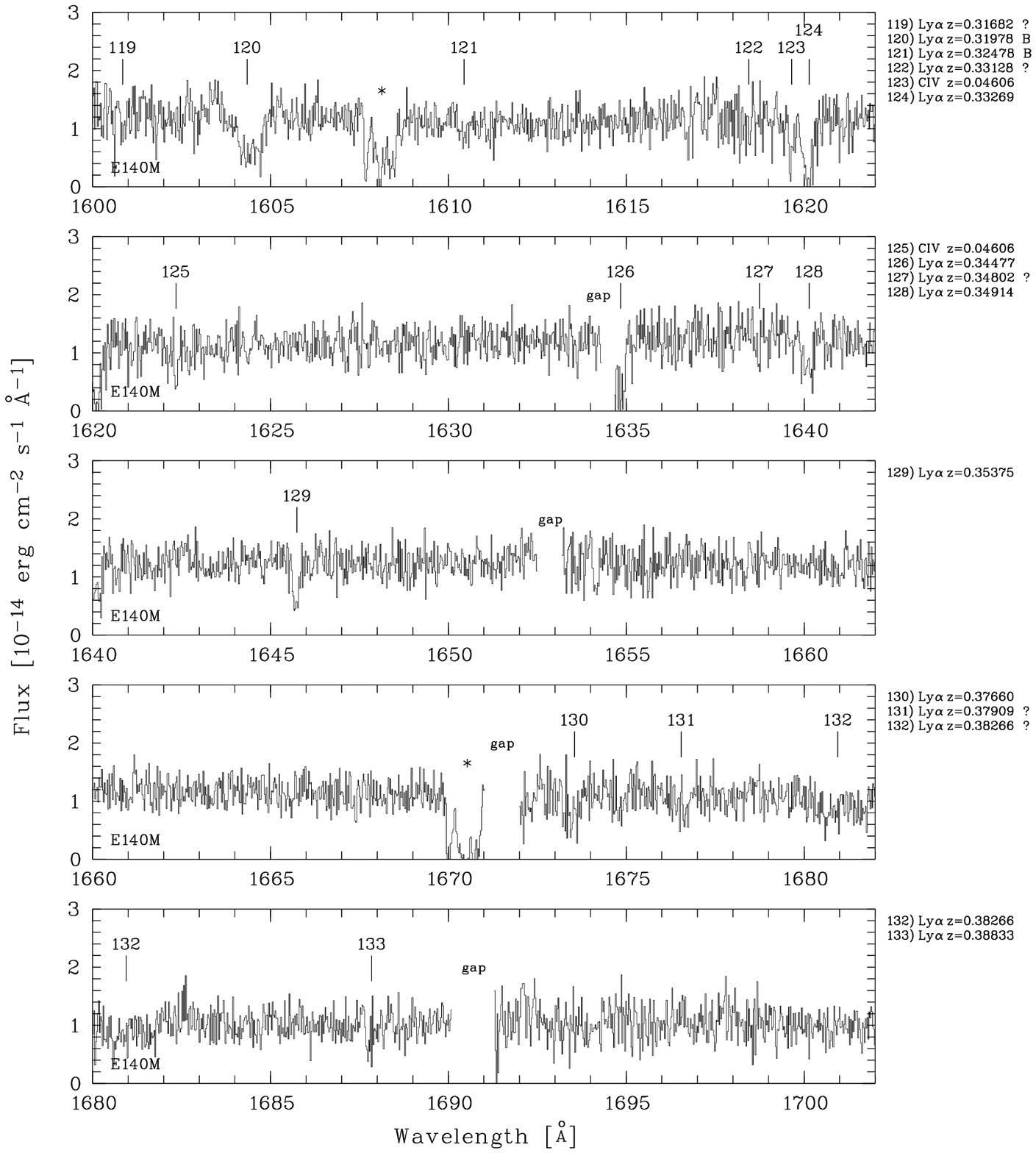}
\figcaption[f11.eps]{FUV spectrum of PG\,1259+593 (see caption to Fig.\,2).}

\clearpage
\newpage
\includegraphics{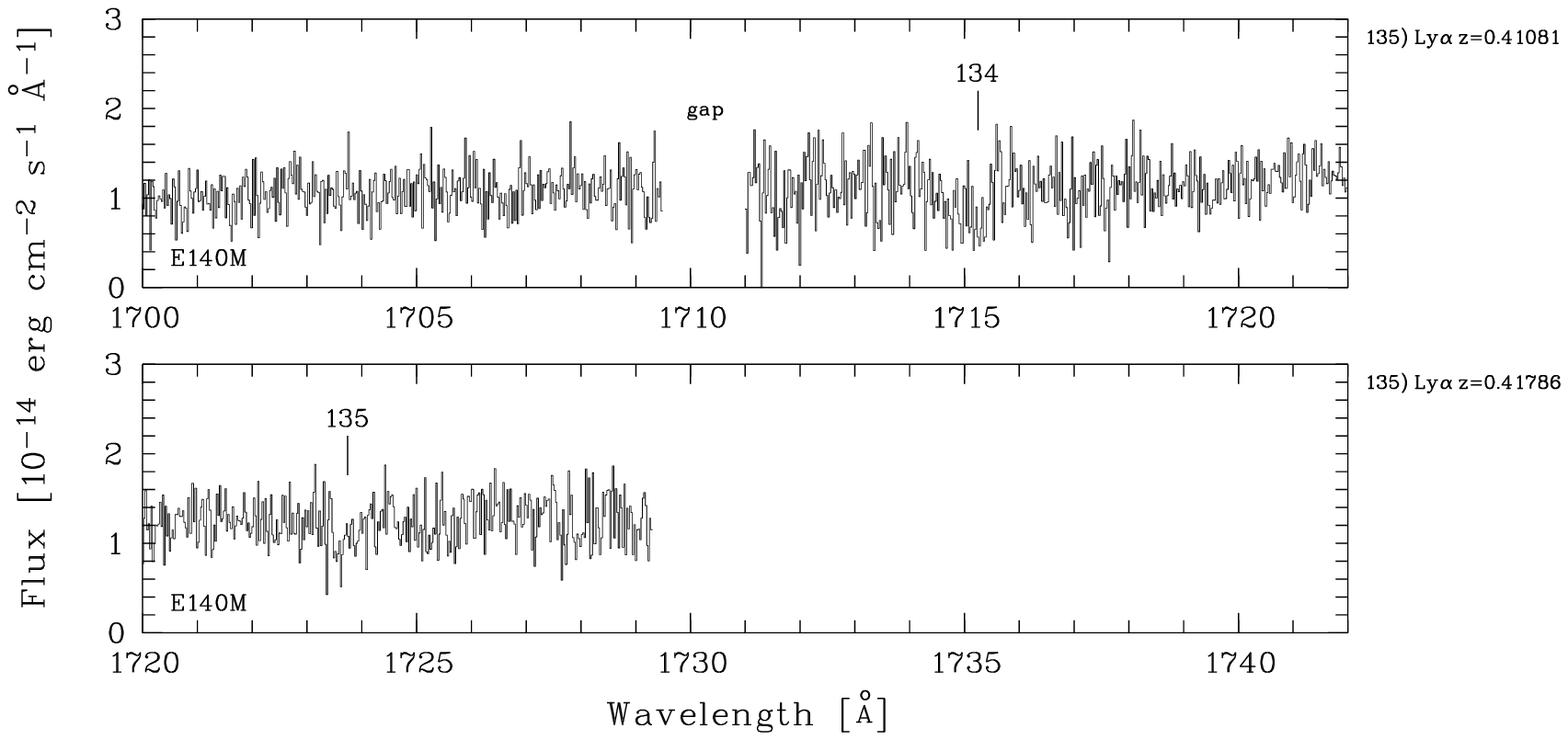}
\figcaption[f12.eps]{FUV spectrum of PG\,1259+593 (see caption to Fig.\,2).}

\clearpage
\newpage
\includegraphics{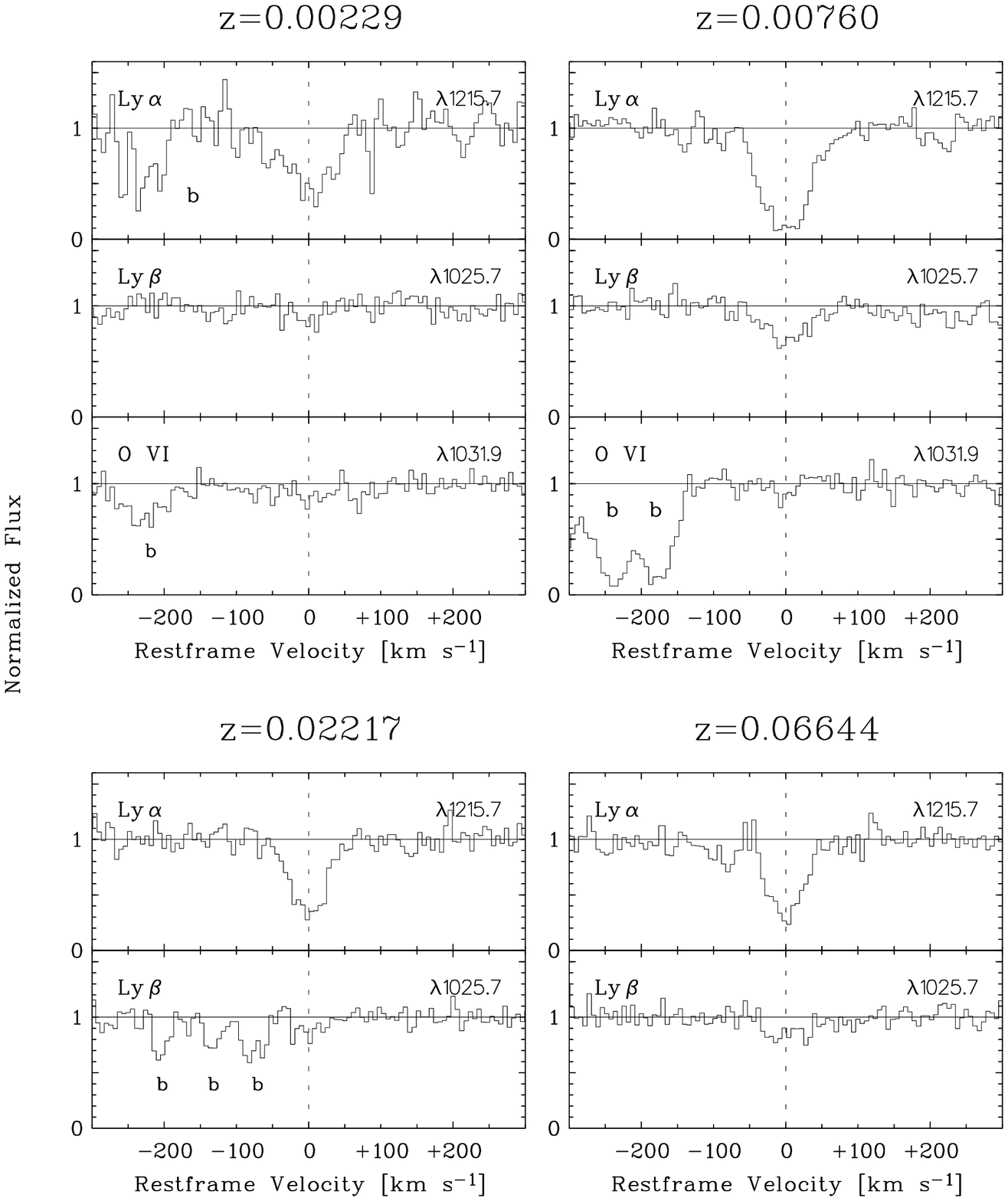}
\figcaption[f13.eps]{Continuum-normalized absorption profiles of multiple-line
IGM absorbers towards PG\,1259+593 plotted versus rest-frame velocities.
For this plot, the data
has been binned to $6$ km\,s$^{-1}$ wide pixels.}

\clearpage
\newpage
\includegraphics{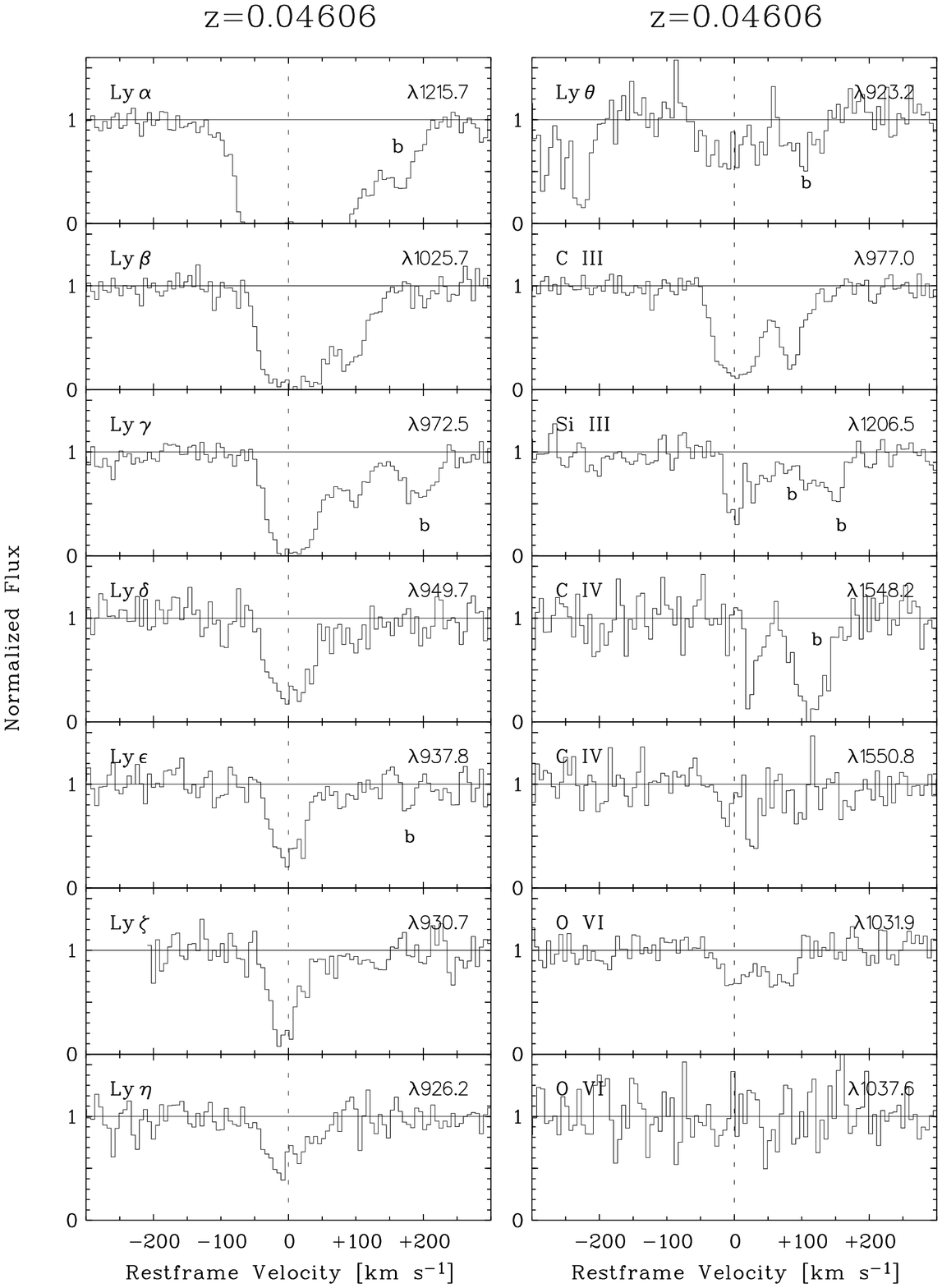}
\figcaption[f14.eps]{Continuum-normalized absorption profiles of multiple-line
IGM absorbers towards PG\,1259+593 (ctd).}

\clearpage
\newpage
\includegraphics{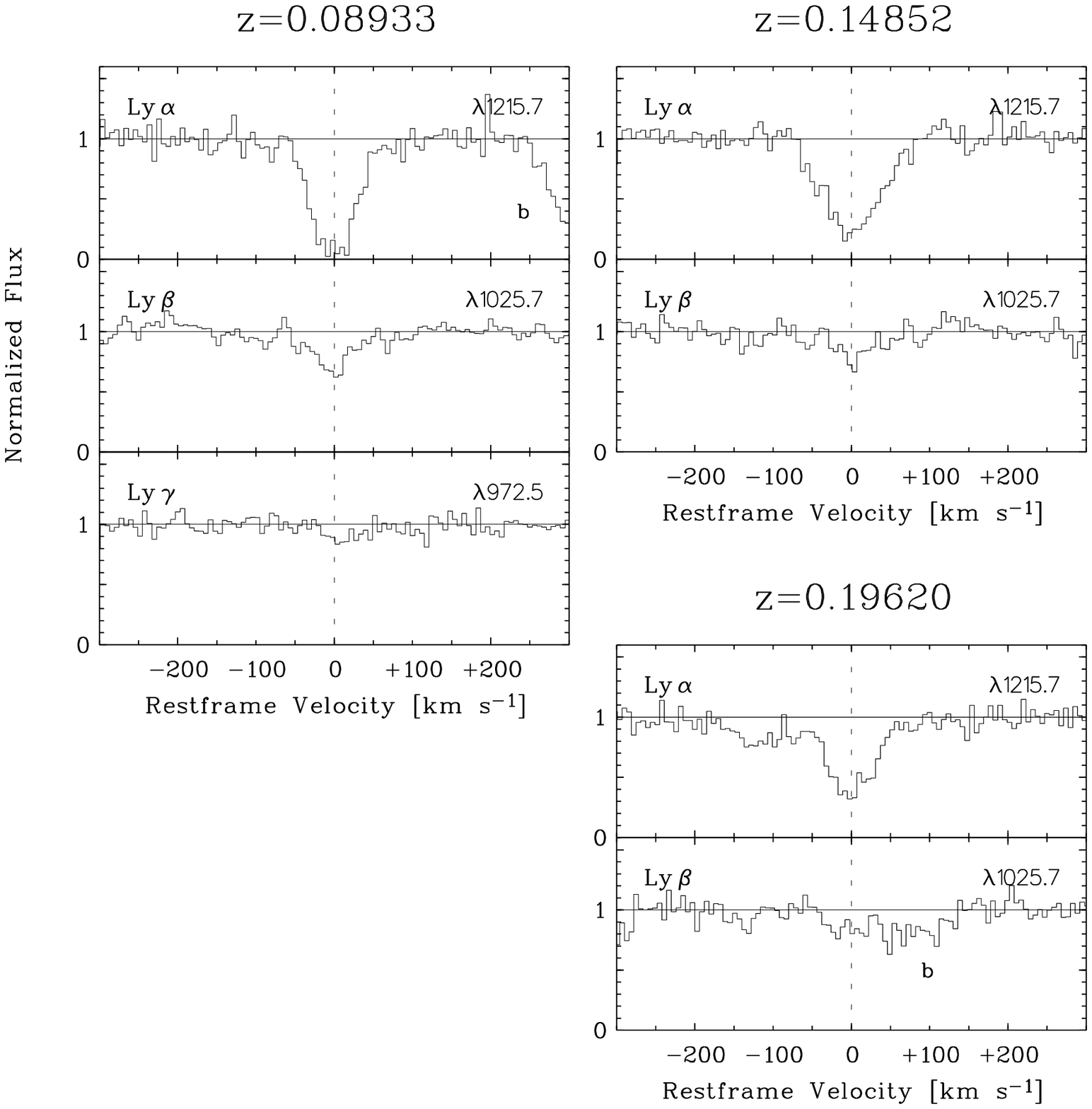}
\figcaption[f15.eps]{Continuum-normalized absorption profiles of multiple-line
IGM absorbers towards PG\,1259+593 (ctd).}

\clearpage
\newpage
\includegraphics{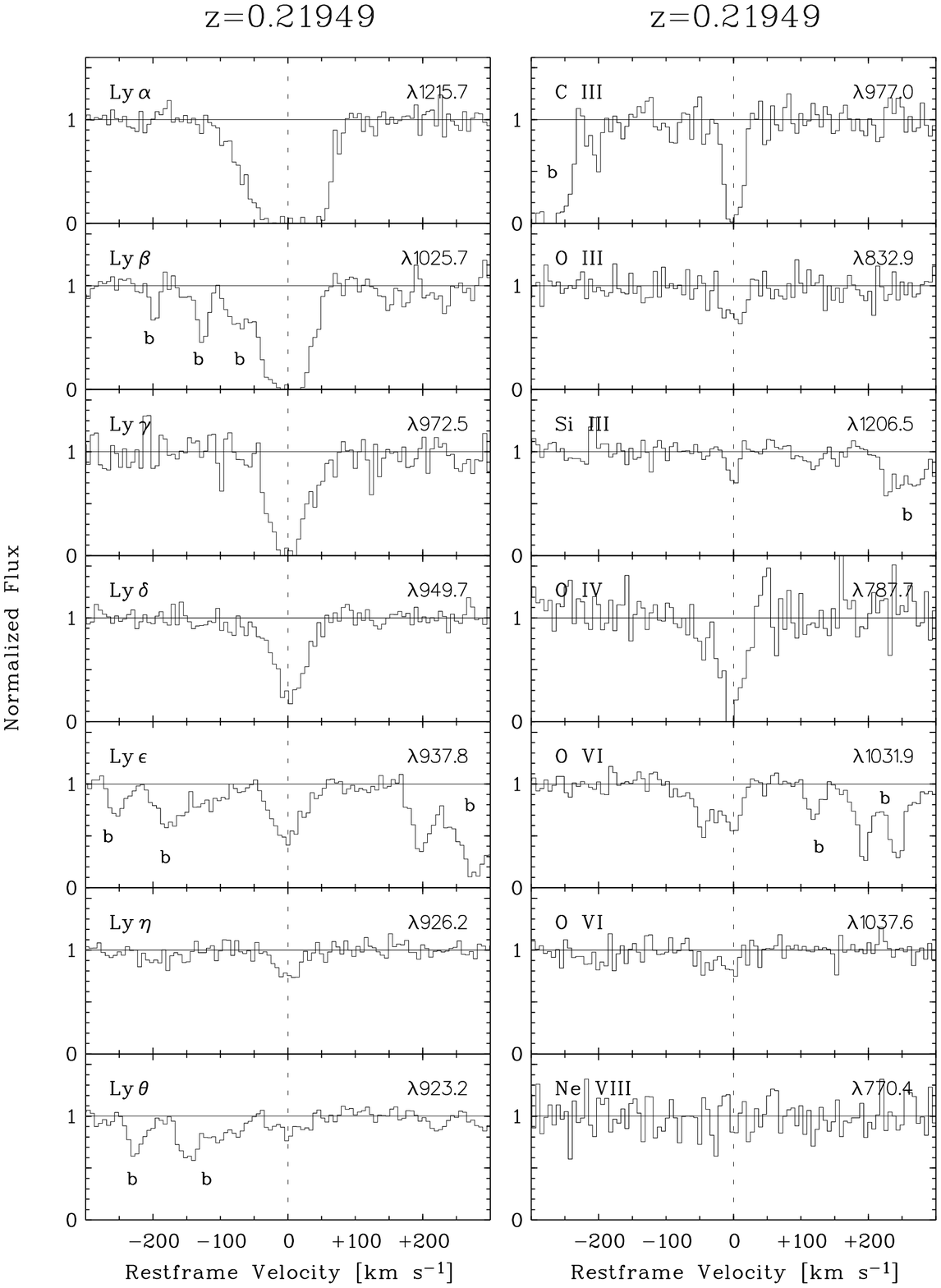}
\figcaption[f16.eps]{Continuum-normalized absorption profiles of multiple-line
IGM absorbers towards PG\,1259+593 (ctd).}

\clearpage
\newpage
\includegraphics{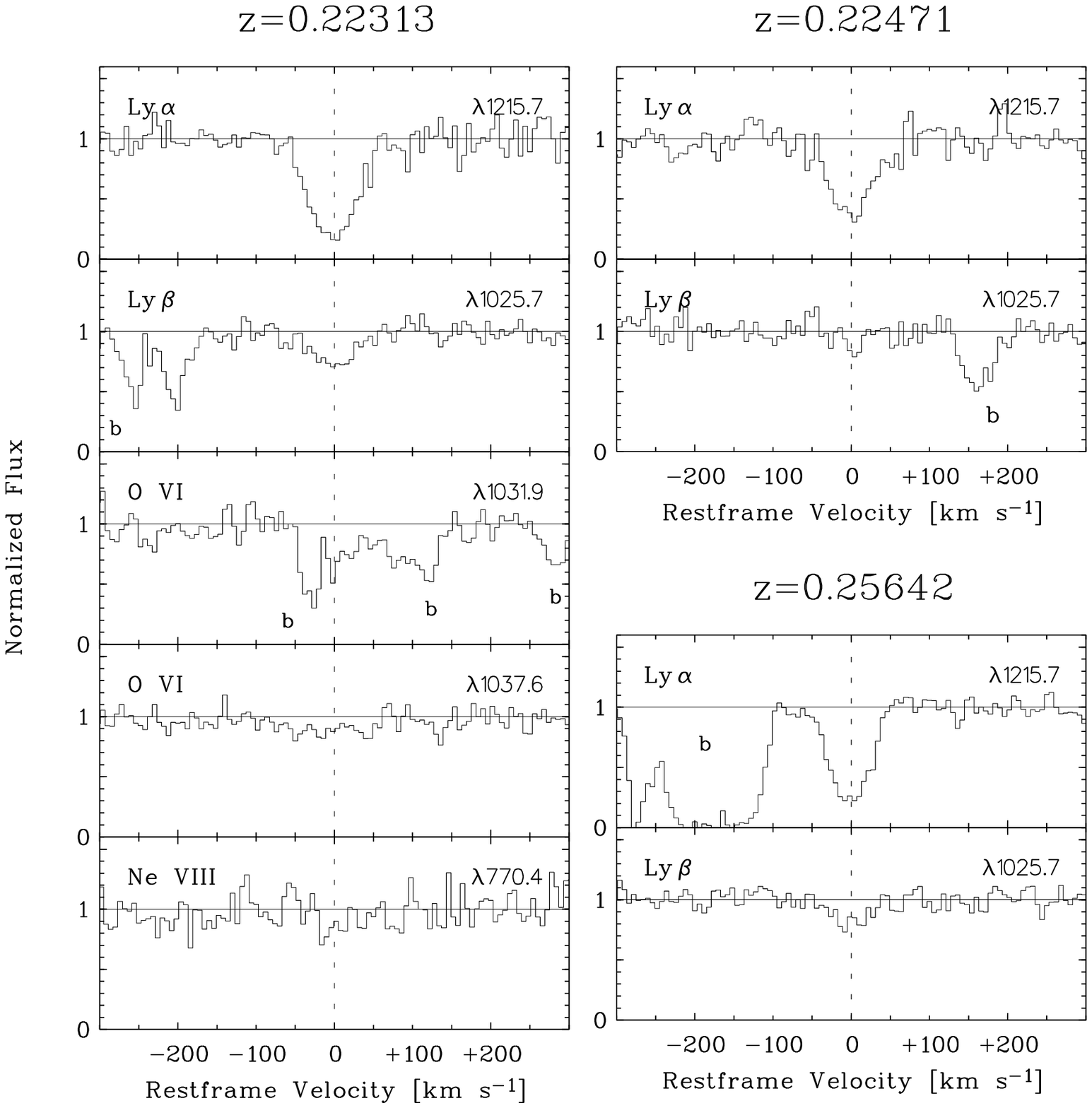}
\figcaption[f17.eps]{Continuum-normalized absorption profiles of multiple-lineIGM absorbers towards
PG\,1259+593 (ctd).}

\clearpage
\newpage
\includegraphics{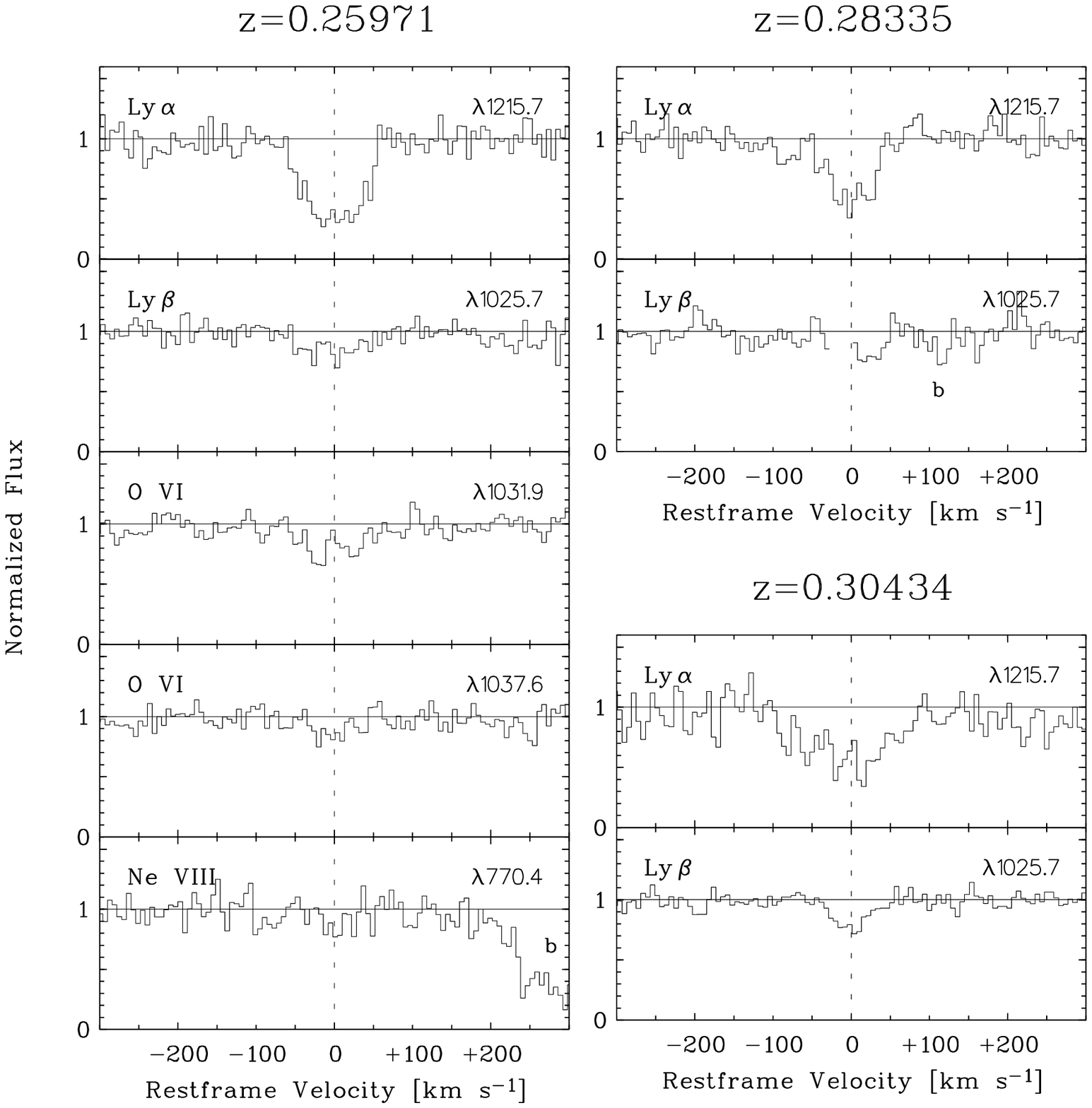}
\figcaption[f18.eps]{Continuum-normalized absorption profiles of multiple-line
IGM absorbers towards PG\,1259+593 (ctd).}

\clearpage
\newpage
\includegraphics{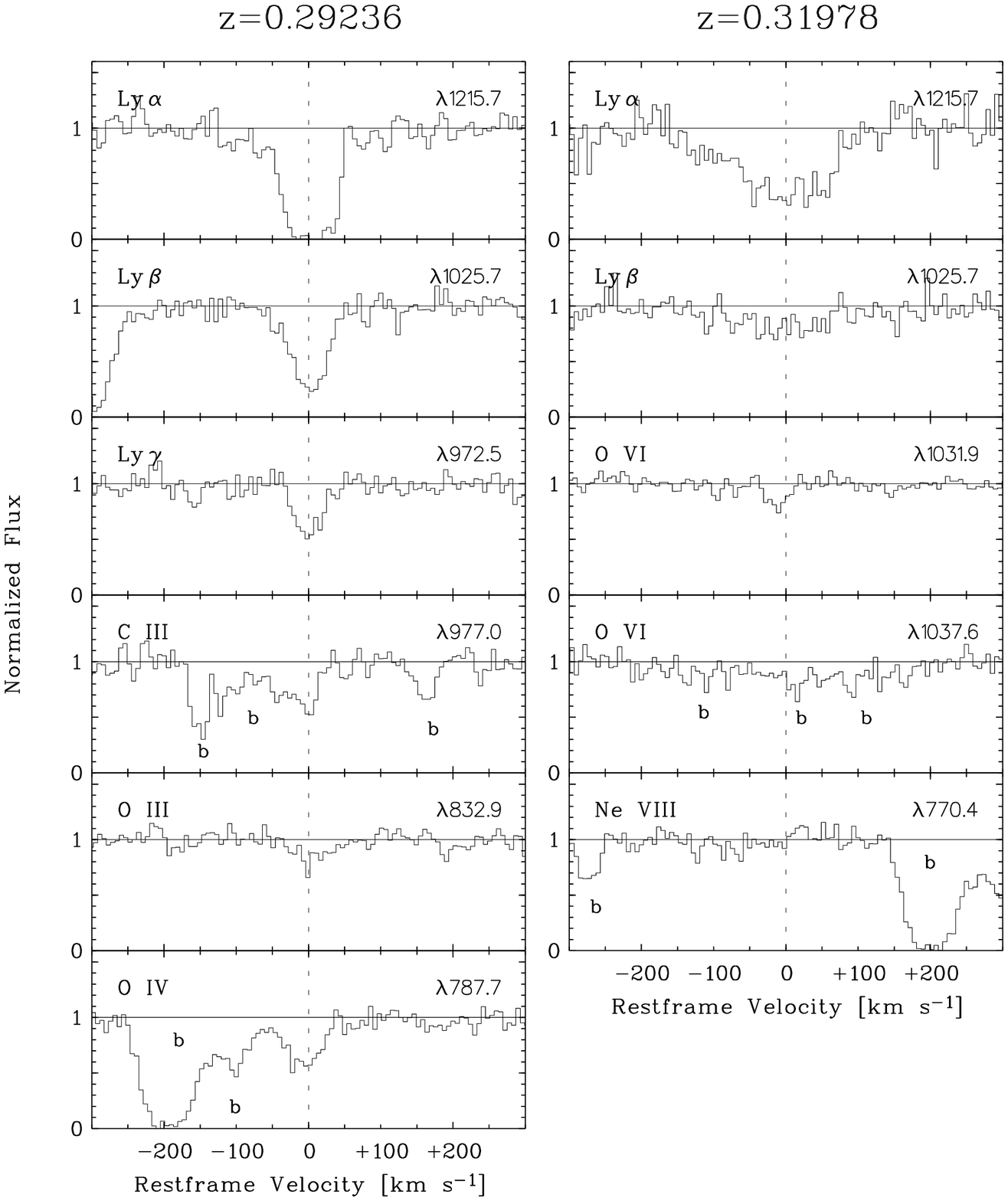}
\figcaption[f19.eps]{Continuum-normalized absorption profiles of multiple-line
IGM absorbers towards PG\,1259+593 (ctd).}

\clearpage
\newpage
\includegraphics{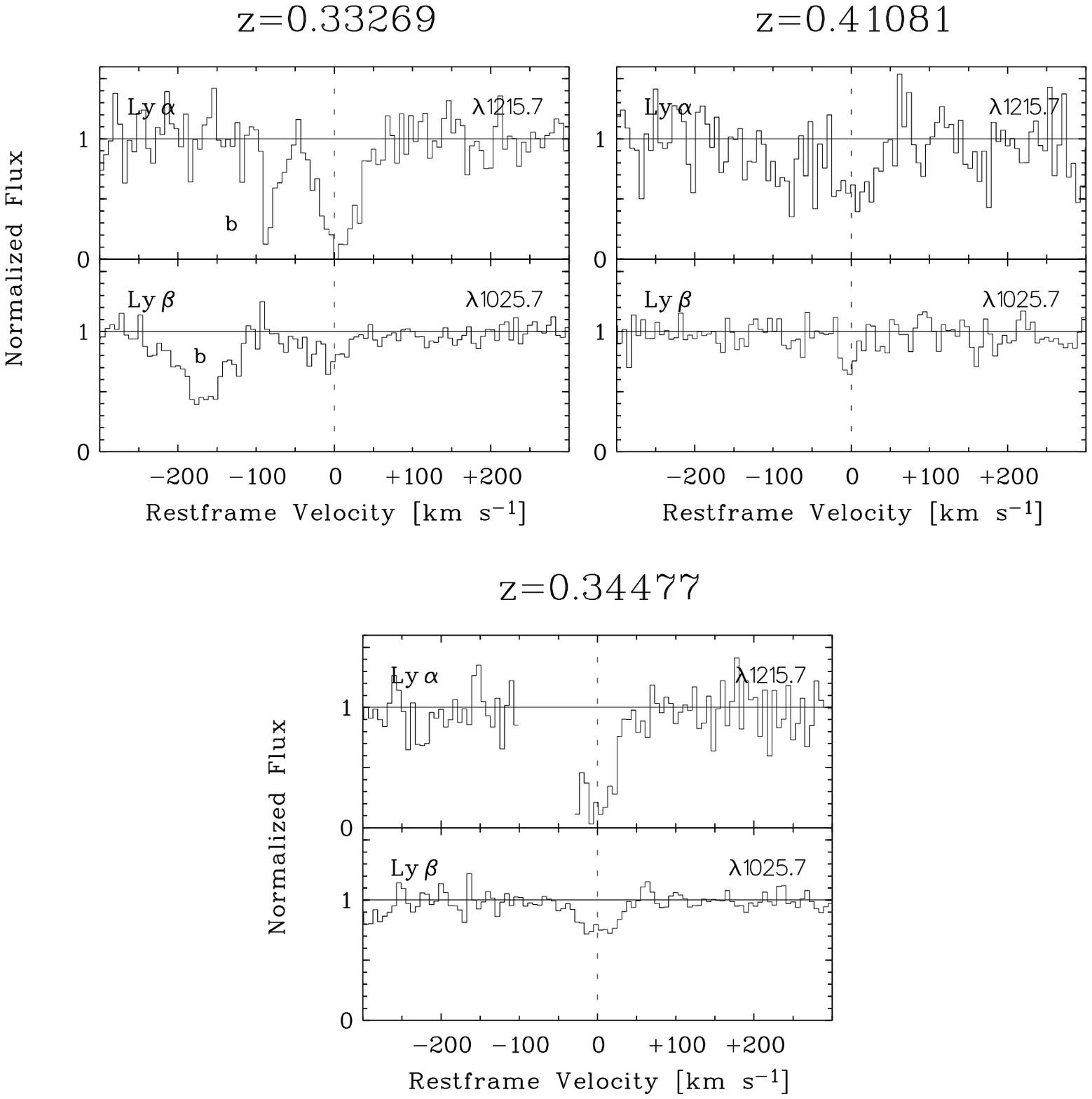}
\figcaption[f20.eps]{Continuum-normalized absorption profiles of multiple-line
IGM absorbers towards PG\,1259+593 (ctd).}

\clearpage
\newpage
\includegraphics{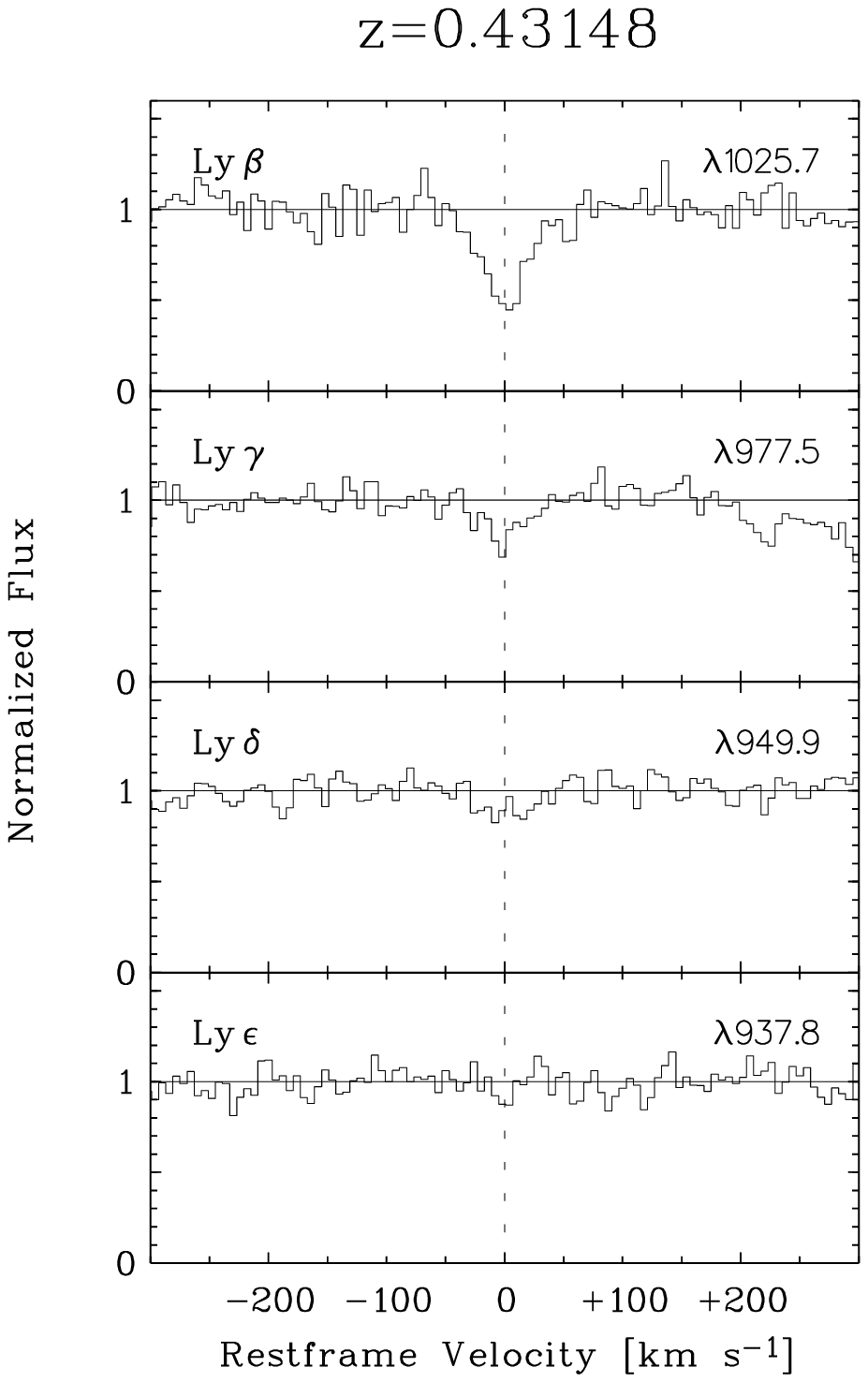}
\figcaption[f21.eps]{Continuum-normalized absorption profiles of multiple-line
IGM absorbers towards PG\,1259+593 (ctd).}

\clearpage
\newpage
\includegraphics{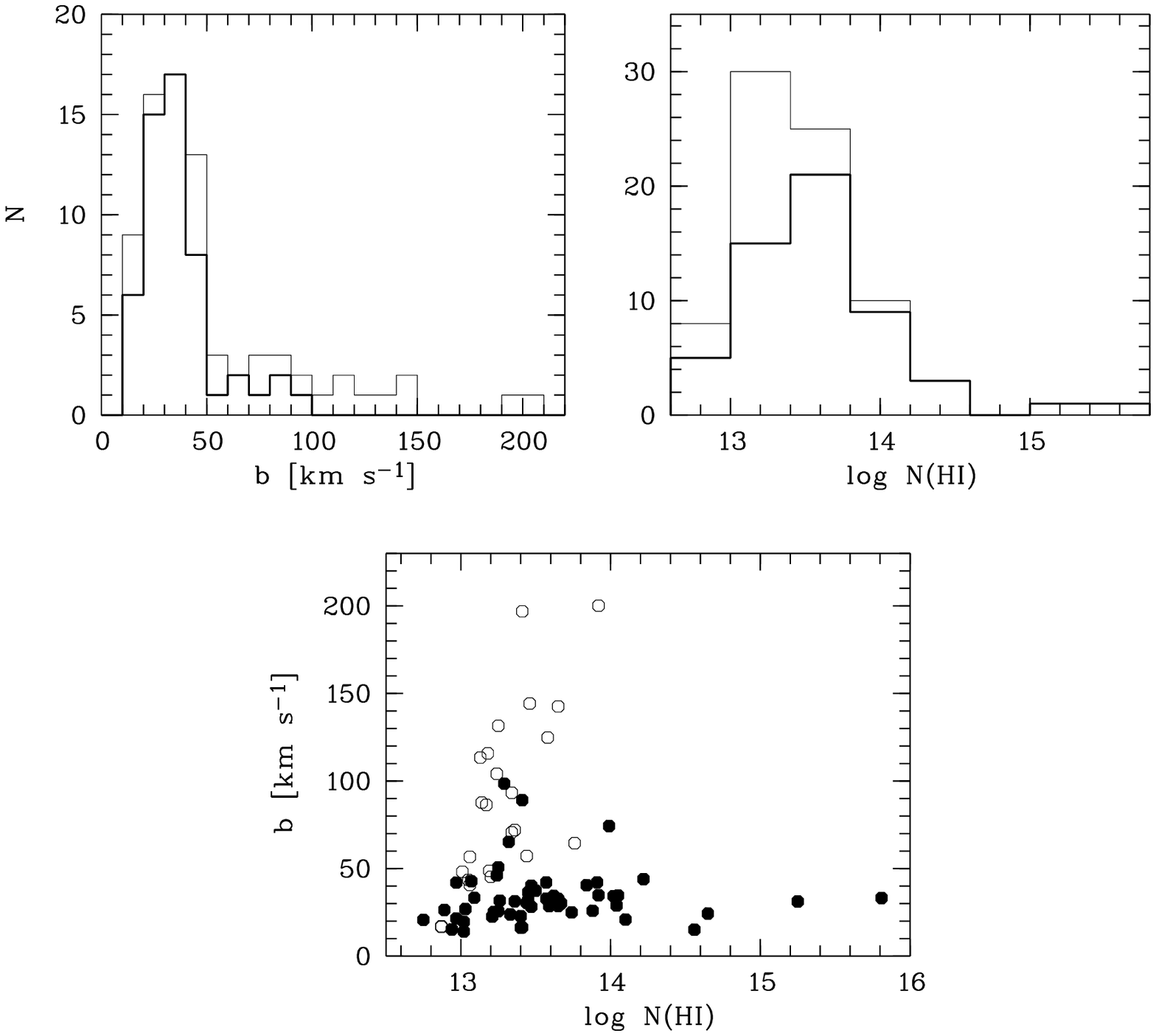}
\figcaption[f22.eps]{{\it Upper two panels:} Number 
Distribution of H\,{\sc i} $b$-values and column densities
for all absorption systems (upper left panel and upper right panel, respectively). 
The thin lines show the distributions for the total sample of available $b$ values
and column densities, whereas the thick lines show the distributions for $b$ and
log $N$ excluding the uncertain cases.
{\it Lower Panel:} 
$b$ value versus column density. Data points based on uncertain estimates for $b$ and
log $N$ (see Table\,5 and \S3.4.2) are plotted as open circles; data points based on
reliable estimates for these parameters are shown as filled circles.}

\clearpage
\newpage
\includegraphics{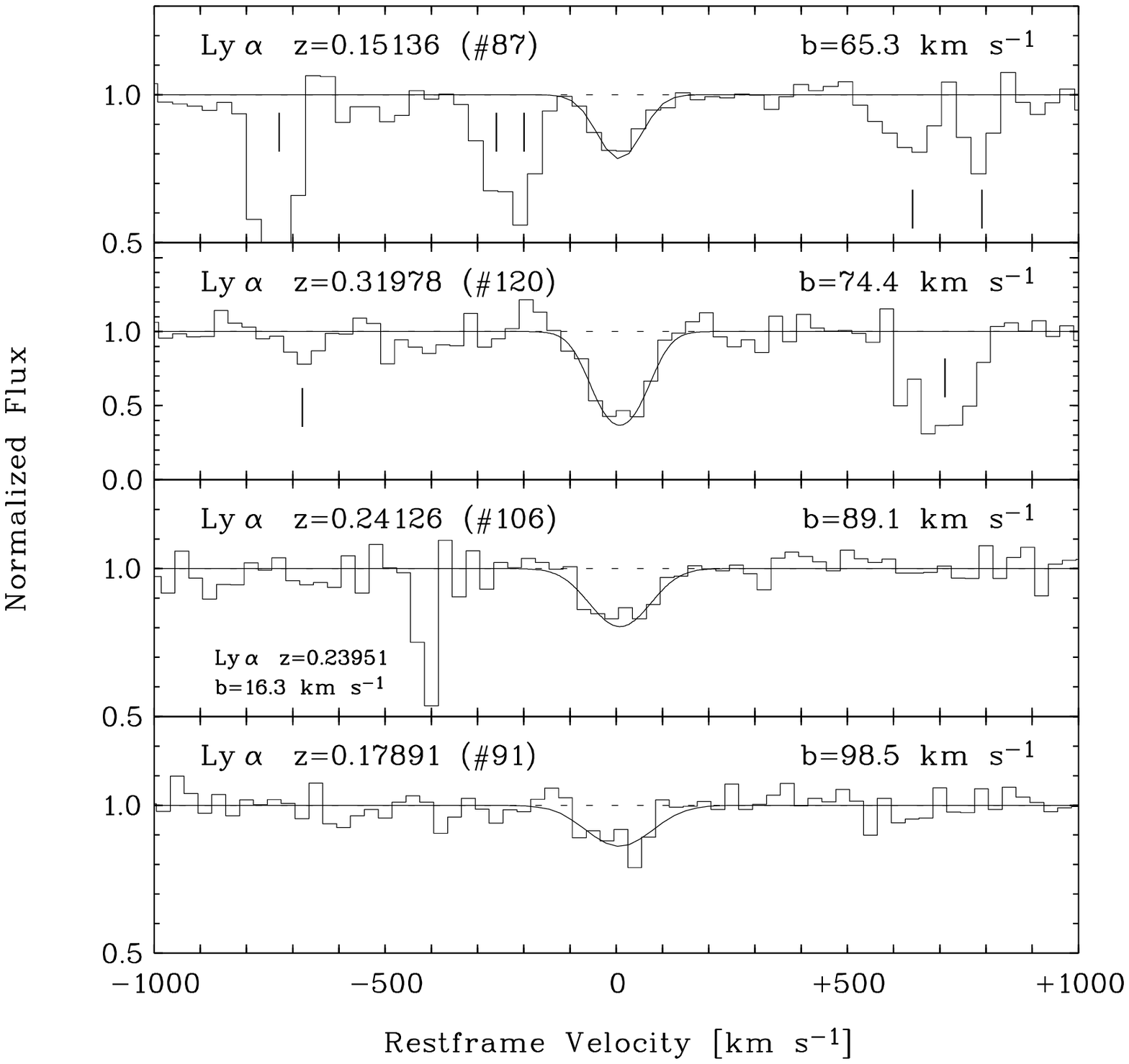}
\figcaption[f23.eps]{Examples for Voigt-profile
fits of broad Ly\,$\alpha$ lines towards PG\,1259+593 (see Table 6).
For this plot, the data has been binned to $30$ km\,s$^{-1}$ wide 
pixels. Note that the fitting was performed using unbinned data. Other
identified absorption lines in the vicinity of the broad absorption
features are labeled in the plot or are indicated with tic marks.}

\clearpage
\newpage
\includegraphics{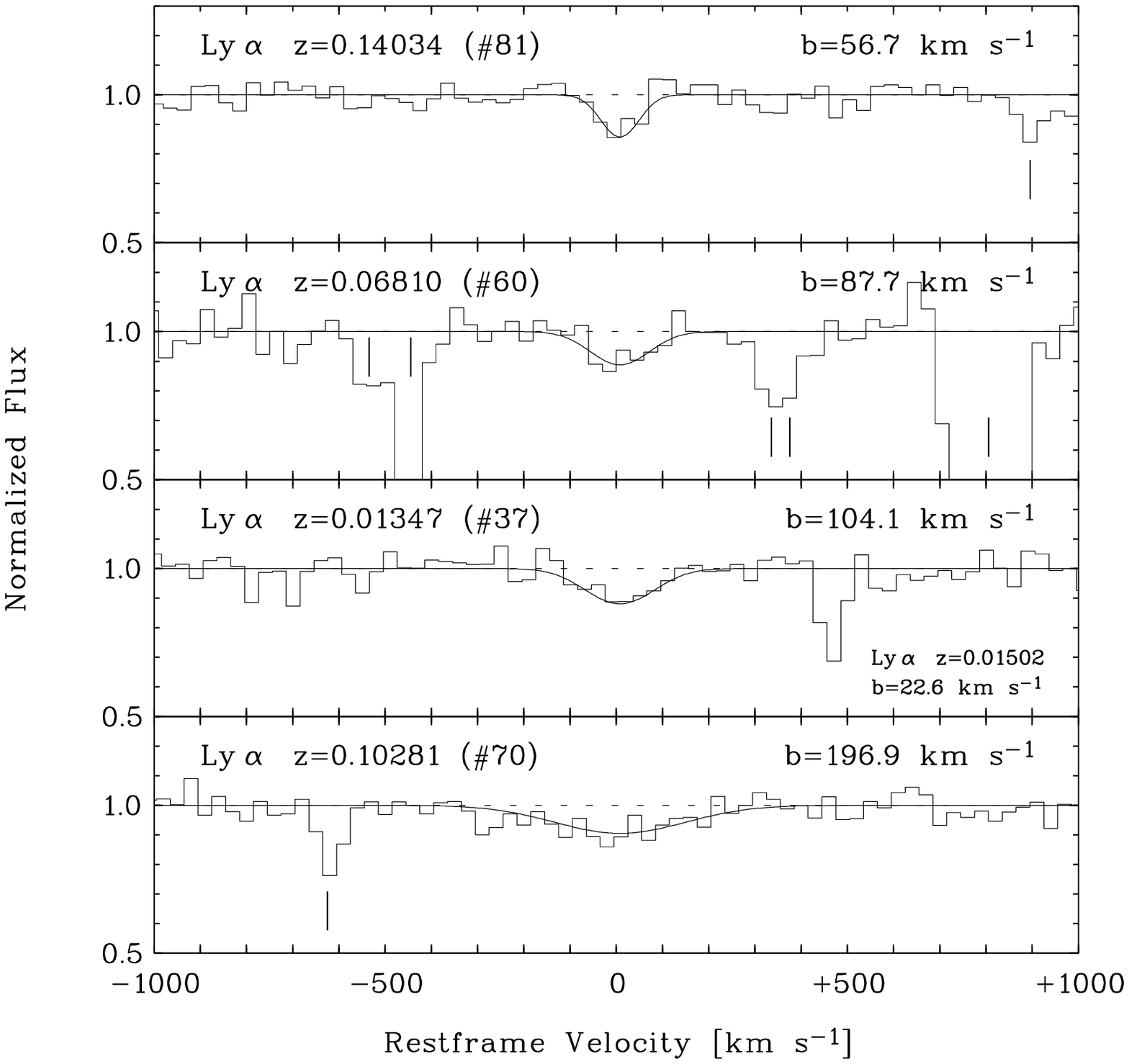}
\figcaption[f24.eps]{Examples for Voigt-profile
fits of broad Ly\,$\alpha$ candidate lines (uncertain cases) towards PG\,1259+593 
(see Tables 3 and 5). For this plot, the data has been binned to $30$ km\,s$^{-1}$ wide
pixels. Note that the fitting was performed using unbinned data.}

\clearpage
\newpage
\includegraphics{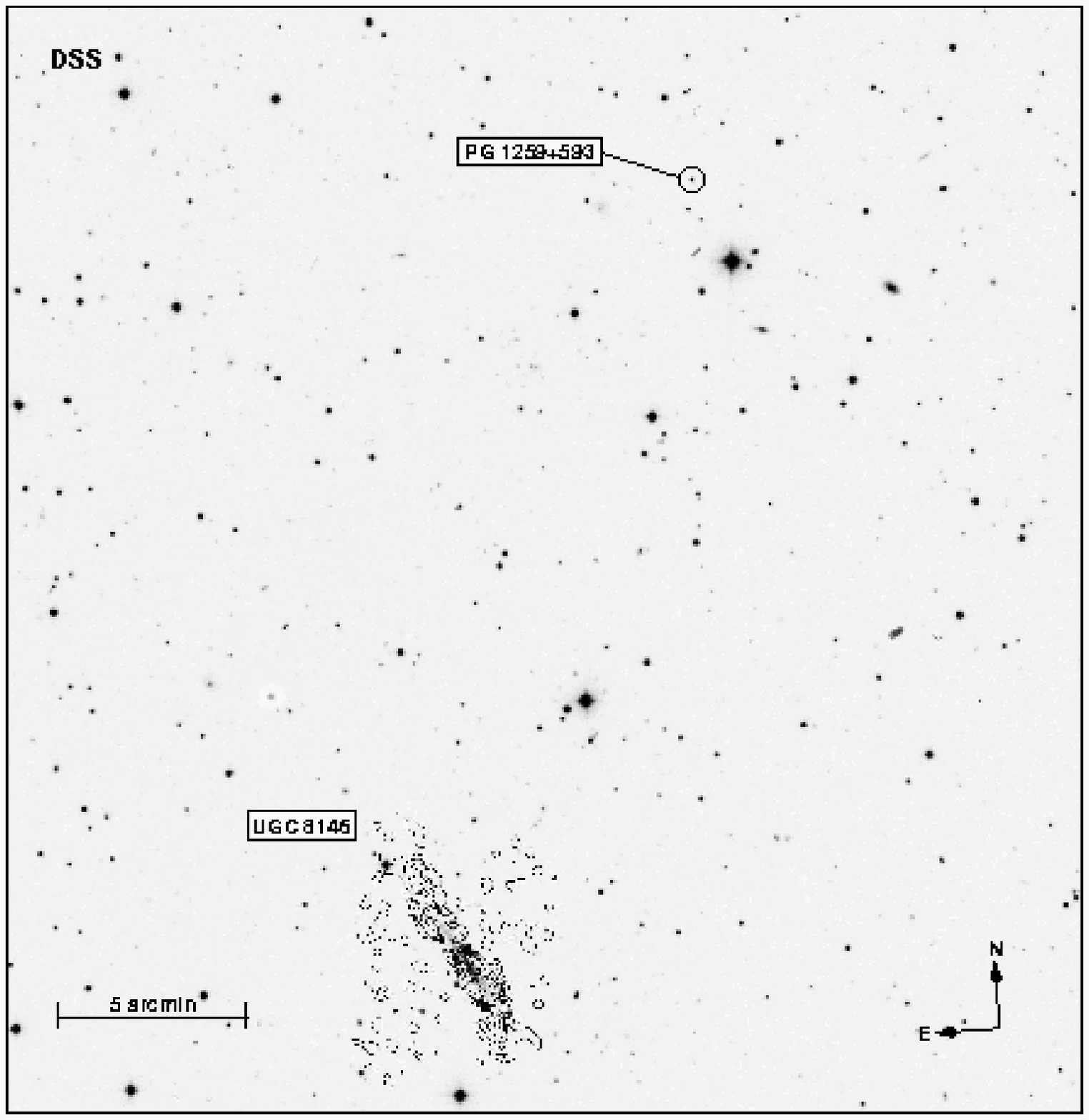}
\figcaption[f25.eps]{Sky positions of PG\,1259+593 and UGC\,08146 on
an image from the {\it Digitized Sky Survey}. H\,{\sc i} contours for 
UGC\,08146 from data of the {\it Westerbork Synthesis Radio Telescope}
(Rhee \& van Albada 1996) are overlayed. A HIGHER RESOLUTION VERSION OF THIS
FIGURE IS AVAILABLE ON REQUEST.}

\clearpage
\newpage
\begin{deluxetable}{lrrrrr} \tabletypesize{\normalsize}
\tablecaption{Log of FUSE and STIS Observations of PG\,1259+593$^{\rm a}$
\label{tbl-2}}
\tablewidth{0pt}
\tablehead{
\colhead{Instrument} & \colhead{Data Set ID} & \colhead{Obs.\,Date} & \colhead{Aperture}
& \colhead{Exp. Time} & \colhead{Notes$^{\rm b}$}\\
\colhead{} & \colhead{} & \colhead{} & \colhead{} & \colhead{(ks)} & \colhead{}
}
\startdata
FUSE		& P1080101 & 2000-Feb-25 & $30\farcs0 \times
30\farcs0$ & 52.4 & 1 \\
FUSE		& P1080102 & 2000-Dec-25 & $30\farcs0 \times
30\farcs0$ & 56.8 & 1,2 \\
FUSE		& P1080103 & 2001-Jan-29 & $30\farcs0 \times
30\farcs0$ & 52.0 & 1,3 \\
FUSE		& P1080104 & 2001-Mar-12 & $30\farcs0 \times
30\farcs0$ & 105.9 & 1 \\
FUSE		& P1080105 & 2001-Mar-14 & $30\farcs0 \times
30\farcs0$ & 104.0 & 1 \\
FUSE		& P1080106 & 2001-Mar-17 & $30\farcs0 \times
30\farcs0$ & 63.3 & 1,3 \\
FUSE		& P1080107 & 2001-Mar-19 & $30\farcs0 \times
30\farcs0$ & 95.7 & 1 \\
FUSE		& P1080108 & 2001-Mar-22 & $30\farcs0 \times
30\farcs0$ & 33.5 & 1 \\
FUSE		& P1080109 & 2001-Mar-28 & $30\farcs0 \times
30\farcs0$ & 32.7 & 1 \\
\\
STIS-E140M		& O63G05	& 2001-Jan-17 & $0\farcs2 \times
0\farcs06$  & 14.4 & \\
STIS-E140M		& O63G06	& 2001-Jan-17 & $0\farcs2 \times
0\farcs06$  & 14.4 & \\
STIS-E140M		& O63G07	& 2001-Jan-18 & $0\farcs2 \times
0\farcs06$  & 14.4 & \\
STIS-E140M		& O63G08	& 2001-Jan-18 & $0\farcs2 \times
0\farcs06$  & 14.4 & \\
STIS-E140M		& O63G09	& 2001-Jan-19 & $0\farcs2 \times
0\farcs06$  & 14.4 & \\
STIS-E140M		& O63G11	& 2001-Jan-19 & $0\farcs2 \times
0\farcs06$  &  9.1 & \\
STIS-E140M		& O63G10	& 2001-Dec-19 & $0\farcs2 \times
0\farcs06$  & 14.4 & \\
\enddata

\tablenotetext{a}{This table summarizes the FUSE and STIS observations of
PG\,1259+593. A more detailed observation log is presented in Sembach et
al.\,(2004).}
\tablenotetext{b}{Notes:\\ 1) Exposure time listed represents the exposure time for the
LiF\,1A channel. The exposure times for the other three channels may be
significantly different (see Table 1 in Sembach et al.\,2004). \\
2) Alignment problems - no LiF\,2, SiC\,1 and SiC\,2 data. \\ 
3) Detector high voltage problems during observations.} 
\end{deluxetable}

\clearpage
\newpage
\begin{deluxetable}{lrr}
\tabletypesize{\normalsize}
\tablecaption{Atomic Data for IGM Absorption Lines$^{\rm a}$ \label{tbl-2}}
\tablewidth{0pt}
\tablehead{
\colhead{ID} & \colhead{$\lambda_{\rm vac}$} & \colhead{log $\lambda f$}\\
\colhead{}   & \colhead{[\AA]} & \colhead{}
}
\startdata
H\,{\sc i} Ly\,$\alpha$ \dotfill	&	1215.670 &	2.704 \\
H\,{\sc i} Ly\,$\beta$  \dotfill	&	1025.722 &	1.909 \\
H\,{\sc i} Ly\,$\gamma$ \dotfill	&	972.537 &	1.450 \\
H\,{\sc i} Ly\,$\delta$ \dotfill	&	949.743 &	1.122 \\
H\,{\sc i} Ly\,$\epsilon$  \dotfill     &	937.803 &	0.864 \\
H\,{\sc i} Ly\,$\zeta$   \dotfill	&	930.748 &	0.651 \\
H\,{\sc i} Ly\,$\eta$    \dotfill	&	926.226 &	0.470 \\
H\,{\sc i} Ly\,$\theta$   \dotfill      &	923.150 &	0.311 \\
H\,{\sc i} Ly\,$\iota$    \dotfill      &	920.963 &	0.170 \\
H\,{\sc i} Ly\,$\kappa$  \dotfill	&	919.351 &	0.043 \\
C\,{\sc iii}      \dotfill	&	977.020 &	2.870 \\
C\,{\sc iv}       \dotfill	&	1550.777 &	2.169 \\
	                        &	1548.203 &	2.470 \\
N\,{\sc iv}       \dotfill	&	765.148 &	2.673 \\
O\,{\sc iii}      \dotfill	&	832.927 &	1.950 \\
O\,{\sc iv}       \dotfill	&	787.711 &	1.942 \\
O\,{\sc vi}       \dotfill	&	1037.617 &	1.835 \\
	                        &	1031.926 &	2.136 \\
Ne\,{\sc viii}      \dotfill	&	780.324 &	1.596 \\
	                        &	770.409 &	1.900 \\
Si\,{\sc iii}       \dotfill	&	1206.500 &	3.304 \\
\enddata
\tablenotetext{a}{Wavelengths and $f$ values are taken from Morton (2003, 1991) 
and Verner et al.\,(1994).} 
\end{deluxetable}

\clearpage
\newpage
\begin{deluxetable}{lrrrrrrrrrrl}
\rotate
\tabletypesize{\scriptsize}
\tablecaption{Multiple-Line IGM Absorbers toward PG\,1259+593$^{\rm a}$
\label{tbl-2}}
\tablewidth{0pt}
\tablehead{
\colhead{ID} & \colhead{No.} & \colhead{$\lambda_{\rm obs}$} & 
\colhead{$\lambda_{\rm vac}$\,$^{\rm b}$}
& \colhead{Instrument}
& \colhead{$W_{\rm r}$} & \colhead{$v_{-}$} & \colhead{$v_{+}$} & 
\colhead{log $N$} &  \colhead{$b$} & \colhead{Method$^{\rm c}$} & \colhead{Comments}\\
\colhead{} & \colhead{} & \colhead{[\AA]} & \colhead{[\AA]} & \colhead{} & \colhead{[m\AA]}
&  \colhead{[km\,s$^{-1}$]} &  \colhead{[km\,s$^{-1}$]} & \colhead{}
& \colhead{[km\,s$^{-1}$]} & \colhead{} & \colhead{}
}
\startdata

\hline
\multicolumn{12}{c}{$z=0.00229\pm0.00003$}\\
\hline

H\,{\sc i}-Ly\,$\alpha$	&  31 & 1218.449 &	1215.670 & STIS E140M
&	$190\pm24$ & $-110$ & $+80$  & $13.57\pm0.10$ & $42.1\pm4.4$ & PF & \\
	&	&		&
&       &	    &		 &	     & $13.67\pm0.05$ &
\nodata  & AOD & \\
	&	&		&
&       &	    &		 &	     & $13.61\pm0.06$ &
$44.0^{+9.4}_{-4.1}$ & COG & \\
H\,{\sc i}-Ly\,$\beta$	&  12 & 1028.058 &	1025.722 & FUSE LiF\,1A
&	$21\pm6$   & $-85$  & $+85$  & $13.51\pm0.11$ & \nodata  & AOD & \\
	&	&	&	&
&	   &        &	     & $13.61\pm0.06$ & $44.0^{+9.4}_{-4.1}$ & COG & \\
	&	& 1028.103 &		& FUSE LiF\,2B
&	$26\pm8$   & $-95$  & $+95$  & $13.64\pm0.12$ & \nodata  & AOD & \\
O\,{\sc vi}	&  14 & 1034.301 &	1031.926 & FUSE LiF\,1A
&	$63\pm22$  & $-110$ & $+120$ & $13.73\pm0.14$ & \nodata  & AOD & \\
	&	&			&		&
&           &        &        & $13.74\pm0.09$ & $44.0$   & COG$^{\rm d}$ & \\
	&	& 1034.312 &		& FUSE LiF\,2B
&	$46\pm24$  & $-130$ & $+150$ & $13.64\pm0.17$ & \nodata  & AOD & \\
O\,{\sc vi}    & \nodata & $\sim 1040.0$   &	1037.617 & FUSE LiF\,1A
& $\leq 51$  & $-110$ & $+150$ & \nodata        & \nodata  & \nodata & \\

\hline
\multicolumn{12}{c}{$z=0.00760\pm0.00002$}\\
\hline

H\,{\sc i}-Ly\,$\alpha$	&  33 & 1224.904 &	1215.670 & STIS E140M
& $301\pm15$ & $-130$ & $+130$ & $14.05\pm0.05$ & $34.6\pm2.0$ & PF & 
pos. velocity wing\\
	&	&	&	&
	&       &     &	        & $14.03\pm0.04$ & \nodata &       AOD & \\
	&	&	&		&
	&            &	   &		& $14.09\pm0.04$ &
$30.6^{+4.9}_{-4.2}$ & COG & \\
H\,{\sc i}-Ly\,$\beta$	&  13 & 1033.502 &	1025.722 & FUSE LiF\,1A
&	$75\pm6$   & $-100$ & $+100$ & $14.07\pm0.04$ & \nodata  & AOD & \\
	&	&	&		&
	&        &	&	   & $14.09\pm0.04$ &
$30.6^{+4.9}_{-4.2}$ & COG & \\
	&	& 1033.628 &		& FUSE LiF\,2B
& $60\pm9$   & $-100$ & $+100$ & $13.98\pm0.06$ & \nodata & AOD & \\
O\,{\sc vi}     &  15 & 1039.771 &      1031.926 & FUSE LiF\,1A
&       $15\pm4$  & $-50$ & $+50$ & $13.06\pm0.04$ & \nodata  & AOD & \\
                &     &          &               & FUSE LiF\,2B
&        $\leq9$  & $-50$ & $+50$ & \nodata        & \nodata  & AOD & \\

\hline 
\multicolumn{12}{c}{$z=0.02217\pm0.00002$}\\ 
\hline

H\,{\sc i}-Ly\,$\alpha$	&  39 &	1242.617 &	1215.670 & STIS E140M
& $160\pm9$  & $-90$  & $+90$  & $13.67\pm0.04$ & $30.2\pm2.2$
&	PF  & \\
	&	&	&	&
	&        &	    &	 & $13.63\pm0.04$ & \nodata & AOD \\
	&	&	&	&
	&     &	 &	    & $13.58\pm0.06$ & $34.1^{+11.1}_{-6.0}$ & COG & \\
H\,{\sc i}-Ly\,$\beta$	&  16 &	1048.421 &	1025.722 & FUSE LiF\,1A
& $20\pm4$   & $-45$  & $+60$  & $13.46\pm0.08$ & \nodata & AOD & \\
	&	&	&	&
	&        &	    &	 & $13.58\pm0.06$ & $34.1^{+11.1}_{-6.0}$ & COG & \\
	&	& $\sim$1048.4 &	& FUSE LiF\,2B
	& $\leq27$   & $-45$  & $+70$  & \nodata	 & \nodata & & \nodata \\

\hline
\multicolumn{12}{c}{$z=0.04606\pm0.00003$}\\
\hline

H\,{\sc i}-Ly\,$\alpha$	&  52 &	1271.812 &	1215.670 & STIS E140M
& $878\pm29$ & $-190$ & $+130$ & \nodata   & \nodata
& \nodata & incl. Comp. A \& B \\
	&	&	&	&
	&       &	&	& $15.58\pm0.21$ &
$47.6\pm12.4$ & PF & Comp.A; left wing only \\
H\,{\sc i}-Ly\,$\beta$	&  18 &	1073.003 &	1025.722 & FUSE LiF\,1A
& $377\pm17$ & $-100$ & $+75$  & $15.81\pm0.08$ & $33.2^{+6.5}_{-5.1}$
&	COG & Comp. A ($v_{\rm c}\approx0$ km\,s$^{-1}$)\\
	&	&	&	&
& $120\pm11$ & $+76$  & $+160$ & $14.43\pm0.09$
&	\nodata  & AOD & Comp. B ($v_{\rm c}\approx+95$ km\,s$^{-1}$)\\
	&	&	&	&
	&	&	   &	& $14.56\pm0.15$ &
$15.1^{+8.4}_{-3.3}$ & COG & Comp. B\\
H\,{\sc i}-Ly\,$\gamma$	&   9 &	1017.331 &	972.537 & FUSE LiF\,1A
& $279\pm18$ & $-70$  & $+70$  & $15.81\pm0.08$ & $33.2^{+6.5}_{-5.1}$ & COG & Comp. A\\
	&	&	&	& FUSE LiF\,2B
& $277\pm24$ & $-70$  & $+70 $ & $15.60\pm0.12$ & $27.6^{+4.8}_{-4.1}$ & COG & Comp. A\\
	&	&		&		& FUSE LiF\,1A
& $75\pm13$  & $+71$  & $+150$ & $14.58\pm0.07$ & \nodata  & AOD
&	Comp. B\\
	&	&	&	&
	&         &	     &	  & $14.56\pm0.15$ &
$15.1^{+8.4}_{-3.3}$ & COG & Comp. B\\
	&	&	&	& FUSE LiF\,2B      & $69\pm14$  & $+71$  &
$+160$ & $14.55\pm0.13$ & \nodata  & AOD & Comp. B\\

H\,{\sc i}-Ly\,$\delta$	&   7 &	993.451 &	949.743 & FUSE SiC\,2A
&	$174\pm15$ & $-80$  & $+70$  & $15.81\pm0.08$ & $33.2^{+6.5}_{-5.1}$
& COG & Comp. A\\
&   &   &  &   &	$28\pm8$	& $+71$  & $+140$ & $14.46\pm0.11$ & \nodata  & AOD &
Comp. B\\
H\,{\sc i}-Ly\,$\epsilon$ &   6 &  980.978 &  937.803 & FUSE SiC\,2A
& $146\pm12$ & $-90$  & $+90$  & $15.81\pm0.08$ & $33.2^{+6.5}_{5.1}$ & COG & Comp. A\\
	&	&  980.998 &	& FUSE SiC\,1B
& $143\pm14$ & $-100$ & $+100$ & $15.60\pm0.12$ & $27.6^{+4.8}_{4.1}$ & COG & Comp. A\\
H\,{\sc i}-Ly\,$\zeta$	&   5 &  973.574 &  930.748 & FUSE SiC\,2A
& $144\pm13$ & $-70$  & $+80$  & $15.81\pm0.08$ & $33.2^{+6.5}_{5.1}$ & COG & Comp. A\\
	&	&  974.412 &	& FUSE SiC\,1B
& $95\pm14$  & $-70$  & $+80$  & $15.60\pm0.12$ & $27.6^{+4.8}_{4.1}$ & COG & Comp. A\\
H\,{\sc i}-Ly\,$\eta$		&   4 &  968.886 &  926.226 & FUSE SiC\,2A
& $107\pm19$ & $-80$  & $+90$  & $15.76\pm0.07$ & \nodata  & AOD & Comp. A\\
	&	&     &		&
&	&	&	& $15.81\pm0.08$ & $33.2^{+6.5}_{-5.1}$ & COG & Comp. A\\
	&	&  969.074 &	& FUSE SiC\,1B
& $132\pm20$ & $-80$  & $+90$  & $15.88\pm0.06$ & \nodata  & AOD & Comp. A\\
	&	&     &		&
&	&	&     & $15.60\pm0.12$ & $27.6^{+4.8}_{-4.1}$ & COG & Comp. A\\
H\,{\sc i}-Ly\,$\theta$	&   3 &  965.683 &  923.150 & FUSE SiC\,2A
& $94\pm24$  & $-70$  & $+70$  & $15.84\pm0.11$ & \nodata  & AOD & Comp. A\\\
	&	&	     &		&
&	&	&	& $15.81\pm0.08$ & $33.2^{+6.5}_{-5.1}$ & COG & Comp. A\\
	&	&  965.707 &		& FUSE SiC\,1B
& $44\pm19$  & $-50$  & $+70$  & $15.70\pm0.15$ & \nodata  & AOD & Comp. A\\
			&		&		     &		&
&	&	&	& $15.60\pm0.12$ & $27.6^{+4.8}_{-4.1}$ & COG & Comp. A\\
H\,{\sc i}-Ly\,$\kappa$	&  & $\sim 961.7$ & 919.351 & FUSE SiC\,2A
& $\leq62$	& $-50$  & $+50$  & \nodata        & \nodata &
\nodata & Comp. A; low S/N\\
C\,{\sc iii}	&  11 & 1022.043 &  977.020 & FUSE LiF\,1A
& $209\pm10$ & $-80$  & $+65$  & $13.80\pm0.04$ & \nodata  & AOD & Comp. A; poss. sat.\\
	&	&       &     	&
&	&	&	& $13.75\pm0.04$ & $33.2$	& COG$^{\rm d}$ & Comp. A\\
	&	& 1022.002 &	& FUSE LiF\,2B
& $217\pm17$ & $-90$  & $+65$  & $13.84\pm0.04$ & \nodata  & AOD & Comp. A\\
	&	& 1022.043 &	& FUSE LiF\,1A
& $111\pm8$  & $+66$  & $+150$ & $13.42\pm0.04$ & \nodata  & AOD & Comp. B\\
	&	&	    &		&
&	&	&	& $13.55\pm0.07$ & $15.1$	& COG$^{\rm d}$ & Comp. B\\
	&	& 1022.002 &	& FUSE LiF\,2B
& $100\pm12$ & $+66$  & $+160$ & $13.43\pm0.05$ & \nodata  & AOD & Comp. B\\
C\,{\sc iv}	& 123 & 1619.643 & 1548.195 & STIS E140M
& $92\pm16$  & $-50$  & $+55$  & $13.61\pm0.08$ & \nodata  & AOD & Comp. A, poss. sat.\\
	&	&	&	&
&	&	&	& $13.73\pm0.41$ & $7.3^{+8.5}_{-4.1}$  & COG & Comp. A\\
& \nodata	& $+56$  & $+130$ & \nodata	     & \nodata  &
\nodata  & Comp. B; blended\\
	& 125 & 1622.350 & 1550.770 & STIS E140M & $66\pm20$  
& $-30$  & $+50$  & $13.66\pm0.12$ & \nodata  & AOD & Comp. A\\
	&	&	&	&
&	&	&	& $13.73\pm0.41$ & $7.3^{+8.5}_{-4.1}$  & COG & Comp. A\\
	&	&	&	&
& $\leq61$	& $+51$  & $+120$ & $\leq13.64$    & \nodata  & AOD & Comp. B\\
O\,{\sc vi}	&  20 & 1079.532 & 1031.926 & FUSE LiF\,1A
& $50\pm7$	& $-70$  & $+40$  & $13.68\pm0.05$ & \nodata  & AOD & Comp. A\\
	&	&	&	&
&	&	&	& $13.64\pm0.06$ & $33.2$	& COG$^{\rm d}$ & Comp. A\\
	&	&	&	&
& $45\pm8$	& $+41$  & $+125$ & $13.63\pm0.06$ & \nodata  & AOD & Comp. B\\
	&	&		&	&
&	&	&	& $13.70\pm0.06$ & $15.1$	& COG$^{\rm d}$ & Comp. B\\
O\,{\sc vi}			&  21 & $\sim$1085.6 & 1037.617 & FUSE SiC\,1A
& $\leq47$ & $-70$  & $+130$ & $\leq 13.93$   & \nodata  & AOD & Comp. A \& B; low S/N\\
Si\,{\sc iii}		&  44 & 1262.060 & 1206.500 & STIS E140M
& $81\pm13$  & $-40$  & $+30$  & $12.74\pm0.07$ & \nodata  & AOD & Comp. A; poss. sat.\\
	&	&		&	&
&	&	&	& $12.63\pm0.08$ & $33.2$	& COG$^{\rm d}$ & Comp. A\\
	&	&		&	&
& \nodata	& $+31$  & $+130$ & \nodata	     & \nodata  &
\nodata & Comp. B; blended\\

\hline
\multicolumn{12}{c}{$z=0.06644\pm0.00002$}\\
\hline

H\,{\sc i}-Ly\,$\alpha$	&  59 & 1296.440 & 1215.670 & STIS E140M
& $162\pm12$ & $-50$  & $+50$  & $13.65\pm0.05$ & $28.7\pm3.3$ & PF  &
Comp. A ($v_{\rm c}\approx0$ km\,s$^{-1}$)\\
	&	&	&	&
&	&	&	& $13.65\pm0.06$ & \nodata      & 
AOD & Comp. A\\
	&	&	&	&
&	&	&	& $13.72\pm0.11$ & $20.4^{+5.7}_{-3.7}$
& COG & Comp. A\\
	&	&	&	&
& $35\pm7$	& $-141$ & $-51$  & $12.86\pm0.08$ & \nodata  & AOD
& Comp. B ($v_{\rm c}\approx -90$ km\,s$^{-1}$)\\

H\,{\sc i}-Ly\,$\beta$	&  22 & 1093.992 & 1025.722 & FUSE LiF\,2A
& $42\pm5$	& $-70$  & $+70$  & $13.80\pm0.05$ & \nodata  & AOD &
Comp. A\\
	&	&	&	&
&	&	&	& $13.72\pm0.11$ & $20.4^{+5.7}_{-3.7}$
& COG & Comp. A\\

\hline
\multicolumn{12}{c}{$z=0.08933\pm0.00002$}\\
\hline

H\,{\sc i}-Ly\,$\alpha$	&  65 & 1324.262 & 1215.670 & STIS E140M
& $273\pm11$ & $-100$ & $+100$ & $14.04\pm0.03$ & $28.9\pm1.7$ & PF  & \\
	&	&	&	&  &
	&       &	& $14.04\pm0.04$ & \nodata  & AOD & \\
	&	&	&	&
&	&	&	& $14.05\pm0.07$ & $27.8^{+1.8}_{-3.2}$ & COG & \\
H\,{\sc i}-Ly\,$\beta$	&  23 & 1117.353 & 1025.722 & FUSE LiF\,2A
& $74\pm12$  & $-90$  & $+90$  & $14.07\pm0.06$ & \nodata  & AOD & \\
	&	&	&	&
&	&	&	& $14.05\pm0.07$ & $27.8^{+1.8}_{-3.2}$ & COG & \\
	&	&	&	& FUSE LiF\,1B
& $68\pm8$	& $-90$  & $+110$ & $14.02\pm0.05$ & \nodata  & AOD & \\
H\,{\sc i}-Ly\,$\gamma$	&  17 & 1059.413 &  972.537 & FUSE LiF\,1A
& $24\pm4$	& $-60$  & $+75$  & $14.02\pm0.04$ & \nodata  & AOD & \\
	&	&	&	&
&	&	&	& $14.05\pm0.07$ & $27.8^{+1.8}_{-3.2}$ & COG & \\
	&	&	&	& FUSE LiF\,2B
& $16\pm5$	& $-65$  & $+75$  & $13.90\pm0.07$ & \nodata  & AOD & \\
H\,{\sc i}-Ly\,$\delta$ & \nodata & $\sim 1034.6$  &  949.743 & FUSE LiF\,1A
& $\leq22$ & $-60$ & $+60$ & \nodata	 & \nodata  & \nodata & \\

\hline
\multicolumn{12}{c}{$z=0.14852\pm0.00003$}\\
\hline

H\,{\sc i}-Ly\,$\alpha$	&  84 & 1396.225 & 1215.670 & STIS E140M
& $254\pm9$  & $-115$ & $+120$ & $13.91\pm0.06$ & $42.1\pm2.4$ & PF  & \\
	&	&	&	&
&	&	&	& $13.88\pm0.04$ & \nodata  & AOD & \\
	&	&	&	&
&	&	&	& $13.96\pm0.05$ & $38.3^{+3.4}_{-3.1}$ & COG & \\
H\,{\sc i}-Ly\,$\beta$	&  29 & 1178.173 & 1025.722 & FUSE LiF\,1B
& $53\pm8$	& $-90$  & $+100$ & $13.90\pm0.06$ & \nodata  & AOD & \\
	&	&	&	&
&	&	&	& $13.96\pm0.05$ & $38.3^{+3.4}_{-3.1}$ & COG & \\

\hline
\multicolumn{12}{c}{$z=0.19620\pm0.00003$}\\
\hline

H\,{\sc i}-Ly\,$\alpha$	&  95 & 1454.186 & 1215.670 & STIS E140M
& $201\pm14$ & $-70$  & $+140$ & $13.65\pm0.05$ & $32.7\pm3.2$
& PF  & Comp. A ($v_{\rm c}\approx0$ km\,s$^{-1}$)\\
	&	&	&	&
&	&	&	& $13.71\pm0.04$ & \nodata      & AOD & Comp. A\\
&     &	    &		  &
&	&	&	& $13.78\pm0.08$ & $28.8^{+2.2}_{-4.4}$ & COG & Comp. A\\
	&	&	&	&
& $76\pm7$	& $-235$ & $-71$  & $13.07\pm0.14$ & $42.8\pm7.2$ &
PF  & Comp. B ($v_{\rm c}\approx -120$ km\,s$^{-1}$)\\
	&	&	&	&
&	&	&	& $13.19\pm0.04$ & \nodata  & AOD & Comp. B\\
H\,{\sc i}-Ly\,$\beta$	&  35 & 1226.996 & 1025.722 & STIS E140M
& $41\pm7$	& $-70$  & $+50$  & $13.79\pm0.07$ & \nodata  & AOD & Comp. A\\
	&	&	&	&
&	&	&	& $13.78\pm0.08$ & $28.8^{+2.2}_{-4.4}$ & COG & Comp. A\\
	&	&	&
&	& $\leq20$   & $-200$ & $-71$  & \nodata     &
\nodata  & \nodata   & Comp. B\\

\hline
\multicolumn{12}{c}{$z=0.21949\pm0.00003$}\\
\hline

H\,{\sc i}-Ly\,$\alpha$	&  100 & 1482.429 & 1215.670 & STIS E140M
& $552\pm26$ & $-160$ & $+120$ & $15.25\pm0.06$ & $31.2^{+4.1}_{2.7}$ & 
COG &  neg. velocity wing\\
H\,{\sc i}-Ly\,$\beta$	&  40 & 1250.844 & 1025.722 & STIS E140M
& $263\pm32$ & $-50$  & $+80$  & $15.08\pm0.08$ & $32.3\pm1.4$ & 
PF  & blended by Galactic S\,{\sc ii}\\
	&	&	&	&
&	&	&	& $15.25\pm0.06$ & $31.2^{+4.1}_{-2.7}$ & COG & \\
H\,{\sc i}-Ly\,$\gamma$	&  30 & 1186.000 &  972.537 & STIS E140M
& $217\pm24$ & $-110$ & $+100$ & $15.32\pm0.12$ & $29.7\pm3.1$ & PF  & \\
	&	&	&	&
&	&	&	& $15.25\pm0.06$ & $31.2^{+4.1}_{-2.7}$ & COG & \\
	&	& 1186.043 &	& FUSE LiF\,2A
& \nodata	& \nodata & \nodata  & \nodata & \nodata    & \nodata
& detector edge\\
H\,{\sc i}-Ly\,$\delta$	&  28 & 1158.204 &  949.743 & FUSE LiF\,2A
& $148\pm10$ & $-100$ & $+100$ & $15.31\pm0.04$ & \nodata      & AOD & \\
&	&	&	&
&	&	&	& $15.25\pm0.06$ & $31.2^{+4.1}_{-2.7}$ & COG & \\
	&  27 & 1158.189 &	& FUSE LiF\,1B
& $153\pm11$ & $-120$ & $+120$ & $15.32\pm0.04$ & \nodata      & AOD & \\
	&	&	&	&
&	&	&	& $15.17\pm0.08$ & $30.5^{+5.1}_{-4.6}$ & COG & \\
H\,{\sc i}-Ly\,$\epsilon$ &   & 1143.618 &  937.804 & FUSE LiF\,2A
&	$88\pm6$  & $-60$  & $+70$  & $15.27\pm0.05$ & \nodata	& AOD & \\
&	&	&	&
&	&	&	& $15.25\pm0.06$ & $31.2^{+4.1}_{-2.7}$ & COG & \\
	&	& 1143.600 &	& FUSE LiF\,1B
&	$64\pm7$  & $-70$  & $+75$  & $15.14\pm0.05$ & \nodata	&
AOD & too weak\\
	&       &	&	&
&	&	&	& $15.17\pm0.08$ & $30.5^{+5.1}_{-4.6}$ & COG & \\
H\,{\sc i}-Ly\,$\eta$	&  26 & 1129.574 &  926.226 & FUSE LiF\,2A
&	$35\pm4$  & $-70$  & $+70$  & $15.22\pm0.05$ & \nodata	& AOD & \\
	&	&	&	&
&	&	&	& $15.25\pm0.06$ & $31.2^{+4.1}_{-2.7}$ & COG & \\
	&	& 1129.459 &	& FUSE LiF\,1B
&	$30\pm4$  & $-70$  & $+75$  & $15.17\pm0.05$ & \nodata	& AOD & \\
	&	&	&	&
&	&	&	& $15.17\pm0.08$ & $30.5^{+5.1}_{-4.6}$ & COG & \\
H\,{\sc i}-Ly\,$\theta$	&  25 & 1125.792 &  923.150 & FUSE LiF\,2A
&	$25\pm6$  & $-45$  & $+60$  & $15.22\pm0.09$ & \nodata	& AOD & \\
	&	&	&	&
&	&	&	& $15.25\pm0.06$ & $31.2^{+4.1}_{-2.7}$ & COG & \\
	&	& 1125.814 &	& FUSE LiF\,1B
&	$20\pm5$  & $-50$  & $+60$  & $15.11\pm0.11$ & \nodata	& AOD & \\
	&	&	&	&
&	&	&	& $15.17\pm0.08$ & $30.5^{+5.1}_{-4.6}$ & COG & \\
H\,{\sc i}-Ly\,$\iota$	&  24 & 1123.122 &  920.963 & FUSE LiF\,2A
&	$17\pm3$  & $-45$  & $+45$  & $15.18\pm0.07$ & \nodata	& AOD & \\
	&	&	&	&
&	&	&	& $15.25\pm0.06$ & $31.2^{+4.1}_{-2.7}$ & COG & \\
	&	& 1123.074 &	& FUSE LiF\,1B
&	$14\pm4$  & $-50$  & $+55$  & $15.13\pm0.11$ & \nodata	& AOD & \\
	&	&       &	&
&	&	&	& $15.17\pm0.08$ & $30.5^{+5.1}_{-4.6}$ & COG & \\
H\,{\sc i}-Ly\,$\kappa$	& \nodata & $\sim 1121.1$ & 919.351 & FUSE LiF\,2A
& $\leq15$ & $-40$ & $+40$ & \nodata	 & \nodata      & \nodata & \\
C\,{\sc iii}	&  31 & 1191.464 &  977.020 & STIS E140M
& $107\pm12$ & $-70$  & $+70$  & $13.62\pm0.13$ & $12.2\pm2.4$ & PF  & narrow; poss. sat.\\
	&	&	&	&
&	&	 &	& $13.64\pm0.05$ & \nodata	& AOD & \\
	&	&	&	&
&	&	&	& $13.33\pm0.06$ & $31.2$	& COG$^{\rm d}$ & \\

N\,{\sc iv}	&   1 &  933.125 &  765.148 & FUSE SiC\,2A
&	$\leq72$  & $-80$ & $+80$	& $\leq13.35$    & \nodata
& AOD & uncertain; low S/N\\
O\,{\sc iii}	&   8 & 1015.790 &  832.927 & FUSE LiF\,1A
&	$36\pm5$ & $-80$ & $+90$  & $13.82\pm0.06$ & \nodata	& AOD & \\
	&	&	&	&
&	&	&	& $13.78\pm0.07$ & $31.2$	& COG$^{\rm d}$ & \\
	&	&	&	&
&	&	&	& $13.85\pm0.08$ & $12.2$	& COG$^{\rm d}$ & \\
	&	& 1015.802 &	& FUSE LiF\,2B
&	$43\pm5$  & $-80$ & $+90$  & $13.95\pm0.05$ & \nodata	& AOD & \\
O\,{\sc iv}	&   2 &  960.569 &  787.711 & FUSE SiC\,2A
& $103\pm13$ & $-80$ & $+90$	& $14.27\pm0.05$ & \nodata	& AOD & poss. sat.\\
	&	&	&	&
&	&	&	& $14.37\pm0.07$ & $31.2$	& COG$^{\rm d}$ & \\
	&	&	&	& FUSE SiC\,1B
&	$65\pm16$ & $-100$ & $+100$ & $14.21\pm0.10$ & \nodata	& AOD & \\
O\,{\sc vi}	&  44 & 1258.330 & 1031.926 & STIS E140M
&	$90\pm7$  & $-120$ & $+60$  & $13.96\pm0.04$ & \nodata	&
AOD & double peak\\
	&	&	&	&
&	&	&	& $13.87\pm0.04$ & $31.2$	& COG$^{\rm d}$ & \\
	&	&	&	&
&	&	&	& $13.68\pm0.06$ & $16.2\pm2.4$ &
PF  & Comp. B ($v_{\rm c} \approx -50$ km\,s$^{-1}$)\\
	&	&	&	&
&	&	&	& $13.68\pm0.05$ & $15.5\pm1.9$ &
PF  & Comp. A ($v_{\rm c} \approx 0$ km\,s$^{-1}$)\\
O\,{\sc vi}	&  49 & 1265.312 & 1037.617 & STIS E140M
&	$27\pm6$  & $-100$ & $+50$  & $13.72\pm0.08$ & \nodata	& AOD & \\
	&	&	&	&
&	&	&	& $13.87\pm0.04$ & $31.2$	 & COG$^{\rm d}$ & \\
Si\,{\sc iii}		&  98 & 1471.328 & 1206.500 & STIS E140M
&	$11\pm3$  & $-50$ & $+50$
& $12.04\pm0.07$ & $7.9\pm2.0$  & PF & narrow\\
	&	&	&	&
&	&	&	& $11.82\pm0.10$ & \nodata	& AOD & \\
Ne\,{\sc viii}	& \nodata & $\sim939.5$ & 770.409 & FUSE SiC\,2A
& $\leq 38$ & $-150$ & $+100$  & $\leq 13.94$  & \nodata    & AOD & \\

\hline
\multicolumn{12}{c}{$z=0.22313\pm0.00002$}\\
\hline

H\,{\sc i}-Ly\,$\alpha$	& 101 & 1486.918 &	1215.670 & STIS E140M
&	$247\pm8$  & $-100$ & $+100$ & $13.92\pm0.04$ & $34.8\pm1.1$ & PF \\
	&	&	&	&
&	&	&	& $13.88\pm0.04$ & \nodata  & AOD & poss. saturated\\
	&	&	&	&
&	&	&	& $14.01\pm0.06$ & $25.6^{+3.8}_{-2.5}$ & COG & \\
H\,{\sc i}-Ly\,$\beta$	&  41 & 1254.602 &	1025.722 & STIS E140M
&	$63\pm7$	& $-100$ & $+100$ & $13.99\pm0.05$ & \nodata  & AOD & \\
	&	&	&      	&
&	&	&	& $14.01\pm0.06$ & $25.6^{+3.8}_{-2.5}$ & COG & \\
O\,{\sc vi}     &  46 & $\sim 1262.2$ & 1031.926 & STIS E140M
&	\nodata  & \nodata & \nodata & \nodata &     \nodata &
\nodata & blended \\
O\,{\sc vi}	&  51 & 1269.367 &	1037.617 & STIS E140M
&   $57\pm14$	& $-120$ & $+120$ & $13.99\pm0.11$ & \nodata  & AOD & \\
	&	&        &	&
&	&       &	& $14.02\pm0.08$ & $25.6$	& COG$^{\rm d}$ & \\
Ne\,{\sc viii}	& \nodata & $\sim939.5$ & 770.409 & FUSE SiC\,2A
&	$\leq 30$  & $-100$ & $+100$ & $\leq 13.75$   & \nodata  & AOD & \\

\hline
\multicolumn{12}{c}{$z=0.22471\pm0.00003$}\\
\hline

H\,{\sc i}-Ly\,$\alpha$	& 102 & 1488.849 &	1215.670 & STIS E140M
&	$169\pm12$ & $-100$ & $+100$ & $13.59\pm0.06$ & $28.9\pm1.7$ & PF & \\
	&	&	&	&
&	&	&	& $13.64\pm0.04$ & \nodata  & AOD & \\
	&	&	&	&
&	&	&	& $13.57\pm0.08$ & $36.1^{+4.0}_{-3.6}$ & COG & \\
H\,{\sc i}-Ly\,$\beta$	&  42 & 1256.198 &	1025.722 & STIS E140M
&	$16\pm4$	& $-90$  & $+90$  & $13.41\pm0.10$ 
& \nodata  & AOD & too weak \\
	&	&	&      	&
&	&	&	& $13.57\pm0.08$ & $36.1^{+4.0}_{-3.6}$ & COG & \\

\hline
\multicolumn{12}{c}{$z=0.25642\pm0.00002$}\\
\hline

H\,{\sc i}-Ly\,$\alpha$	& 107 & 1527.390 &	1215.670 & STIS E140M
& $184\pm6$  & $-80$ & $+90$	& $13.74\pm0.03$ & $25.0\pm0.9$ & PF & \\
	&	&	&	&
	&       &	&	& $13.74\pm0.04$ & \nodata  & AOD & \\
        &	&	&	&
	&       &	&	& $13.73\pm0.05$ & $27.1^{+3.0}_{-2.4}$ & COG & \\
H\,{\sc i}-Ly\,$\beta$	&  57 & 1288.741 &	1025.722 & STIS E140M
& $36\pm6$   & $-80$  & $+80$  & $13.75\pm0.06$ & \nodata  & AOD \\
	&	&       &	&
	&       &	&	& $13.73\pm0.05$ & $27.1^{+3.0}_{-2.4}$ & COG & \\

\hline
\multicolumn{12}{c}{$z=0.25971\pm0.00003$}\\
\hline

H\,{\sc i}-Ly\,$\alpha$	& 109 &	1531.396 &	1215.670 & STIS E140M
 & $249\pm10$ & $-120$ & $+120$ & $13.84\pm0.12$ & $40.5\pm4.9$ & PF & \\
	&	&	&	&
	&       &	&       & $13.80\pm0.04$ & \nodata  &
AOD & poss. saturated\\
	&	&	&	&
	&       &	&       & $13.95\pm0.05$ & $28.2^{+1.8}_{-2.5}$ & COG & \\
H\,{\sc i}-Ly\,$\beta$	&  58 &	1292.154 &	1025.722 & STIS E140M
	& $56\pm7$   & $-100$ & $+120$ & $13.93\pm0.05$ & \nodata  & AOD & \\
	&	&	&	&
	&       &       &	& $13.95\pm0.05$ & $28.2^{+1.8}_{-2.5}$ & COG & \\
O\,{\sc vi}	&  61 &	1299.933 &	1031.926 & STIS E140M
& $76\pm12$  & $-110$ & $+110$ & $13.84\pm0.07$ & \nodata  & AOD & \\
	&	&	&      &
&       &       &       & $13.83\pm0.07$ & $28.2$   & COG$^{\rm d}$ & \\
O\,{\sc vi}	&  62 &	1307.067 &	1037.617 & STIS E140M
 & $34\pm10$  & $-100$ & $+100$ & $13.79\pm0.11$ & \nodata  & AOD & \\
	&	&	&	&
	&       &	&	& $13.83\pm0.07$ & $28.2$   & COG$^{\rm d}$ \\
Ne\,{\sc viii}        & \nodata & $\sim970.5$ & 770.409 & FUSE SiC\,2A
&	$\leq 43$  & $-90$ & $+90$   & $\leq 13.96$ & \nodata  & AOD & \\

\hline
\multicolumn{12}{c}{$z=0.28335\pm0.00003$}\\
\hline

H\,{\sc i}-Ly\,$\alpha$	& 111 &	1560.135 &	1215.670 & STIS E140M
& $158\pm9$  & $-150$ & $+100$ & $13.59\pm0.10$ & $37.0\pm5.2$
&	PF  & neg. vel. wing ? \\
	&	&	&	&
&       &       &       & $13.59\pm0.04$ & \nodata  & AOD & \\
	&	&	&	&
	&       &	&       & $13.61\pm0.07$ & $28.6^{+5.8}_{-4.2}$ & COG & \\
H\,{\sc i}-Ly\,$\beta$	&  64 &	1366.551 &	1025.722 & STIS E140M
& $27\pm8$   & $-90$  & $+90$  & $13.61\pm0.13$ & \nodata  & 
AOD &	uncertain; det. feature\\
	&	&	&	&
&        &        &	 & $13.61\pm0.07$ & $28.6^{+5.8}_{-4.2}$ & COG & \\

\hline
\multicolumn{12}{c}{$z=0.29236\pm0.00003$}\\
\hline

H\,{\sc i}-Ly\,$\alpha$	& 113 &	1571.050 &	1215.670
&	STIS E140M      & $404\pm19$ & $-150$ & $+150$ & $14.65\pm0.09$ &
$24.3\pm2.8$ & PF  & \\
&	&	&
&       &       &	 &     & $14.50\pm0.04$ & $31.2^{+3.8}_{-3.1}$ & COG & \\
H\,{\sc i}-Ly\,$\beta$	&  66 & 1325.600 &	1025.722 & STIS E140M
& $160\pm7$   & $-90$  & $+90$ & $14.47\pm0.05$ & $26.2\pm1.6$ & PF  & \\
	&	&	&	&
	&       &	&	& $14.52\pm0.04$ & \nodata  & AOD & \\
	&	&	&
	&       &       &	&     &
$14.50\pm0.04$ & $31.2^{+3.8}_{-3.1}$ & COG & \\
H\,{\sc i}-Ly\,$\gamma$	&  43 & 1256.869 &	972.537 & STIS E140M
& $59\pm5$   & $-80$  & $+80$  & $14.50\pm0.05$ & \nodata  & AOD & \\
	&	&	&	&
&	&        &	& $14.50\pm0.04$ & $31.2^{+3.8}_{-3.1}$ & COG & \\
H\,{\sc i}-Ly\,$\delta$	&  36 & $\sim$1227.3 &	949.743 & STIS E140M
& \nodata	& \nodata & \nodata & \nodata & \nodata & \nodata &
present, but blended \\
C\,{\sc iii}	&  47 & 1262.683 &	977.020 & STIS E140M
& $79\pm14$ & $-50$ & $+50$    	& $13.18\pm0.07$ & \nodata  & AOD &
uncertain, blended\\
	&	&	&	&
&	&       &	& $13.17\pm0.09$ & $31.2$	& COG$^{\rm d}$ & \\
O\,{\sc iii}		&  19 & 1076.439 &	832.927 & FUSE LiF\,1A
& $37\pm10$  & $-60$ & $+60$   & $13.80\pm0.06$ & \nodata  & AOD & \\
	&	&	&
&       &       &       &    & $13.79\pm0.05$ & $31.2$   & COG$^{\rm d}$ & \\
O\,{\sc iv}	&  10 & 1017.979 &	787.711 & FUSE LiF\,1A
& $72\pm6$	& $-50$ & $+50$   & $14.16\pm0.04$ & \nodata  & AOD & \\
	&	&	&
&       &       &       &       & $14.08\pm0.06$ & $31.2$   & COG$^{\rm d}$ & \\
	&	& 1017.945 &	& FUSE LiF\,2B
& $62\pm8$	& $-55$ & $+55$   & $14.10\pm0.05$ & \nodata  & AOD & \\

\hline
\multicolumn{12}{c}{$z=0.30434\pm0.00003$}\\
\hline

H\,{\sc i}-Ly\,$\alpha$	& 116 & 1585.630 &	1215.670 & STIS E140M
& $240\pm17$ & $-150$ & $+150$ & $13.76\pm0.14$ & $64.5\pm9.6$ & PF & 
Component A ($v_c\approx 0$ km\,s$^{-1}$); broad\\
	&	&	&	&
&	&       &	& $13.76\pm0.04$ & \nodata  & AOD & neg. velocity wing ?\\
	&       &	&	&
&	&       &	& $13.74\pm0.05$ & $54.2^{+7.8}_{-8.6}$ & COG & \\
&       &	&	&
& $192\pm44$ & $+151$ & $+400$ & $13.48\pm0.21$ & $124.8\pm21.5$ & 
PF & Component B ($v_c\approx +290$ km\,s$^{-1}$);\\
& & & & & & & &  \nodata & \nodata & \nodata & broad; uncertain\\
H\,{\sc i}-Ly\,$\beta$	&  57 & 1337.887 &	1025.722 & STIS E140M
& $38\pm4$	& $-90$  & $+90$  & $13.92\pm0.04$ & $34.7\pm9.6$ & PF & narrow; line width \\
&       &       &       & $13.79\pm0.05$ & \nodata  & AOD &
inconsistent with Ly\,$\alpha$\\
	&       &	&	 &
&	&	&	& $13.74\pm0.05$ & $54.2^{+7.8}_{-8.6}$ & COG & \\

\hline
\multicolumn{12}{c}{$z=0.31978\pm0.00003$}\\
\hline

H\,{\sc i}-Ly\,$\alpha$	& 120 & 1604.411 &	1215.670 & STIS E140M
& $389\pm16$ & $-220$ & $+150$ & 
$13.98\pm0.06$ & $74.4\pm8.7$ & PF & very broad; neg. velocity wing ?\\
H\,{\sc i}-Ly\,$\beta$	&  72 & 1353.713 &	1025.722 & STIS E140M
& $110\pm13$ & $-180$ & $+130$ & $14.07\pm0.09$ & $62.7\pm12.1$  & PF & \\
O\,{\sc iv}     &  \nodata & $\sim1039.6$  &  787.711 & FUSE LiF\,1A
& $\leq14$      & $-50$ & $+50$  & $\leq13.30$ & \nodata  & AOD & \\
O\,{\sc vi}	&  74 & 1361.865 &	1031.926 & STIS E140M
& $25\pm4$	& $-65$  & $+60$  & $13.49\pm0.07$ & $19.3\pm4.2$ & PF & \\
	&       &	&		 &
&	&	&	& $13.44\pm0.06$ & \nodata  & AOD & \\
O\,{\sc vi}	&  78 & $\sim1369.4$ & 1037.617 & STIS E140M
& \nodata	& \nodata & \nodata  & \nodata & \nodata  & \nodata & blended \\

Ne\,{\sc viii}	& \nodata & $\sim1016.8$ & 770.409 & FUSE LiF\,1A
& $\leq 15$ & $-90 $ & $+90$  & $\leq 13.57$   & \nodata  & AOD & \\

\hline
\multicolumn{12}{c}{$z=0.33269\pm0.00003$}\\
\hline

H\,{\sc i}-Ly\,$\alpha$	& 124 & 1620.107 &	1215.670 & STIS E140M
& $202\pm19$ & $-50$  & $+90$  & $13.88\pm0.08$ & $25.9\pm3.2$ & PF  & \\
	&	 &	&	 &
&	&	&	& $13.76\pm0.04$ & \nodata  & AOD &
poss. saturated\\
	&	 &	&	 &
&	&	&	& $13.98\pm0.11$ & $19.3^{+5.1}_{-3.6}$ & COG & \\
H\,{\sc i}-Ly\,$\beta$	&  77 & 1367.029 &	1025.722 & STIS E140M
& $58\pm10$  & $-80$  & $+90$  & $13.94\pm0.07$ & \nodata  & AOD & \\
	&	 &      &	 &
&	&	&	& $13.98\pm0.11$ & $19.3^{+5.1}_{-3.6}$ & COG & \\

\hline
\multicolumn{12}{c}{$z=0.34477\pm0.00005$}\\
\hline

H\,{\sc i}-Ly\,$\alpha$	& 126 & 1634.834 &	1215.670 & STIS E140M
& \nodata	& \nodata & \nodata & \nodata & \nodata & \nodata &
detector gap\\
H\,{\sc i}-Ly\,$\beta$	&  80 & 1379.357 &	1025.722 & STIS E140M
& $52\pm6$	& $-110$  & $+90$   & $14.02\pm0.08$ & $34.3\pm4.4$ &
PF & \\
& & & & & & & & $13.91\pm0.05$ & \nodata & AOD & \\

\hline
\multicolumn{12}{c}{$z=0.41081\pm0.00004$}\\
\hline

H\,{\sc i}-Ly\,$\alpha$	& 134 & 1715.089 &	1215.670 & STIS E140M
& $152\pm34$ & $-100$ & $+100$ & \nodata	    & \nodata        &
PF & low S/N; neg. velocity wing ?\\
	&	&	&	&
&	&	&	& $13.56\pm0.12$ & \nodata   & AOD & \\
	&	&	&	&
&	&	&	& $13.57\pm0.08$ & $32.8^{+6.4}_{-4.2}$ & COG & \\
H\,{\sc i}-Ly\,$\beta$	& 94  & 1447.101 &	1025.722 & STIS E140M
& $25\pm5$	& $-60$  & $+60$  & $13.60\pm0.05$ & \nodata & AOD & \\
	&	&	&	&
&	&	&	& $13.57\pm0.08$ & $32.8^{+6.4}_{-4.2}$ & COG & \\

\hline
\multicolumn{12}{c}{$z=0.43148\pm0.00003$}\\
\hline

H\,{\sc i}-Ly\,$\beta$	&  97 & 1468.305 &	1025.722 & STIS E140M
& $79\pm5$	& $-100$ & $+100$ & $14.10\pm0.06$ & $20.9\pm4.3$  &
PF & \\
	&	&	&	&
&	&	&	& $14.15\pm0.04$ & \nodata  & AOD & \\
	&	&	&	&
&	&	&	& $14.09\pm0.07$ & $29.0^{+4.9}_{-3.8}$ & COG & \\
H\,{\sc i}-Ly\,$\gamma$	&  83 & 1392.146 &	972.537 & STIS E140M
& $22\pm4$	& $-80$  & $+80$  & $14.04\pm0.07$ & \nodata  & AOD & \\
	&	&	&	&
&	&	&	& $14.09\pm0.07$ & $29.0^{+4.9}_{-3.8}$ & COG & \\
H\,{\sc i}-Ly\,$\delta$	&  73 & 1359.501 &	949.743 & STIS E140M
& $13\pm4$	& $-70$  & $+70$  & $14.14\pm0.11$ & \nodata  & AOD & \\
	&	&	&	&
&	&	&	& $14.09\pm0.07$ & $29.0^{+4.9}_{-3.8}$ & COG & \\
H\,{\sc i}-Ly\,$\epsilon$ &  71 & 1342.494 &	937.804 & STIS E140M
& $10\pm4$	& $-50$  & $+50$  & $14.09\pm0.07$ & $29.0^{+4.9}_{-
3.8}$ & COG & \\
C\,{\sc iii}	&  86 & 1398.715 &	977.020 & STIS E140M
& \nodata & \nodata & \nodata & \nodata & \nodata & \nodata &
blended\\

\hline
\multicolumn{12}{c}{$z=0.43569\pm0.00003$}\\
\hline

H\,{\sc i}-Ly\,$\beta$	& 99 & 1472.614 & 1025.722 & STIS E140M
& $79\pm9$	& $-90$ & $+90$  & $14.22\pm0.10$ & $44.0\pm3.9$ & PF
& aligned with FOS Ly\,$\alpha$\,$^{\rm e}$ \\
& & & & & & & & $14.13\pm0.15$ & \nodata & AOD & \\
\enddata

\tablenotetext{a}{Equivalent widths, velocities, column densities, 
and $b$ values refer to the rest frame. Errors are $1\sigma$ estimates (see \S2.3), upper
limits are $3\sigma$.}
\tablenotetext{b}{Vacuum wavelengths from Morton (2003, 1999) and Verner et al.\,(1994).}
\tablenotetext{c}{PF = profile fit; AOD = apparent optical depth
method; COG = curve of growth.} 
\tablenotetext{d}{Fixed $b$ value (derived from H\,{\sc i}) is used to determine
log $N$ with the COG method. Note that the $1\sigma$ error given for log $N$ therefore
does not include the intrinsic uncertainty for $b$.}
\tablenotetext{e}{See Bahcall et al.\,(1993).} 
\end{deluxetable}

\clearpage
\newpage
\begin{deluxetable}{lrrrrrlrr} \tabletypesize{\normalsize} 
\tablecaption{Overview Column Densities for Multiple-Line IGM Absorbers$^{\rm a}$
\label{tbl-2}}
\tablewidth{0pt}
\tablehead{
\colhead{$z$} & \colhead{H\,{\sc i}} & \colhead{C\,{\sc iii}} & \colhead{C\,{\sc iv}}
& \colhead{O\,{\sc iii}} & \colhead{O\,{\sc iv}} & \colhead{O\,{\sc vi}}
& \colhead{Si\,{\sc iii}} &  \colhead{Ne\,{\sc viii}}
}
\startdata
0.00229	&	13.6 & \nodata & \nodata & n.a.	& n.a.	& (13.7)$^{\rm b}$
& \nodata & n.a. \\
0.00760	&	14.1 & \nodata & \nodata & n.a.	& n.a.	& (13.1)
& \nodata & n.a. \\
0.02217	&	13.7 & \nodata & \nodata & n.a.	& n.a.	&
\nodata & \nodata & n.a. \\
0.04606\,A &	15.8 & 13.8    & 13.6	& n.a.	& n.a.	& 13.7
& 12.7    & n.a. \\
0.04606\,B &	14.5 & 13.4    & $\leq	13.6$ & n.a. & n.a. & 13.6
& bld.    & n.a. \\
0.06644\,A &	13.7 & \nodata & \nodata & n.a.	& n.a.	&
\nodata & \nodata & n.a. \\
0.06644\,B &	12.9 & \nodata & \nodata & n.a.	& n.a.	&
\nodata & \nodata & n.a. \\
0.08933	&	14.0 & \nodata & \nodata & n.a.	& n.a.	&
\nodata & \nodata & n.a. \\
0.14852	&	13.9 & \nodata & n.a.	& \nodata & n.a. &
\nodata & \nodata & n.a. \\
0.19620\,A &	13.7 & \nodata & n.a.	& \nodata & \nodata &
\nodata & \nodata & \nodata \\
0.19620\,B &	13.1 & \nodata & n.a.	& \nodata & \nodata &
\nodata & \nodata & \nodata \\
0.21949	&	15.2 & 13.6    & n.a.	& 13.8	& 14.3	&
14.0$^{\rm c}$ & 12.0    & $\leq 13.9$\\
0.22313	&	13.9 & \nodata & n.a.	& \nodata & \nodata & (14.0)
& \nodata & $\leq 13.8$ \\
0.22471	&	13.6 & \nodata & n.a.	& \nodata & \nodata &
\nodata & \nodata & \nodata \\
0.25642	&	13.7 & \nodata & n.a.	& \nodata & \nodata &
\nodata & \nodata & \nodata \\
0.25971	&	13.8 & \nodata & n.a.	& \nodata & \nodata & 13.8
& \nodata & $\leq 14.0$ \\
0.28335	&	13.6 & \nodata & n.a.	& \nodata & \nodata &
\nodata & \nodata & \nodata \\
0.29236	&	14.5 & 13.2    & n.a.	& 13.8	& 14.2	&
\nodata & \nodata & \nodata \\
0.30434\,A &	13.7 & \nodata & n.a.	& \nodata & \nodata &
\nodata & \nodata & \nodata \\
0.30434\,B &	13.6 & \nodata & n.a.	& \nodata & \nodata &
\nodata & \nodata & \nodata \\
0.31978 &	14.0 & \nodata & n.a.	& \nodata & \nodata & (13.4)
& \nodata & $\leq 13.6$ \\
0.33269	&	13.9 & \nodata & n.a.	& \nodata & \nodata &
\nodata & \nodata & \nodata \\
0.34477    & 13.9 & \nodata & n.a.	 & \nodata & \nodata &
\nodata & \nodata & \nodata \\
0.41081    & 13.6 & \nodata & n.a.	 & \nodata & \nodata &
\nodata & \nodata & \nodata \\
0.43148    & 14.1 & bld.	 & n.a.	 & \nodata & \nodata &
\nodata & \nodata & \nodata \\
0.43569    & 14.1 & \nodata & n.a.	 & \nodata & \nodata &
\nodata & n.a.    & \nodata \\
\enddata
\tablenotetext{a}{The following abbreviations are used: ...$=$ species not detected; n.a.$=$
species not available in the FUSE and STIS bandpass; bld.$=$ blending problems.}
\tablenotetext{b}{Column densities for tentative detections are given in brackets.}
\tablenotetext{c}{Total O\,{\sc vi} column density for the two components (see \S3.2.4).}
\end{deluxetable}

\clearpage
\newpage
\begin{deluxetable}{lrrlllrl}
\tabletypesize{\normalsize}
\tablecaption{Probable and Possible Single-Line Ly\,$\alpha$ Absorbers towards PG\,1259+593$^{\rm a}$
\label{tbl-2}}
\tablewidth{0pt}
\tablehead{
\colhead{No.} & \colhead{$z$} & \colhead{$\lambda_{\rm obs}$} & \colhead{$W_{\lambda}$} & 
\colhead{log $N$\,$^{\rm b}$} & \colhead{$b$\,$^{\rm b}$} & \colhead{Status$^{\rm c}$} & \colhead{Comments} \\
\colhead{} & \colhead{} & \colhead{[\AA]} & \colhead{[m\AA]} & \colhead{} & 
\colhead{[km\,s$^{-1}$]} & \colhead{}
}
\startdata
33A & 0.00440 & 1221.024 & $215\pm24:$ & $13.65\pm0.14:$ & $142.6\pm29.4:$ & UC & Component A\\
33B & 0.00505 & 1221.812 & $133\pm20:$ & $13.44\pm0.10:$ & $57.2\pm8.9:$ & UC & Component B\\
37 &	0.01347 &	1232.046 &  $90\pm17:$ & $13.24\pm0.14:$ & $104.1\pm17.5:$ & UC & \nodata \\
38 &	0.01502 &	1233.924 &  $73\pm6$   & $13.21\pm0.06$ & $22.6\pm4.4$ & OK & \nodata \\
48 &	0.03924 &	1263.372 &  $41\pm5$   & $12.94\pm0.05$ & $15.3\pm2.8$ & OK & \nodata \\
50 &	0.04250 &	1267.334 &  $58\pm7:$  & $13.06\pm0.06:$ & $40.7\pm4.9:$ & UC & \nodata \\
53 &	0.05112 &	1277.819 & $168\pm9$   & $13.62\pm0.07$ & $34.4\pm2.5$ & OK & pos. vel. wing$^{\rm d}$ \\
54 &	0.05257 &	1279.576 &  $28\pm6$   & $12.75\pm0.06$ & $20.7\pm3.0$ & OK & \nodata \\
55 &	0.05376 &	1281.024 & $119\pm7$   & $13.44\pm0.04$ & $30.5\pm1.9$ & OK & \nodata \\
56 &    0.05586 &       1283.572 &  $149\pm20:$ & $13.46\pm0.14:$ & $144.2\pm27.9:$ & UC & \nodata \\
60 &	0.06810 &	1298.456 &  $72\pm10:$ & $13.14\pm0.09:$ & $87.7\pm9.2:$ & UC & poss. mult. comp.$^{\rm e}$ \\
63 &	0.08041 &	1313.423 &  $45\pm7$   & $12.97\pm0.10$ & $42.0\pm4.5$ & OK & \nodata \\
67 &	0.09196 &	1327.468 &  $72\pm22:$ & $13.13\pm0.21:$ & $113.4\pm29.2:$ & UC & neg. vel. wing$^{\rm f}$ \\
68 &	0.09591 &	1332.267 &  $45\pm5$   & $12.97\pm0.03$ & $21.5\pm2.3$ & OK & \nodata \\
70 &	0.10281 &	1340.654 & $136\pm18:$ & $13.41\pm0.17:$ & $196.9\pm31.2:$ & UC & poss. mult. comp.$^{\rm e}$ \\
75 &	0.12188 &	1363.834 &  $53\pm8$   & $13.03\pm0.07$ & $26.9\pm4.2$ & OK & \nodata \\
76 &	0.12387 &	1366.252 & $124\pm10$  & $13.47\pm0.06$ & $28.2\pm3.0$ & OK & \nodata \\
79 &	0.13351 &	1377.977 &  $53\pm11:$ & $13.01\pm0.08:$ & $48.1\pm4.3:$ & UC & \nodata \\
81 &	0.14034 &	1386.276 &  $60\pm10:$ & $13.06\pm0.07:$ & $56.7\pm5.4:$ & UC & \nodata \\
82 &	0.14381 &	1390.498 &  $80\pm15:$ & $13.18\pm0.13:$ & $115.8\pm9.6:$ & UC & \nodata \\
85A &   0.15029 &       1398.405 &  $82\pm9$   & $13.25\pm0.11$ & $25.7\pm4.3$ & OK & Component A \\
85B &   0.15058 &       1398.722 & $124\pm14$  & $13.45\pm0.13$ & $32.0\pm5.1$ & OK & Component B \\
87 &	0.15136 &	1399.668 & $106\pm9$   & $13.32\pm0.09$ & $65.3\pm5.5$ & OK & \nodata \\
88 &	0.15435 &	1403.311 &  $77\pm4$   & $13.22\pm0.04$ & $25.2\pm1.9$ & OK & \nodata \\
89 &	0.16647 &	1418.044 & $113\pm13:$ & $13.34\pm0.08:$ & $93.3\pm8.9:$ & UC & poss. neg. vel. comp.$^{\rm g}$ \\
90 &	0.17148 &	1424.130 &  $94\pm19:$ & $13.25\pm0.16:$ & $131.5\pm17.4:$ & UC & \nodata \\
91 &	0.17891 &	1433.167 & $101\pm10$  & $13.29\pm0.10$ & $98.5\pm9.1$ & OK & \nodata \\
92 &	0.18550 &	1441.174 &  $77\pm11:$ & $13.17\pm0.12:$ & $86.4\pm10.0:$ & UC & \nodata \\
93 &	0.18650 &	1442.390 &  $50\pm4$   & $13.02\pm0.03$ & $19.6\pm1.4$ & OK & \nodata \\
96 &	0.19775 &	1456.070 &  $92\pm8$   & $13.33\pm0.06$ & $23.9\pm2.8$ & OK & pos. vel. wing$^{\rm d}$ \\
103 & 0.22861 &         1493.587 & $133\pm10$  & $13.47\pm0.05$ & $40.3\pm2.9$ & OK & \nodata \\
104 & 0.23280 &         1498.676 & $140\pm12$  & $13.50\pm0.07$ & $37.4\pm3.2$ & OK & \nodata \\
105 & 0.23951 &         1506.835 &  $92\pm5$   & $13.40\pm0.03$ & $16.3\pm1.3$ & OK & \nodata \\
106 & 0.24126 &         1508.963 & $130\pm12$  & $13.41\pm0.09$ & $89.1\pm6.9$ & OK & \nodata \\
108 & 0.25875 &         1530.230 & $35\pm10:$  & $12.87\pm0.09:$ & $16.6\pm4.1:$ & UC & \nodata \\
110 & 0.28014 &         1556.227 &  $36\pm6:$  & $12.87\pm0.05:$ & $17.1\pm1.8:$ & UC & \nodata \\
112 & 0.28853 &         1566.428 & $111\pm16:$ & $13.34\pm0.11:$ & $70.7\pm8.5:$ & UC & \nodata \\
114 & 0.29847 &         1578.515 &  $61\pm9$   & $13.09\pm0.10$ & $33.3\pm2.9$ & OK & \nodata \\
115 & 0.30164 &         1582.366 &  $86\pm12$  & $13.26\pm0.14$ & $31.7\pm4.7$ & OK & \nodata \\
117 & 0.30906 &         1591.389 &  $78\pm9:$  & $13.20\pm0.10:$ & $45.3\pm5.0:$ & UC & \nodata \\
118 & 0.31070 &         1593.370 & $103\pm9$   & $13.40\pm0.07$ & $22.8\pm2.9$ & OK & \nodata \\
119 & 0.31682 &         1600.816 &  $62\pm14:$ & $13.05\pm0.17:$ & $43.5\pm6.4:$ & UC & \nodata \\
121 & 0.32478 &         1610.496 &  $86\pm12$  & $13.24\pm0.15$ & $46.1\pm10.2$ & OK & \nodata \\
122 & 0.33128 &         1618.397 &  $45\pm13:$ & $12.98\pm0.24:$ & $18.6\pm3.5:$ & UC & \nodata \\
127 & 0.34802 &         1638.750 &  $78\pm14:$ & $13.19\pm0.14:$ & $48.7\pm6.2:$ & UC & \nodata \\
128 & 0.34914 &         1640.112 & $103\pm11$  & $13.36\pm0.09$ & $31.3\pm4.8$ & OK & \nodata \\
129 & 0.35375 &         1645.719 &  $94\pm8$   & $13.41\pm0.06$ & $16.4\pm1.8$ & OK & \nodata \\
130 & 0.37660 &         1673.489 & $127\pm15$  & $13.45\pm0.13$ & $36.4\pm6.2$ & OK & \nodata \\
131 & 0.37909 &         1676.519 & $116\pm19:$ & $13.36\pm0.15:$ & $72.0\pm8.9:$ & UC & neg. vel. wing$^{\rm f}$ \\
133 & 0.38266 &         1680.862 & $403\pm77:$ & $13.92\pm0.42:$ & $200.1\pm22.8:$ & UC & poss. mult. comp.$^{\rm f}$ \\
134 & 0.38833 &         1687.753 &  $47\pm13$  & $13.02\pm0.19$ & $14.1\pm3.8$ & OK & \nodata \\
135 & 0.41786 &         1723.654 &  $88\pm10$  & $13.25\pm0.08$ & $50.7\pm3.9$ & OK & pos. vel. wing.$^{\rm d}$ \\
\enddata

\tablenotetext{a}{We list in this table all unidentified statistically significant
absorption lines with $\lambda >1218$ \AA\, observed in the spectrum of
PG\,1259+593. Most of these lines probably are Ly\,$\alpha$ absorbers
although the lines are too weak to confirm with a detection of Ly\,$\beta$.
The very  broad and weak lines listed could be undulations in
the continuum of PG\,1259+593, broad Ly\,$\alpha$ absorbers, or weak
complex multi-component Ly\,$\alpha$ absorbers.} 
\tablenotetext{b}{Values for log $N$ and $b$ from lines that are flagged as uncertain (UC; see below)
are marked with a colon behind the numbers.}
\tablenotetext{c}{The status of a line is either uncertain (UC) or OK.
The flag UC is given to those lines for which we see evidence
that effects such as continuum undulations, unresolved line blending, fixed-pattern
features, etc. are possibly influencing the significance of the detection of a true IGM
feature or add an additional uncertainty for determining log $N$ and $b$ that
is not accounted for in the formal $1\sigma$ error estimate listed (see also 
\S2.3).}
\tablenotetext{d}{Line shows a positive-velocity wing.} 
\tablenotetext{e}{Line shows evidence for a multi-component structure.}
\tablenotetext{f}{Line shows a negative-velocity wing.} 
\tablenotetext{g}{Line shows evidence for an additional component at negative velocities.}
\end{deluxetable}

\clearpage
\newpage
\begin{deluxetable}{rlrrrrrr}
\tabletypesize{\normalsize}
\tablecaption{Broad Ly\,$\alpha$ Absorbers towards PG\,1259+593$^{\rm a}$
\label{tbl-2}}
\tablewidth{0pt}
\tablehead{
\colhead{No.} & \colhead{$z$} & \colhead{$\lambda_{\rm obs}$} & \colhead{log $N$(H\,{\sc i})} & 
\colhead{$b$(H\,{\sc i})} & \colhead{$T$\,$^{\rm b}$} &
\colhead{log $f_{\rm H}$\,$^{\rm c}$} & \colhead{log $N$(H$^0+$H$^+$)\,$^{\rm d}$} \\
\colhead{} & \colhead{} & \colhead{[\AA]} & \colhead{} & \colhead{[km\,s$^{-1}$]} & 
\colhead{[$10^5$ K]} & \colhead{} & \colhead{}
}
\startdata
\hline
\multicolumn{8}{c}{Broad Ly\,$\alpha$ systems without O\,{\sc vi}
detection}\\
\hline
103 &	0.22861 &	1493.587 & $13.47\pm0.05$ & $40.3\pm2.9$ &    $\leq0.98$ & $\leq4.83$ & $\leq18.30$\\
63  &	0.08041 &	1313.423 & $12.97\pm0.10$ & $42.0\pm4.5$ &    $\leq1.06$ & $\leq4.90$ & $\leq17.87$\\
84  &	0.14852 &	1396.225 & $13.91\pm0.06$ & $42.1\pm2.4$ &    $\leq1.07$ & $\leq4.91$ & $\leq18.82$\\
95B & 	0.19620 & 	1454.186 & $13.07\pm0.14$ & $42.8\pm7.2$ &    $\leq1.10$ & $\leq4.94$ & $\leq18.01$\\
99$^{\rm e}$  & 0.43569 & 1472.614$^{\rm e}$ & $14.22\pm0.10$  & $44.0\pm3.9$ &  $\leq1.16$ & $\leq4.99$ & $\leq19.21$\\
121 &	0.32478 & 	1610.496 & $13.24\pm0.15$ & $46.1\pm10.2$&    $\leq1.28$ & $\leq5.07$ & $\leq18.31$\\
87  &	0.15136 &	1399.668 & $13.32\pm0.09$ & $65.3\pm5.5$ &    $\leq2.56$ &  $\leq5.65$ & $\leq18.97$\\
106 &	0.24126 &	1508.963 & $13.41\pm0.09$ & $89.1\pm6.9$ &    $\leq4.76$ &  $\leq6.12$ & $\leq19.53$\\
91  &	0.17891 &	1433.167 & $13.29\pm0.10$ & $98.5\pm9.1$ &    $\leq5.83$ &  $\leq6.26$ & $\leq19.55$\\
\hline
\multicolumn{8}{c}{Broad Ly\,$\alpha$ systems with O\,{\sc vi} detection}\\
\hline
32  &	0.00229 &	1218.449 & $13.57\pm0.10$ & $42.1\pm4.4$ &    $\leq1.07$ & $\leq4.91$ & $\leq18.48$\\
120 &	0.31978 &	1604.410 & $13.99\pm0.09$ & $74.3\pm11.5$ &   $\leq3.32$ & $\leq5.91$ & $\leq19.90$\\
\enddata

\tablenotetext{a}{Broad Ly\,$\alpha$ absorbers (including one Ly\,$\beta$ line)
without evidence for blending and sub-component structure are listed.}
\tablenotetext{b}{A limit for $T$ is listed assuming the line broadening is dominated by
thermal Doppler broadening.}
\tablenotetext{c}{Log $f_{\rm H}=$ log (H$^+/$H$^0$) assuming the gas
is in collisional ionization equilibrium at the estimated temperature.}
\tablenotetext{d}{Estimated total hydrogen column density in each system.}
\tablenotetext{e}{Broad Ly\,$\beta$ line. Corresponding Ly\,$\alpha$ absorption
is redwards of the available STIS wavelength range.} 
\end{deluxetable}

\clearpage
\newpage
\begin{deluxetable}{lrllrrrr}
\tabletypesize{\small}
\tablecaption{Galaxies and Galaxy Groups in the General Direction of PG\,1259+593$^{\rm a}$
\label{tbl-2}}
\tablewidth{0pt}
\tablehead{
\colhead{ID} & \colhead{Type$^{\rm b}$} & \colhead{$\alpha$ (2000)} & \colhead{$\delta$ (2000)}
& \colhead{Ang.\,Sep.} & \colhead{$z_{\rm gal}$} &
\colhead{$z_{\rm near. abs.}$}  & \colhead{$\rho_{75}$}\\ 
\colhead{} & \colhead{} & \colhead{} & \colhead{} & \colhead{[arcmin]} & 
\colhead{} & \colhead{} & \colhead{[kpc]}
}
\startdata
PG\,1259+593             & QSO  & 13 01 12.9  & +59 02 06  & \nodata & 0.47780 & \nodata & \nodata \\
1259+5920 (+0270$-$0313) & G    & 13 01 16.4  & +59 01 35  &  0.7    & 0.19670 & 0.19620 &  120\\
1259+5920 ($-$0234+0685) & G    & 13 01 09.8  & +59 03 15  &  1.2    & 0.24120 & 0.24126 &  248\\
CGCG 294-006             & G    & 12 59 26.1  & +59 01 06  & 13.8    & 0.04683 & 0.04606 &  701\\
UGC 08146                & G    & 13 02 08.1  & +58 42 05  & 21.3    & 0.00224 & 0.00229 &  55\\
Mrk 0233                 & G    & 12 58 29.7  & +59 07 57  & 21.8    & 0.02762 & 0.02217 &  671\\
SBS 1256+596             & G    & 12 58 25.8  & +59 21 53  & 29.1    & 0.02840 & 0.02217 &  922\\
Mrk 0232                 & G    & 12 57 17.5  & +59 04 02  & 30.3    & 0.02152 & 0.02217 &  735\\
SBS 1304+594             & G    & 13 06 07.1  & +59 13 03  & 39.3    & 0.03219 & 0.03924 &  1403\\
CGCG 294$-$015           & G    & 13 06 36.9  & +58 45 51  & 44.9    & 0.02805 & 0.02217 &  1405\\
FGC 1524                 & G    & 12 55 47.1  & +59 21 52  & 46.1    & 0.04295 & 0.04606 &  2165\\ 
UGC 08046                & G    & 12 55 04.0  & +58 47 25  & 49.8    & 0.00858 & 0.00760 &  491\\
CGCG 239$-$042           & G    & 12 54 49.8  & +58 52 56  & 50.2    & 0.00861 & 0.00760 &  496\\
WBL 425                  & GGr  & 12 54 38.0  & +58 49 45  & 52.4    & 0.00850 & 0.00760 &  512\\
UGC 08040                & G    & 12 54 43.5  & +58 46 36  & 52.6    & 0.00839 & 0.00760 &  507\\
SBS 1252+591             & G    & 12 54 22.5  & +58 53 42  & 53.6    & 0.00828 & 0.00760 &  509\\
Mahtessian 185           & GGr  & 12 54 56.2  & +58 32 46  & 56.9    & 0.00953 & 0.00760 &  622\\ 
\enddata

\tablenotetext{a}{Based on a search in the NED data archive available at
{\tt http://nedwww.ipac.caltech.edu} with a search radius of $60\farcm0$ centered on
PG\,1259+593.}
\tablenotetext{b}{QSO = quasar; G = galaxy; GGr = galaxy group.}
\end{deluxetable}


\begin{thebibliography}{}

\bibitem[]{}
Allende Prieto, C., Lambert,  D. L.,  \& Asplund, M. 2001, \apj, 556, L63
\bibitem[]{}
Allende Prieto, C., Lambert,  D. L.,  \& Asplund, M. 2002, \apj, 573, L137
\bibitem[]{}
Anders F., \& Grevesse, N. 1989, Geochim.\,Cosmochim.\,Acta, 53, 197
\bibitem[]{}
Bahcall, J.N., et al.\,1993, \apjs, 87, 1
\bibitem[]{}
Bergeron, J., Aracil, B., Petitjean, P., \& Pichon, C. 2002, \aap, 396, L11
\bibitem[]{}
Bowers, C.W., et al.\,1998, Proc.\,SPIE, 3356, 401
\bibitem[]{}
Carswell, R., Schaye, J., \& Kim, T.-S. 2002, \apj, 578, 43
\bibitem[]{}
Cen, R., \& Ostriker, J. 1999, \apj, 514, 1
\bibitem[]{}
Cen, R., Tripp, T.M., Ostriker, J.P., \& Jenkins, E.B. 2001, \apj, 559, L5
\bibitem[]{}
Chen, H.-W., \& Prochaska, J.X. 2000, \apj, 543, L9
\bibitem[]{}
Chen, H.-W., Lanzetta, K.M., Webb, J.K., \& Barcons, X. 2001, \apj, 559, 654
\bibitem[]{}
Chen, X., Weinberg, D.H., Katz, N., \& Dav\'e, R. 2003, \apj, 594, 42
\bibitem[]{}
Collins, J.A., Shull, J.M., \& Giroux, M.L. 2002, \apj, 585, 336
\bibitem[]{}
Dav\'e, R., \& Tripp, T.M. 2001, \apj, 553, 528
\bibitem[]{}
Fang, T., \& Bryan, G.L. 2001, \apj, 561, L31
\bibitem[]{}
Fang, T., Marshall, H.L., Lee, J.C., Davis, D.S., \& Canizares, C.R. 2002,
\apj, 572, L127
\bibitem[]{}
Ferland, G. 1996, Hazy, a Brief Introduction to CLOUDY 90, Univ. Kentucky Phys. Dept. Int. Rep.
\bibitem[]{}
Fox, A.J., Savage, B.D., Wakker, B.P., Richter, P., Sembach K.R., \& Tripp, T.M. 2004, \apj, 602, 738
\bibitem[]{}
Haardt, F. \& Madau, P. 1996, ApJ, 461, 20
\bibitem[]{}
Janknecht, E., Baade, R., \& Reimers, D. 2002, \aap, 391, L11
\bibitem[]{}
Kim, T.-S., Carswell, R.F., Cristiani, S., D'Odorico, S., \& Giallongo, E. 2002, MNRAS, 335, 555
\bibitem[]{}
Kimble, R.A., et al.\,1998, \apj, 492, L83
\bibitem[]{}
Landsman, W., \& Bowers, C.W. 1997, in HST Calibration Workshop with
a new generation of Instruments, ed. S. Casertano, R. Jedrzejewski, C.D.
Keyes, \& M. Stevens (Baltimore: STScI), 132
\bibitem[]{}
Leitherer, C., et al.\,2002, STIS Instrument Handbook, v6.0, (Baltimore: STScI)
\bibitem[]{}
Lu, L., Sargent, W.L.W., Womble, D.S., \& Takada-Hidai, M. 1996, \apj, 472, 509
\bibitem[]{}
Moos, H.W., et al. 2000, \apj, 538, L1
\bibitem[]{}
Morton, D.C. 1991, \apjs, 77, 119
\bibitem[]{}
Morton, D.C. 2003, in preparation
\bibitem[]{}
Nicastro, F., et al.\,2002, \apj, 573, 157
\bibitem[]{}
Oegerle, W.R., et al.\,2000, \apj, 538, L23
\bibitem[]{}
Penton, S.V., Shull, J.M., \& Stocke, J.T. 2000, \apj, 544, 150
\bibitem[]{}
Rasmussen, A., Kahn, S.M., \& Paerels, F. 2003, astro-ph/0301183 
\bibitem[]{}
Rauch, M., et al.\,1997, \apj, 489, 7
\bibitem[]{}
Rauch, M. 1998, ARA\&A, 36, 267
\bibitem[]{}
Rhee, M.-H., \& van Albada, T.S. 1996, \aaps, 115, 407
\bibitem[]{}
Richter, P., Savage, B.D., Wakker, B.P., Sembach, K.R., \& Kalberla, P.M.W. 2001a, \apj, 549, 281
\bibitem[]{}
Richter, P., et al. 2001b, \apj, 559, 318
\bibitem[]{}
Sahnow, D.J., et al. 2000, \apj, 538, L7
\bibitem[]{}
Savage, B.D., \& Sembach, K.R. 1991, \apj, 379, 245
\bibitem[]{}
Savage, B.D., Sembach, K.R., Tripp, T.M., \& Richter, P. 2002, \apj, 564, 631
\bibitem[]{}
Sembach, K.R., Howk, J.C., Savage, B.D., Shull, J.M., \& Oegerle, W.R. 2001,
\apj, 561, 573
\bibitem[]{}
Sembach, K.R., et al.\,2004, \apjs, 150, 387
\bibitem[]{}
Shull, J. M. 2003, in The IGM/Galaxy Connection: The Distribution of
Baryons at $z=0$, eds. J. L. Rosenberg and M. E. Putman (Dordrecht:
Kluwer Academic Pub.), 1
\bibitem[]{}
Stil, J.M., \& Israel, F.P. 2002, \aap, 389, 29
\bibitem[]{}
Stocke, J.T., Shull, J.M., Penton, S., Donahue, M., \& Carilli, C. 1995, \apj,
451, 24
\bibitem[]{}
Sutherland, R.S., \& Dopita, M.A. 1993, \apjs, 88, 253
\bibitem[]{}
Tripp, T.M., Savage, B.D., \& Jenkins, E.B. 2000, \apj, 534, L1
\bibitem[]{}
Tripp, T.M., \& Savage, B.D. 2000, \apj, 542, 42
\bibitem[]{}
Tripp, T.M., Giroux, M.L., Stocke, J.T., Tumlinson, J., \& Oegerle, W.R.
2001, \apj, 563, 724
\bibitem[]{}
Tripp, T.M., et al.\,2002, \apj, 575, 697 
\bibitem[]{}
Tucker, D.L., et al.\,2002, \apjs, 130, 237
\bibitem[]{}
Verner, D., Barthel, P., \& Tytler, D. 1994, \aaps, 108, 287
\bibitem[]{}
Weinberg, D.H., Miralda-Escud\'e, J., Hernquist, L., \& Katz, N. 1997,
\apj, 490, 564
\bibitem[]{}
Weymann, R.J., Jannuzi, B.T., Lu, L., et al.\,1998, \apj, 506, 1
\bibitem[]{}
Woodgate, B.E., et al.\,1998, PASP, 110, 1183
\end{thebibliography}
\end{document}